\documentclass[final]{siamart1116}
\usepackage{etex}
\usepackage[spanish,english]{babel}
\usepackage{amsfonts} 
\usepackage{amsmath}
\usepackage{amssymb}
\usepackage{booktabs}
\usepackage{epstopdf}
\usepackage{graphicx}
\graphicspath{{./Figures/}}
\usepackage{subfigure}
\usepackage{colortbl}
\usepackage{pst-all}
\usepackage{wasysym}
\usepackage{mathrsfs}
\usepackage{float}
\usepackage{verbatim}
\usepackage{upgreek}
\usepackage{todonotes}
\usepackage{color}
\usepackage[applemac]{inputenc}

\usepackage{array}
\newcolumntype{L}[1]{>{\raggedright\let\newline\\\arraybackslash\hspace{0pt}}m{#1}}
\newcolumntype{C}[1]{>{\centering\let\newline\\\arraybackslash\hspace{0pt}}m{#1}}
\newcolumntype{R}[1]{>{\raggedleft\let\newline\\\arraybackslash\hspace{0pt}}m{#1}}

\usepackage{tikz}
\usetikzlibrary{arrows,shapes,backgrounds}

\newcommand{\pgftextcircled}[1]{                                                                    
    \setbox0=\hbox{#1}%
    \dimen0\wd0%
    \divide\dimen0 by 2%
    \begin{tikzpicture}[baseline=(a.base)]%
        \useasboundingbox (-\the\dimen0,0pt) rectangle (\the\dimen0,1pt);
        \node[circle,draw,outer sep=0ex,inner sep=0.1ex] (a) {#1};
    \end{tikzpicture}
}

\let\oldsqrt\sqrt
\def\sqrt{\mathpalette\DHLhksqrt}
\def\DHLhksqrt#1#2{%
\setbox0=\hbox{$#1\oldsqrt{#2\,}$}\dimen0=\ht0
\advance\dimen0-0.2\ht0
\setbox2=\hbox{\vrule height\ht0 depth -\dimen0}%
{\box0\lower0.4pt\box2}}

\newcommand{\vectorr}[1]{\mathbf{#1}}
\newcommand{\vecol}[2]{\left( \begin{array}{c} 
\displaystyle#1 \\
\\[-1.7ex]
\displaystyle#2
\end{array}\right)}
\newcommand{\mados}[4]{\left(\begin{array}{cc}
\displaystyle#1 &\displaystyle #2 \\
\\[-1.7ex]
\displaystyle#3 & \displaystyle#4
\end{array}\right)}

\newcommand{\bsub}{\begin{subequations}}
\newcommand{\esub}{\end{subequations}$\!$}
\renewcommand{\theequation}{\arabic{section}.\arabic{equation}}
\renewcommand{\thesection}{\arabic{section}}

\newcommand{\eps}{\varepsilon}
\newcommand{\lam}{\lambda}

\DeclareMathOperator{\diag}{diag}

\graphicspath{{Figures/}}

\newcommand{\TheTitle}{Spot dynamics in a reaction-diffusion model of plant root hair
initiation}

\newcommand{\TheAuthors}{D. Avitabile, V.~F.~Bre\~na--Medina, M.~J.~Ward}
\headers{Spot dynamics in a reaction--diffusion model}{\TheAuthors}

\title{{\TheTitle}}

\author{
  Daniele Avitabile
  \thanks{Centre for Mathematical Medicine and Biology, School of
  Mathematical Sciences, University of Nottingham, University Park, Nottingham, NG7
  2RD, 
    \email{daniele.avitabile@nottingham.ac.uk},
    ORCID ID \url{http://orcid.org/0000-0003-3714-7973}.
  }
  \and
  Victor F.~Bre\~na--Medina
  \thanks{Centro de Ciencias Matem\'aticas, UNAM, Morelia, Michoac\'an, M\'exico. Present address: Departamento Académico de Matemáticas, ITAM, Ciudad de México, México, 
  \email{victor.brena@itam.mx}.
  }
  \and
  Michael J.~Ward
  \thanks{Department of Mathematics, UBC, Vancouver, Canada,
  \email{ward@math.ubc.ca}.
  }
}

\ifpdf
\hypersetup{pdftitle={\TheTitle},pdfauthor={\TheAuthors}}
\fi

\begin{document}
\maketitle

  
\begin{abstract}
  We study pattern formation in a 2-D reaction-diffusion (RD)
  sub-cellular model characterizing the effect of a spatial gradient
  of a plant hormone distribution on a family of G-proteins associated
  with root-hair (RH) initiation in the plant cell \emph{Arabidopsis
    thaliana}. The activation of these G-proteins, known as the Rho of
  Plants (ROPs), by the plant hormone auxin, is known to promote
  certain protuberances on root hair cells, which are crucial for both
  anchorage and the uptake of nutrients from the soil. Our
  mathematical model for the activation of ROPs by the auxin gradient
  is an extension of the model of Payne and Grierson [PLoS
    ONE, {\bf 12}(4), (2009)], and consists of a two-component
  Schnakenberg-type RD system with spatially heterogeneous
  coefficients on a 2-D domain. The nonlinear kinetics in this RD
  system model the nonlinear interactions between the active and
  inactive forms of ROPs. By using a singular perturbation analysis to
  study 2-D localized spatial patterns of active ROPs, it is shown
  that the spatial variations in the nonlinear reaction kinetics,
  due to the auxin gradient, lead to a slow spatial
  alignment of the localized regions of active ROPs along the
  longitudinal midline of the plant cell. Numerical bifurcation
  analysis, together with time-dependent numerical simulations of the
  RD system are used to illustrate both 2-D localized patterns in the
  model, and the spatial alignment of localized structures.
\end{abstract}


\section[Introduction]{Introduction} 
\label{int}

We examine the effect of a spatially-dependent plant hormone
distribution on a family of proteins associated with root hair (RH)
initiation in a specific plant cell. This process is modeled by a
generalized Schnakenberg reaction-diffusion (RD) system on a 2-D
domain with both source and loss terms, and with a spatial gradient
modeling the spatially inhomogeneous distribution of the plant hormone
auxin. This system is an extension of a model proposed by Payne and
Grierson in~\cite{payne01}, and analyzed in a 1-D context in the
companion articles~\cite{bacw,bcwg}. The new goal of this
  paper, in comparison with \cite{bacw,bcwg}, is to analyze 2-D
  localized spot patterns in the RD system~\cref{eq:spotsystem}, and
  how these 2-D patterns are influenced by the spatially inhomogeneous
  auxin distribution.

\begin{figure}[t!]
 \begin{center}
 \centering
 \includegraphics[height=0.23\textheight]{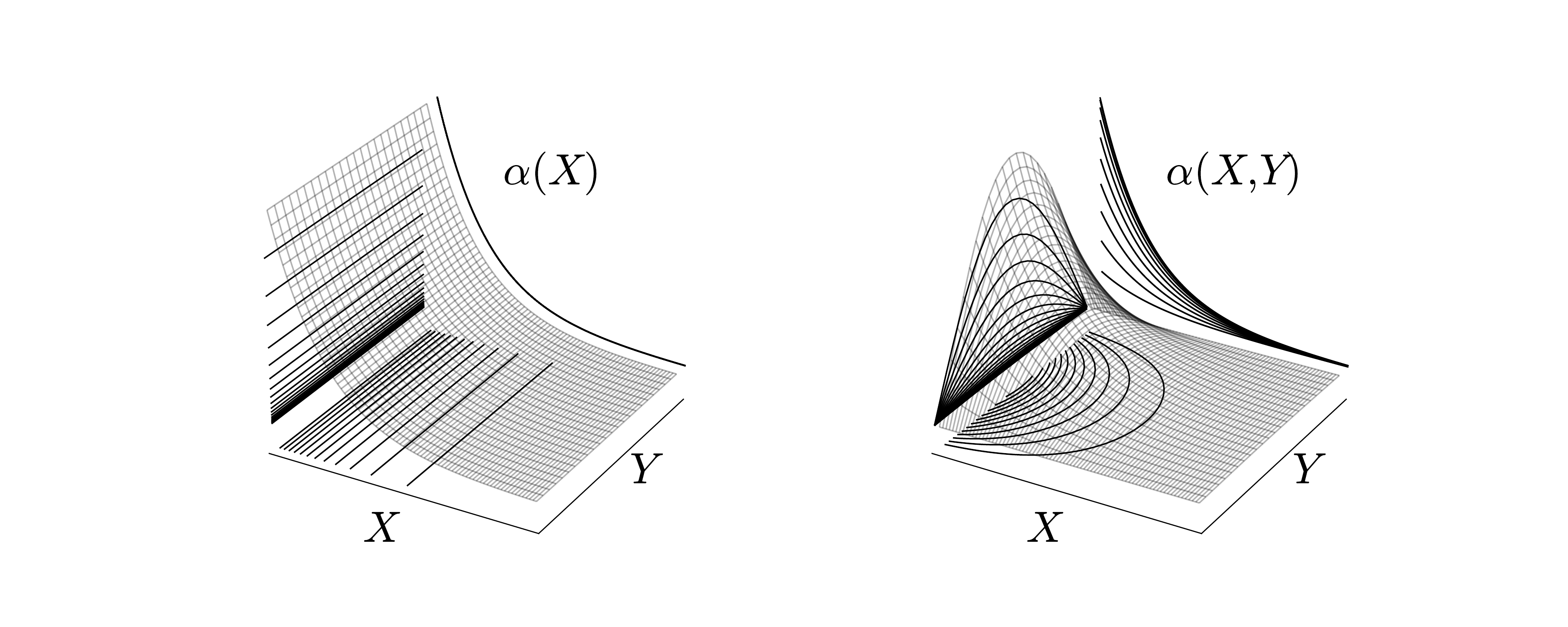}
	\end{center}
	\caption{Sketches of the spatially-dependent auxin gradient
          forms: constant in the transverse direction (left-hand
          panel), and decreasing on either side of the middle of the
          cell (right-hand panel).}
	\label{fig:auxinflows}
\end{figure}

We now give a brief description of the biology underlying the RD
model.  In this model, an on-and-off switching process of a small
G-protein subfamily, called the Rho of Plants (ROPs), is assumed to
occur in a RH cell of the plant {\em Arabidopsis thaliana}. ROPs are
known to be involved in RH cell morphogenesis at several distinct
stages (see \cite{dolan01,mjones02} for details). Such a biochemical
process is believed to be catalyzed by a plant hormone called auxin
(cf.~\cite{payne01}). Typically, auxin-transport models are formulated
to study polarization events between cells
(cf.~\cite{Draelants:2015kv}). However, little is known about specific
details of auxin flow within a cell. In \cite{bacw,bcwg,payne01} a
simple transport process is assumed to govern the auxin flux through a
RH cell, which postulates that auxin diffuses much faster than ROPs in
the cell, owing partially to the in- and out-pump mechanism that
injects auxin into RH cells from both RH and non-RH cells
(cf.~\cite{grien01,heisler01}). Recently, in \cite{krupinski01}, a
model based on the one proposed in \cite{payne01} has been used to
describe patch location of ROP GTPases activation along a 2-D root
epidermal cell plasma membrane. This model is formulated as a
two-stage process, one for ROP dynamics and another for auxin
dynamics. The latter process is assumed to be described by a constant
auxin production at the source together with passive diffusion and a
constant auxin bulk degradation away from the source. In
\cite{krupinski01} the auxin gradient is included in the ROP
finite-element numerical simulation after a steady-state is
attained. Since we are primarily interested in the biochemical
interactions that promote RHs and key ingredients that geometrical
features have on the RHs initiation dynamics, rather than providing a
detailed model of auxin transport in the cell, we shall hypothesize
specific, but qualitatively reasonable in the sense described below,
time-independent spatially varying expressions for the auxin
distribution in a 2-D azimuthal projection of a 3-D RH cell wall. In
this way, we capture crucial ingredients of the
spatially-dependent auxin distribution in the cell. Qualitatively,
since a RH cell is an epidermal cell, influx and efflux biochemical
gates are distributed along the cell membrane, and the auxin flux is
known to be primarily directed from the tip of the cell towards the
elongation zone (cf.~\cite{grien01,jones01}), leading to a
longitudinally spatially decreasing auxin distribution. Our specific
form for the auxin distribution will incorporate such a longitudinal
spatial dependence. However, as an extension to the auxin model
developed in~\cite{bacw,bcwg}, here we also can take into account
non-RH lateral contributions of auxin flux in RH cells 
by considering an auxin distribution with a transverse spatial dependence. This
assumption arises as non-RH cells are believed to promote auxin
transversal flux into RH cells. However, although auxin is believed to
follow pathways either through the cytoplasm or through cell walls of
adjacent cells, we are primarily interested in the former pathway that
is directed within the cell, and which induces the biochemical
switching process (cf.~\cite{mitchinson}). In 
this new 2-D analysis of
the RD model, we will allow for an auxin distribution that has an
arbitrary spatial dependence. However, in illustrating our 2-D theory,
and motivated by the biological framework discussed above, we will
focus on two specific forms for the auxin distribution, as shown in
\cref{fig:auxinflows}: a 1-D form that has the spatially
decreasing longitudinal dependence of \cite{bacw,bcwg}, and a
modification of this 1-D form that allows for a transverse dependence.
We will show that these specific auxin gradients induce 
an
  alignment of localized regions of elevated ROP concentration,
  referred to here as spots. The RD system under consideration exhibits spot-pinning
  phenomena, in the sense that the position of spots in the stationary patterns is
  determined by the spatial gradient of auxin. As we shall see below, spots slowly
  drift until they become aligned with the longitudinal axis of the cell. Their
  steady-state locations along this cell-midline are at locations determined by the
  auxin gradient, which effectively ``pins'' the spots. A similar pinning phenomenon
  is found for spatially localized structures in other systems: an example is the
  vortex-pinning phenomenon in the Ginzburg--Landau model of superconductivity in an
  heterogeneous medium, see~\cite{aftalion}.
  
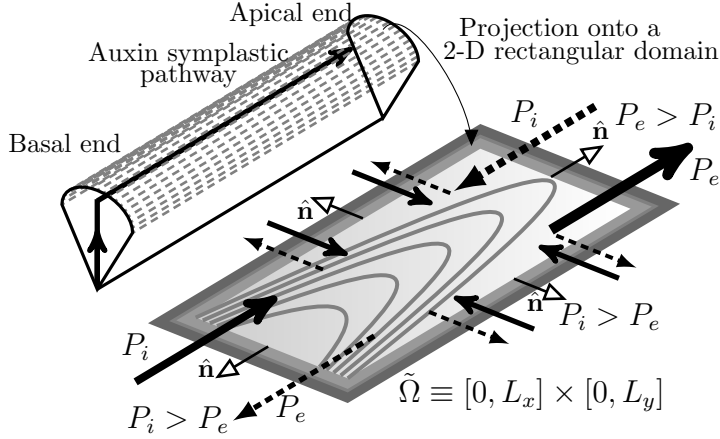
\begin{figure}[t!]
\begin{center}
\begin{tikzpicture}[scale=0.47]
	\begin{scope}[yshift=-180,yslant=0.6,xslant=-1]
		\fill [join=square,black!60] (-3.2,-3.2)  rectangle (6.2,2.45);
		\fill [join=square,black!50] (-2.9,-2.9)  rectangle (5.9,2.15);
		\fill [join=square,black!40] (-2.8,-2.8)  rectangle (5.8,2.05);
		\shade [right color = white, left color = black!20, join=square] (-2.5,-2.5)  rectangle (5.5,1.75);
		\path
		coordinate (start1) at (-2.5,1.8)
		coordinate (c11) at +(8,0)
		coordinate (c21) at +(8,-0.75)
		coordinate (slut1) at (-2.5,-2.5);
		\draw[thick,color=gray,line width=1.5pt] (start1) .. controls (c11) and (c21) .. (slut1);
		\path
		coordinate (start2) at (-2.5,1.6)
		coordinate (c12) at +(6,0)
		coordinate (c22) at +(6,-0.75)
		coordinate (slut2) at (-2.5,-2.3);
		\draw[thick,color=gray,line width=1.5pt] (start2) .. controls (c12) and (c22) .. (slut2);
		\path
		coordinate (start3) at (-2.5,1.4)
		coordinate (c13) at +(4,0)
		coordinate (c23) at +(4,-0.75)
		coordinate (slut3) at (-2.5,-2.1);
		\draw[thick,color=gray,line width=1.5pt] (start3) .. controls (c13) and (c23) .. (slut3);
		\path
		coordinate (start4) at (-2.5,1.2)
		coordinate (c14) at +(2,0)
		coordinate (c24) at +(2,-0.75)
		coordinate (slut4) at (-2.5,-1.9);
		\draw[thick,color=gray,line width=1.5pt] (start4) .. controls (c14) and (c24) .. (slut4);
		\path
		coordinate (start5) at (-2.5,1)
		coordinate (c15) at +(0,0)
		coordinate (c25) at +(0,-0.75)
		coordinate (slut5) at (-2.5,-1.7);
		\draw[thick,color=gray,line width=1.5pt] (start5) .. controls (c15) and (c25) .. (slut5);
		\draw (5.5,-0.25) -- (7,-0.25) [-open triangle 45,thick,line width=1pt]; 
		\node at (7.4,0.25) {\textcolor{black}{$\hat{\vectorr n}$}};
		\draw (-4,-0.25) -- (-2.5,-0.25) [open triangle 45-,thick,line width=1pt]; 
		\node at (-3.8,0.2) {\textcolor{black}{$\hat{\vectorr n}$}};
		\draw (2.2,1.77) -- (2.2,3.27) [-open triangle 45,thick,line width=1pt]; 
		\node at (1.8,3) {\textcolor{black}{$\hat{\vectorr n}$}};
		\draw (2.2,-2.5) -- (2.2,-4) [-open triangle 45,thick,line width=1pt]; 
		\node at (1.8,-3.5) {\textcolor{black}{$\hat{\vectorr n}$}};
		\draw (-1,1) -- (-5,1) [stealth'-,thick,color=black,line width=3pt]; 
		\node at (-4,2) {\textcolor{black}{\large $P_i$}};
		\draw (8,1) -- (4,1) [-latex,thick,densely dashed,color=black,line width=3pt]; 
		\node at (7,2) {\textcolor{black}{\large $P_i$}};
		\draw (3.4,3.27) -- (3.4,1) [-stealth',thick,color=black,line width=2pt]; 
		\draw (1,3.27) -- (1,1) [-stealth',thick,color=black,line width=2pt]; 
		\draw (0.3,1) -- (0.3,3.27) [-latex,thick,densely dashed,color=black,line width=1.5pt];
		\draw (3.9,1) -- (3.9,3.27) [-latex,thick,densely dashed,color=black,line width=1.5pt];
		\draw (3.4,-4.27) -- (3.4,-2) [-stealth',thick,color=black,line width=2pt]; 
		\draw (1,-4.27) -- (1,-2) [-stealth',thick,color=black,line width=2pt];
		\draw (0.3,-2) -- (0.3,-4.27) [-latex,thick,densely dashed,color=black,line width=1.5pt];
		\draw (3.9,-2) -- (3.9,-4.27) [-latex,thick,densely dashed,color=black,line width=1.5pt];
		\draw (-1,-1.75) -- (-5,-1.75) [-latex,densely  dashed,color=black,line width=2pt]; 
		\node at (-4,-2.4) {\large $P_e$};
		\draw (4,-1.75) -- (8,-1.75) [-stealth',thick,color=black,line width=4pt]; 
		\node at (7.4,-2.7) {\large $P_e$};
		\node at (8.5,-0.5) {\large $P_e>P_i$}; 
		\node at (-5.6,-0.8) {\large $P_i>P_e$}; 
		\node at (2.2,-5.2) {\large $P_i>P_e$}; 
		\node at (-0.8,-6) {\large $\tilde{\Omega}\equiv[0,L_x]\times[0,L_y]$};
	\end{scope}
	\begin{scope}[yshift=-30,yslant=0.6,yslant=-1]
		\def\h{3.5}
		\draw[thick,color=black,line width=2pt] (1,0.1,6*\h) -- (1,2.55,6*\h);
		\draw[-stealth',thick,color=black,line width=2pt] (1,0.1,6*\h) -- (1,1.7,6*\h);
		\draw[-stealth',thick,color=black,line width=2pt] (1,2.5,6*\h) -- (1,2.5,0.64*\h);
		\foreach \t in {0,10,...,180}
			\draw[gray,densely dashed,line width=1pt] ({1+cos(\t)},{2+0.8*sin(\t)},0)
		--({1+cos(\t)},{2+0.8*sin(\t)},{6*\h});
			\draw[black,very thick] (1,0,0) 
		\foreach \t in {0,5,...,180}
			{--({1+cos(\t)},{2+0.8*sin(\t)},0)}--cycle;
			\draw[black,very thick] (1,0,{6*\h}) 
		\foreach \t in {0,10,...,180}
			{--({1+cos(\t)},{2+0.8*sin(\t)},{6*\h})}--cycle;
			\draw[black,very thick] (1,0,6*\h) -- (1,0,0); 
		\node at (-1.6,1.9) {\textcolor{black}{Apical end}};
		\node at (-8,-4.3) {\textcolor{black}{Basal end}};
		\node at (-4.5,-0.4) {\textcolor{black}{Auxin symplastic}};
		\node at (-4.5,-1) {\textcolor{black}{pathway}};
		\draw[-latex, -triangle 45]
		(1,2.5) to[out=60,in=90] (3.5,0.2);
		\node at (6,4.5) {\textcolor{black}{Projection onto a}};
		\node at (6.6,4.1) {\textcolor{black}{2-D rectangular domain}};
	\end{scope}
\end{tikzpicture}
\end{center}
\caption{Sketch of an idealized 3-D RH cell with longitudinal and
  transversal spatially-dependent auxin flow. The auxin gradient is
  shown as a consequence of in- and out-pump mechanisms. Influx $P_i$
  and efflux $P_e$ permeabilities are respectively depicted by the
  direction of the arrow; the auxin symplastic pathway, indicated by
  black solid arrows in the 3-D RH cell, is the auxin pathway between
  cells. Here, the cell membrane is depicted by heavily dashed gray
  lines. The gradient is coloured as a light gray shade, and auxin
  gradient level curves are plotted in gray. Cell wall and cell
  membrane are coloured in dark and light gray respectively. Modified
  figure reproduced from~\cite{bcwg}.}
\label{fig:auxincell3D}
\end{figure}

In \cref{fig:auxincell3D}, a sketch of an idealized 3-D RH cell is
shown. In this figure, heavily dashed gray lines represent the RH cell
membrane. The auxin flux is represented by black arrows, which
schematically illustrates a longitudinal decreasing auxin pathway
through adjacent cells. As ROP activation is assumed to occur near the
cell wall and transversal curvature of the domain is not included in
our model, we consider a projection onto a 2-D rectangular domain of
this cell, as is also shown in \cref{fig:auxincell3D}.

In \cite{bcwg} we considered the 1-D version of the RD model given in
\cref{eq:spotsystem}, where the RH cell is assumed to be slim and
flat, and where the auxin distribution depends only on the
longitudinal spatial direction. There it was shown that 1-D localized
steady-state patterns, representing active-ROP proteins, slowly drift
towards the interior of the domain and asymmetrically get pinned to
specific spatial locations by the auxin gradient. This pinning effect,
induced by the spatial distribution of auxin, has been experimentally
observed in, for instance,~\cite{dolan01,fischer01}. Biologically,
although multiple swelling formation in transgenic RH cells may occur,
not all of these will undergo the transition to tip growth
(cf.~\cite{mjones01}). In \cite{bcwg} a linear stability analysis
showed that multiple-active-ROP 1-D spikes may be linearly stable on
$\mathcal{O}(1)$ time-scales to 1-D perturbations, with the stability
properties depending on an inversely proportional relationship between
length and auxin catalytic strength. This dynamic phenomenon is a
consequence of an auxin catalytic process sustaining cell wall
softening as the RH cell grows. For further discussion of this
correlation between growth and biochemical catalysis see~\cite{bcwg}.

In \cite{bacw} the linear stability of 1-D stripe patterns of elevated
ROP concentration to 2-D spatially transverse perturbations was
analyzed.  There it was shown that single interior stripes and
multi-stripe patterns are unstable under small transverse
perturbations.  The loss of stability of the stripe pattern was shown
to lead to more intricate patterns consisting of spots, or a boundary
stripe together with multiple spots (see Figure~9
of~\cite{bacw}). Moreover, it was shown that a single boundary stripe
can also lose stability and lead to spot formation in the small
diffusivity ratio limit. 

Our main new goal in this paper is to analyze these 2-D spot patterns,
which results from the breakup of a stripe, and characterize their
slow evolution towards a true steady-state of the RD
system. In particular, we will examine 
how specific forms of the
auxin gradient lead to a steady-state spot pattern where the spots are
aligned with the direction of the auxin gradient.
  
In the
RD model below, $U(\vectorr X,T)$ and $V(\vectorr X,T)$ denote
active- and inactive-ROP densities, respectively, at time $T>0$ and
point $\vectorr X=(X,Y) \in\tilde\Omega\subset\mathbb R^2$, where
$\tilde{\Omega}$ is the rectangular domain
$\tilde{\Omega}=[0,L_x]\times[0,L_y]$. The on-and-off switching
interaction is assumed to take place on the cell membrane, which is
coupled through a standard Fickian-type diffusive process for both
densities (cf.~\cite{bcwg,payne01}). Although RH cells are flanked by
non-RH cells (cf.~\cite{griersonRH}), there is in general no exchange
of ROPs between them. Therefore, no-flux boundary conditions are
assumed on $\partial\tilde\Omega$. On the other hand, we will consider
two specific forms for the spatially-dependent dimensionless auxin
distribution $\alpha$: (i)~a steady monotone decreasing longitudinal
gradient, which is constant in the transverse direction, and (ii)~a
steady monotone decreasing longitudinal gradient, which decreases on
either side of the midline $Y={L_y/2}$; see the left- and right-hand
panels in \cref{fig:auxinflows}, respectively. These two forms are
modelled by
\begin{gather}\label{mod:aux}
    \textrm{(i)} \,\,\, \alpha(\vectorr X)= \exp\left(-\frac{\nu X}{L_x}\right)\,, \quad \textrm{(ii)} \,\,\, \alpha(\vectorr X)= 
\exp\left(-\frac{\nu X}{L_x}\right)\sin\left(\frac{\pi Y}{L_y}\right)\,, 
\quad \nu=1.5\,.
\end{gather}

As formulated in \cite{bacw} and \cite{bcwg}, the basic dimensionless
RD model is
\begin{subequations}\label{eq:spotsystem}
  \begin{align}
    U_t & = \varepsilon^2\Delta U  +\alpha\left(\vectorr x\right) U^2V- U + (\tau\gamma)^{-1} V, 
        & & \vectorr x\in \Omega\,, \label{eq:spotsystem1}\\
 \tau V_t & = D\Delta V -V+1-\tau\gamma\left(\alpha\left(\vectorr x\right) U^2V -U\right)-\beta\gamma U, 
        & & \vectorr x\in \Omega\,, \label{eq:spotsystem2} 
  \end{align}
\end{subequations}
with homogeneous Neumann boundary conditions, and where
$\Omega=[0,1]\times[0,d_y]$.  Here $\Delta$ is the 2-D Laplacian, and
the domain aspect ratio $d_y\equiv {L_y/L_x}$ is assumed to satisfy
$0<d_y<1$, which represents a cell as is shown in~\cref{fig:auxincell3D}. The other dimensionless
parameters in \cref{eq:spotsystem} are defined in terms of the
original parameters, as in~\cite{bacw,bcwg}, by
\begin{subequations}\label{eq:newpargabe}
 \begin{gather}\label{eq:newpar}
		\varepsilon^2 \equiv \frac{D_1}{L_x^2(c+r)}\,, 
\qquad D \equiv \frac{D_2}{L_x^2k_1} \,, \qquad 
\tau \equiv \frac{c+r}{k_1}\,, \qquad \beta \equiv\frac{r}{k_1}\,,
	\end{gather}
and the primary bifurcation parameter $\gamma$ in this system is given by
\begin{gather}\label{eq:gabe}
		\gamma \equiv \frac{(c+r)k_1^2}{k_{20} b^2} \,.
\end{gather}
\end{subequations}
In the original dimensional model of \cite{bcwg}, $D_1 \ll D_2$ are
the diffusivities for $U$ and $V$, $b$ is the rate of production of
inactive ROP (source parameter), $c$ is the rate constant for
deactivation, $r$ is the rate constant describing active ROPs being
used up in cell softening and subsequent hair formation (loss
parameter), and the activation step is assumed to be proportional to
$k_1V+k_{20}\alpha(\vectorr x)U^2V$. The activation and overall auxin
level within the cell, initiating an autocatalytic reaction, is
represented by $k_1>0$ and $k_{20}>0$ respectively.  The parameter
$k_{20}$, arising in the bifurcation parameter $\gamma$ of~\cref{eq:gabe}, will play a key role for pattern formation (see
\cite{bcwg,payne01} for more details on the model formulation).

We will examine the role that geometrical features, such as the 2-D
domain and the auxin-gradient, play on the dynamics of localized
regions, referred to as spots or patches, where the active-ROP has an elevated
concentration.  This analysis is an extension to the analysis
performed in \cite{bacw}, where it was shown that a 1-D localized
stripe pattern of active-ROP in a 2-D domain will, generically,
exhibit breakup into spatially localized spots. To analyze the
subsequent dynamics and stability of these localized spots in the
presence of the auxin gradient, we will extend the hybrid
asymptotic-numerical methodology developed in
\cite{kolok02,rozada02} for prototypical RD systems
with spatially homogeneous coefficients. This analysis will lead to a
novel finite-dimensional dynamical system characterizing slow spot
evolution. Although in our numerical simulations we will focus on the
two specific forms for the auxin gradient in \cref{mod:aux}, our
analysis applies to an arbitrary gradient.

The outline of the paper is as follows. In \cref{sec:spots} we
introduce the basic scaling for \cref{eq:spotsystem}, and we use
singular perturbation methods to construct an $N$-spot quasi
steady-state solution. In \cref{sec:dynspot} we derive a differential
algebraic system (DAE) of ODEs coupled to a nonlinear algebraic
system, which describes the slow dynamics of the centers of a
collection of localized spots. In particular, we explore the role that
the auxin gradient (i) in \cref{mod:aux} has on the ultimate spatial
alignment of the localized spots that result from the breakup of an
interior stripe. For the auxin gradient (i) of \cref{mod:aux}, in
\cref{sec:instab2D} we study the onset of spot self-replication
instabilities and other ${\mathcal O}(1)$ time-scale instabilities of
quasi steady-state solutions. In \cref{chap:c07} we examine localized
spot patterns for the more biologically realistic model auxin (ii) of
\cref{mod:aux}. To illustrate the theory, throughout the paper we
perform various numerical simulations and numerical bifurcation
analyses using the parameter sets in \cref{tab:tab}, which are
qualitatively close to the biological parameters
(see~\cite{bcwg,payne01}). Finally, a brief discussion is given in
\cref{sec:sum06}.

\begin{table}[t]
	\begin{center}
		\scalebox{0.97}{
		\begin{tabular}{L{2.5cm} L{3.4cm} | L{2.5cm} L{3.4cm}}
 & {\bf Parameter Set A} & & {\bf Parameter Set B} \\[0.2ex]
			\toprule
			{\bf\itshape Original} & {\bf\itshape Re-scaled} & {\bf\itshape Original} & {\bf\itshape Re-scaled} \\[0.2ex]
			\midrule
$D_1=0.075$ & $\varepsilon^2 =1.02\times10^{-4}$ & $D_1=0.1$ & 
  $\eps^2 =3.6\times10^{-4}$\\
$D_2 =20$ & $D=0.51$ &$D_2 =10$ & $D=0.4$ \\ 
$k_1 = 0.008$   &  $\tau = 18.75$ & $k_1 = 0.01$   &  $\tau = 11$\\ 
$b = 0.008$     &  $\beta = 6.25$ & $b = 0.01$     &  $\beta = 1$\\  
$c = 0.1$      &  & $c = 0.1$      &  \\
$r = 0.05$     &  & $r = 0.01$     &  \\ 
$k_{20} \in[10^{-3}, 2.934]$ &  $\gamma \in[ 150, 0.051]$ & 
$k_{20} \in [0.01, 1.0]$ &  $\gamma \in[11, 0.11]$\\
	$L_x= 70$ & & $L_x=50$ \\
	$L_y=29.848$ & $d_y=0.4211$ & $L_y=20$ & $d_y=0.40$\\[1ex]
			\bottomrule
		\end{tabular}
		}
	\end{center} 
\caption{Two parameter sets in the original and dimensionless
  re-scaled variables. The fundamental units of length and time are
  $\mu\textrm{m}$ and sec, and concentration rates are measured by an
  arbitrary datum (con) per time unit; $k_{20}$ is measured by
  $\textrm{con}^2/\textrm{s}$, and diffusion coefficients units are
  $\mu \textrm{m}/\textrm{s}^2$. }\label{tab:tab}
\end{table}


\section{Asymptotic Regime for Spot Formation}
\label{sec:spots}

We shall assume a shallow, oblong cell, which will be modelled as a 2-D
flat rectangular domain, as shown in~\cref{fig:auxincell3D}. In this domain, the biochemical
interactions leading to an RH initiation process are assumed to be
governed by the RD model \cref{eq:spotsystem}.

For a RD system, a spatial pattern in 2-D consisting of spots is
understood as a collection of structures where, at least, one of its
solution components is highly localized. These structures typically
evolve in such a manner that the spatial locations of the spots change
slowly in time. However, depending on the parameter regime, these
localized patterns can also exhibit fast ${\mathcal O}(1)$ time-scale
instabilities, leading either to spot creation or destruction.  For
prototypical RD systems such as the Gierer--Meinhardt, Gray--Scott,
Schnakenberg, and Brusselator models, the slow dynamics of localized
solutions and the possibility of self-replication and competition
instabilities, leading either to spot creation or destruction,
respectively, have been studied using a hybrid asymptotic-numerical
approach in \cite{kolo01,rozada02}.  In addition, in the large
inhibitor diffusion limit, a leading-order-in-${-1/\log\eps}$ analysis, shows
that the linear stability problem for localized spot patterns
characterizing competition instabilities can be reduced to the study
of classes of nonlocal eigenvalue problems (NLEPs)
(cf.~\cite{wei03,wei01}).  In a 1-D context, rigorous results
  for the spectral properties of ``far-from-equilibrium'' periodic 1-D
  patterns have been obtained recently in \cite{rijk01}.

Our previous studies in \cite{bcwg} of 1-D spike patterns and in
\cite{bacw} of 1-D stripe patterns have extended these previous NLEP
analyses of prototypical RD systems to the more complicated, and
biologically realistic, plant root-hair system~\cref{eq:spotsystem}.
In the analysis below we extend the 2-D hybrid asymptotic-numerical
methodology developed in \cite{kolo01,rozada02} for prototypical RD
systems to \cref{eq:spotsystem}. The primary new feature in
\cref{eq:spotsystem}, not considered in these previous works, is to
analyze the effect of a spatial gradient in the right-hand sides
of~\cref{eq:spotsystem}, which represents variations in the auxin
distribution. This spatial gradient is central to the spatial
alignment of the localized structures.

To initiate our hybrid approach for \cref{eq:spotsystem}, we first
need to rescale the RD model \cref{eq:spotsystem} in $\Omega$ in order
to determine the parameter regime where localized spots exist. To do
so, we integrate the steady-state version of the RD model
\cref{eq:spotsystem} over the domain to show, as in Proposition~2.2 of
\cite{bcwg} for the 1-D problem, that for any steady-state solution
$U_0(\vectorr x)$ of \cref{eq:spotsystem} we must have
\begin{gather}\label{eq:prom01}
  \int_{\Omega}U_0(\vectorr x)\;d\vectorr x =\frac{|\Omega|}{\beta\gamma}\,,
\end{gather}
where $|\Omega|=d_y$ is the area of $\Omega$. Since the parameters
$d_y$, $\beta$ and $\gamma$ are $\mathcal O(1)$, the constraint
\cref{eq:prom01} implies that the average of $U$ across the whole
domain is also ${\mathcal O}(1)$. In other words, if we seek localized
solutions such that $U\to 0$ away from localized ${\mathcal O}(\eps)$
regions near a collection of spots, then from \cref{eq:prom01} we
must have that $U=\mathcal O\left(\varepsilon^{-2}\right)$ near each
spot.

\subsection{A Multiple-Spot Quasi Steady-State Pattern}
\label{subsec:multispots}

In order to derive a finite-dimensional  dynamical system for
the slow dynamics of active-ROP spots, we must first construct a quasi
steady-state pattern describing ROP aggregation in $N$ small
distinct spatial regions.

To do so, we seek a quasi steady-state solution where $U$ is localized
within an $\mathcal O\left(\varepsilon\right)$ region near each spot
location $\vectorr x_j$ for $j=1,\ldots,N$. From the conservation law
\cref{eq:prom01} we get that $U=\mathcal
O\left(\varepsilon^{-2}\right)$ and, as a consequence,
$V=\mathcal{O}(\varepsilon^2)$ near each spot. In this way, if we
replace $U=\varepsilon^{-2} U_j$ near each spot, and define
$\boldsymbol \xi=\varepsilon^{-1}\left(\vectorr x-\vectorr
x_j\right)$, we obtain from \cref{eq:prom01} that
$\sum_{j=1}^N\int_{\mathbb R^2}U_j\;d\boldsymbol \xi \sim
{d_y/(\beta\gamma)}$.  This scaling law motivates the introduction of
the new variables $u$ and $v$ defined by
\begin{gather}\label{eq:newvar}
 U=\varepsilon^{-2}u\,, \quad V=\varepsilon^{2}v\,. 
\end{gather}
In terms of these new variables, \cref{eq:spotsystem} with
$\partial_n u =\partial_n v =0$ on $\vectorr x \in \partial\Omega$ becomes
\begin{subequations}\label{eq:uvspots}
\begin{align}
 u_t & = \varepsilon^2\Delta u  +\alpha(\vectorr x)u^2v- u + \varepsilon^4(\tau\gamma)^{-1}v\,, 
     & & \vectorr x \in \Omega\,, 
\label{eq:uvspots1}\\
 \tau \varepsilon^2 v_t & = D_0\Delta v -\varepsilon^2v+1- \varepsilon^{-2}\left[\tau\gamma\left(\alpha\left(\vectorr x\right)u^2v -u\right)+\beta\gamma u\right]\,, 
     & & \vectorr x \in \Omega\,, \label{eq:uvspots2} 
\end{align}
\end{subequations}
where $D=\varepsilon^{-2}D_0$ comes from balancing terms in~\cref{eq:uvspots2}.

In the inner region near the $j$-th spot, the leading-order terms in
the inner expansion are locally radially symmetric. We expand this
inner solution as
\begin{equation}\label{eq:asympexpan}
	u=u_{0j}\left(\rho\right)+\varepsilon u_{1j}+\cdots\,, 
\qquad v=v_{0j}\left(\rho\right)+\varepsilon v_{1j}+\cdots\,, \quad 
\textrm{$\rho=|\boldsymbol\xi|\equiv \varepsilon^{-1} |\vectorr x
- \vectorr x_j|\,$}.
\end{equation}
Substituting \cref{eq:asympexpan} into the steady-state problem
for \cref{eq:uvspots}, we obtain the following
leading-order radially symmetric problem on $0<\rho<\infty$:
\begin{equation}\label{eq:uvcore}
 \Delta_\rho u_{0j} + \alpha\left(\vectorr x_j\right) u_{0j}^2 v_{0j}-u_{0j}=0
\,, \hspace{1.5ex}  D_0\Delta_\rho v_{0j} -
\tau\gamma\left(\alpha\left(\vectorr x_j\right)u_{0j}^2 
 v_{0j}-u_{0j}\right)-\beta\gamma u_{0j}=0\,, 
\end{equation}
where $\Delta_\rho\equiv \partial_{\rho\rho}+\rho^{-1}\partial_{\rho}$
is the Laplacian operator in polar co-ordinates. As a remark, in our
derivation of the spot dynamics in \cref{sec:dynspot} we will need to
retain the next term in the Taylor series of the auxin distribution,
representing the auxin gradient in the regions where active-ROP is
localized. This is given by
\begin{equation}\label{eq:gradexp}
	\alpha\left(\vectorr x\right)= \alpha\left(\vectorr
        x_j\right)+\varepsilon\nabla\alpha\left(\vectorr
        x_j\right)\cdot\boldsymbol\xi+\cdots\,, \qquad \textrm{where
          $\boldsymbol \xi \equiv \varepsilon^{-1}\left(\vectorr
          x-\vectorr x_j\right)$.}
\end{equation}
These higher-order terms are key to deriving a dynamical system for spot
dynamics.

We consider \cref{eq:uvcore} together with the boundary conditions
\begin{equation}\label{eq:bcond}
 u_{0j}^\prime(0)=v_{0j}^\prime(0)=0\,; \quad u_{0j}\to0\,, \quad
v_{0j}\sim S_j\log\rho+\chi\left(S_{j}\right)+ o(1)\,, 
\quad \textrm{as $\rho\to\infty\,,$}
   \end{equation}
where $S_j$ is called the \emph{source parameter}. The
  logarithmic far-field behavior for $v_{0j}$ arises since, owing to
  the fact that there is no $-v_{0j}$ term in the second equation of
  (\ref{eq:uvcore}), we cannot assume that $v_{0j}$ is bounded as
  $\rho\to \infty$.  The correction term $\chi_j=\chi(S_j)$ in the
far-field behavior is determined from
\begin{gather}\label{eq:chi}
 \lim\limits_{\rho\to\infty}\left(v_{0j}-S_j\log\rho\right)=\chi(S_j)\,.
\end{gather}
By integrating the equation for $v_{0j}$ in \cref{eq:uvcore}, and
using the limiting behavior $v_{0j}\sim S_j\log\rho$ as
$\rho\to\infty$, we obtain the identity
\begin{gather}\label{eq:ident}
 S_j=\frac{\beta\gamma}{D_0}\int_0^\infty\left[\frac{\tau}{\beta}\left(
\alpha\left(\vectorr x_j\right)u_{0j}^2v_{0j}-u_{0j}\right)+ u_{0j}\right]
\rho\;d\rho\,.
\end{gather}

In \cref{eq:uvcore}, we introduce the change of variables
\begin{gather}\label{eq:corechvar}
 u_{0j}\equiv\sqrt{\frac{D_0}{\beta\gamma\alpha\left(\vectorr x_j\right)}}
u_c\,, \qquad v_{0j}\equiv\sqrt{\frac{\beta\gamma}
{D_0\alpha\left(\vectorr x_j\right)}}v_c\,,
\end{gather}
to obtain what we refer to as the {\em shape canonical core problem}
(SCCP), which consists of
\begin{subequations}\label{full:core}
\begin{equation}\label{eq:core}
 \Delta_\rho u_{c}  +u_{c}^2v_{c}- u_{c}=0\,,\qquad
 \Delta_\rho v_{c} -\frac{\tau}{\beta}\left(u_{c}^2v _{c}-u_{c}\right)- 
 u_{c}=0\,,
\end{equation}
on $0<\rho<\infty$, together with the following boundary conditions
where $S_{cj}$ is a parameter:
\begin{equation}\label{eq:corebc}
	u_c^\prime(0)=v_c^\prime(0)=0, \quad u_c\to0\,, 
\quad v_{c}\sim S_{cj}\log\rho +\chi_c\left(S_{cj}\right) + o(1)\,, 
\quad \textrm{as $\rho\to\infty\,.$}
\end{equation}
\end{subequations}
Upon substituting~\eqref{eq:corechvar} into the
  identity~\eqref{eq:ident}, we obtain
\begin{subequations}\label{eq:scjxcj}
\begin{gather}\label{eq:scjxcjA}
 S_{j}\equiv\sqrt{\frac{\beta \gamma}{D_0 \alpha\left(\vectorr x_j\right)}}S_{cj}\,, \qquad S_{cj}=\int_0^\infty\left[\frac{\tau}{\beta}\left(u_{c}^2v_{c}-u_{c}\right)+ u_{c}\right] \rho\;d\rho\,,
\end{gather}
which, from comparing~\eqref{eq:corechvar} and~\eqref{eq:bcond}, yields
\begin{gather}\label{eq:scjxcjB}
\chi(S_j) \equiv\sqrt{\frac{\beta\gamma}{D_0\alpha\left(\vectorr x_j\right)}}\chi_{c}\,, \quad \textrm{where} \quad \chi_{c}=\chi_{c}\left(S_{cj}\right)\,.
\end{gather}
\end{subequations}

\begin{figure}[t!]
 \begin{center}
 \centering
 \subfigure[]{\includegraphics[height=0.2\textheight]{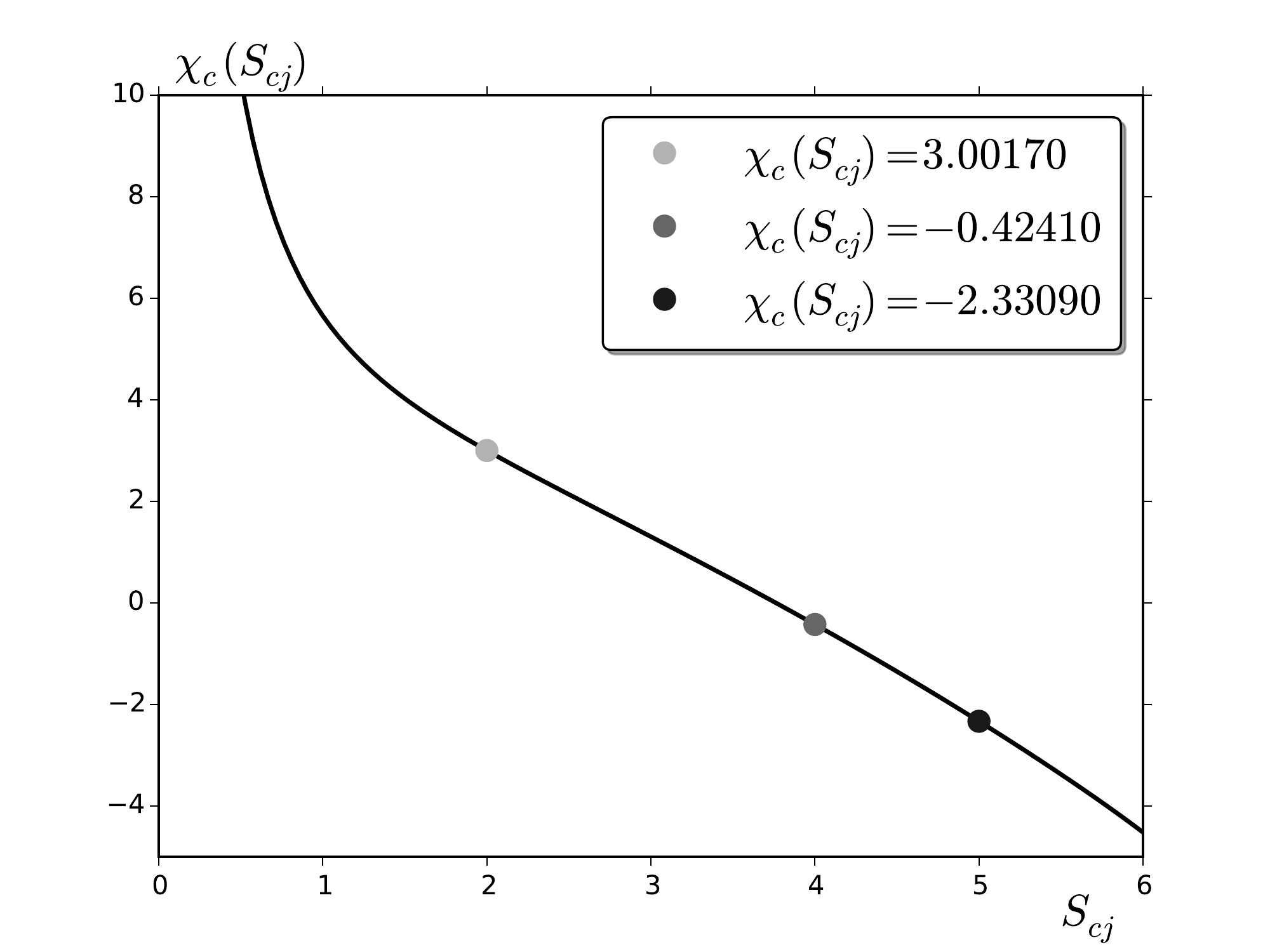}\label{sf:ucvcsca}}
 \centering
 \subfigure[]{\includegraphics[height=0.2\textheight]{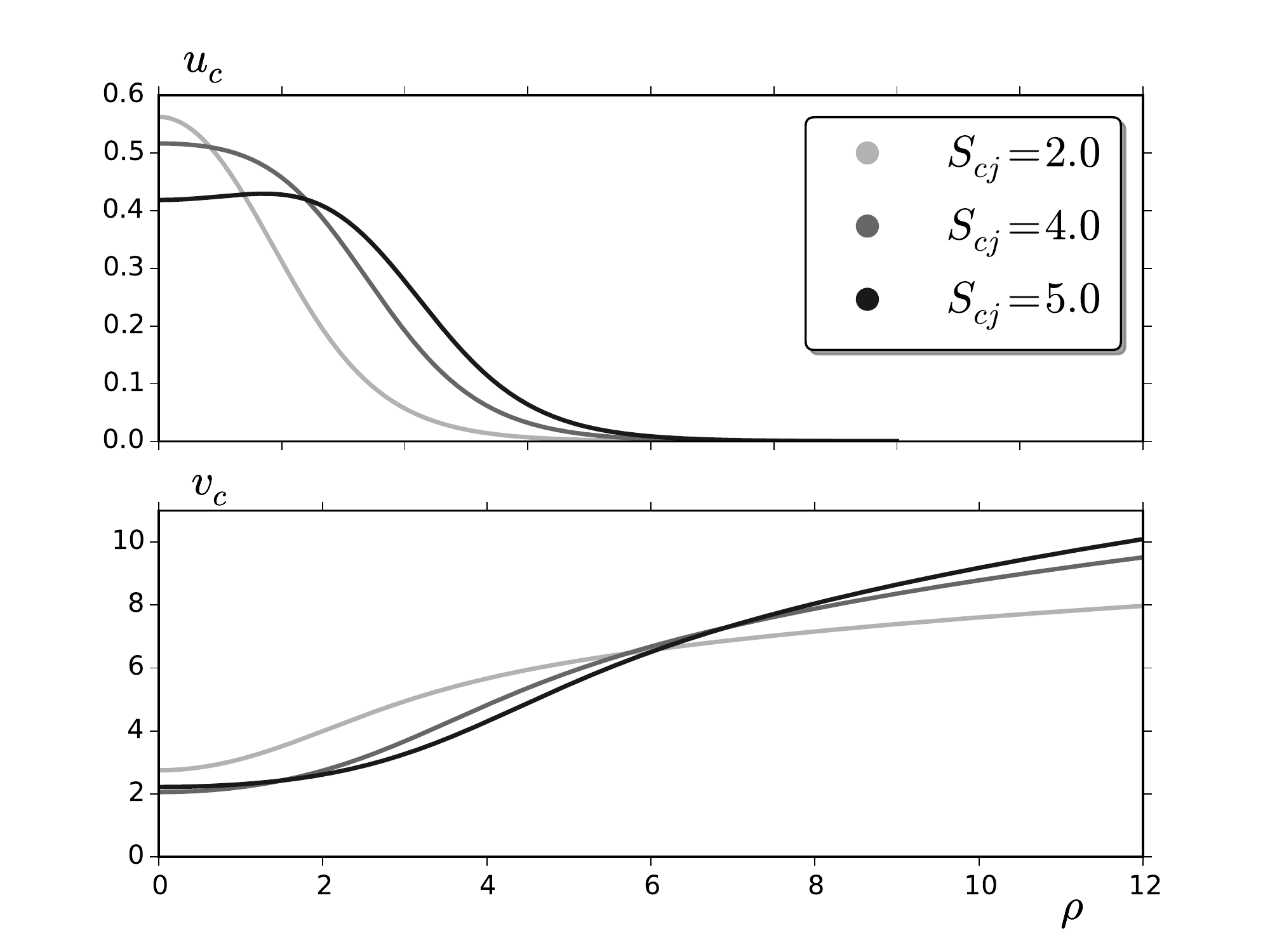}\label{sf:ucvcscb1}}
 \end{center}
\caption{(a) The constant $\chi_c\left(S_{cj}\right)$ versus
    the source parameter $S_{cj}$, computed numerically
    from~\cref{full:core}. (b) Radially symmetric solution $u_c$
    (upper panel) and $v_c$ (bottom panel), respectively, at the
    values of $S_{cj}$ shown in the legend. Parameter Set A from
  \cref{tab:tab}, where ${\tau/\beta}=3$.} \label{fig:ucvcsc}
\end{figure}

To numerically compute solutions to \cref{full:core} we use the
  \textsc{MATLAB} code \textsc{BVP4C} and solve the resulting BVP on
  the large, but finite, interval $[0,\rho_0]$, for a range of values
  of the parameter $S_{cj}$. The constant $\chi_c\left(S_{cj}\right)$
  is identified from $v_c\left(\rho_0\right)-S_{c}\log\rho_0=\chi_c$,
  where $v_c$ is the numerically computed solution, and so it depends
  only on $S_{cj}$ and the specified ratio ${\tau/\beta}$ in
  \cref{full:core}.
  We solve the boundary-value problem on $0<\rho_0<12$ with a tolerance of
    $10^{-4}$. Increasing $\rho_0$ did not change the computational results shown in
    \cref{fig:ucvcsc}.
  The results of our computation are shown in \cref{sf:ucvcsca} where we plot
$\chi_c\left(S_{cj}\right)$ versus $S_{cj}$. In \cref{sf:ucvcscb1} we
also plot $u_c$ and $v_c$ at a few values of $S_{cj}$.  We observe
that $u_c(\rho)$ has a volcano-shaped profile when $S_{cj}=5$, which
suggests, by means of the identity \cref{eq:ident}, that this
parameter value should be near where spot self-replication will occur
(see \cref{sec:split}).

We remark that the SCCP \cref{full:core} depends only on the ratio
${\tau/\beta}$, which characterizes the deactivation and removal rates
of active-ROP (see \cref{eq:newpar}). This implies that the
SCCP~\cref{full:core} only describes the aggregation process. On the
other hand, notice that \cref{eq:corechvar} and \cref{eq:scjxcj}
reveal the role that the auxin plays on the activation process.  Since
$\alpha(\vectorr x)$ decreases longitudinally for any of the
two forms in~\cref{mod:aux}, the scaling~\cref{eq:corechvar} is such
that the further from the left boundary active-ROP localizes, the
larger will be the solution amplitude. This feature will be confirmed
by numerical simulations in \cref{subsec:spobabydrop}.  In addition,
the scaling of the source-parameter in \cref{eq:scjxcj} indicates that
the auxin controls the inactive-ROP distribution along the cell. That
is, while $S_{cj}$ characterizes how $v$ interacts with $u$ in the
region where a spot occurs, the parameter $S_j$ will be determined by
the inhomogeneous distribution of source points, which are also
governed by the auxin gradient. In other words, the gradient controls
and catalyzes the switching process, which leads to localized
elevations of concentration of active-ROP.

Next, we examine the outer region away from the spots centered at
$\vectorr x_1,\ldots,\vectorr x_N$, which are locations where $u$ is
localized. To leading-order, the steady-state problem in
\cref{eq:uvspots1} yields that $u_0\sim {\varepsilon^4
  v_0/(\beta\gamma)}$ as obtained from balancing terms. Therefore, from
this order in the reaction terms of \cref{eq:uvspots2}, we obtain in
the outer region that
\begin{equation}\label{eq:vout}
 -\varepsilon^2v_0+1-\varepsilon^{-2}\left[\tau\gamma\left(\alpha\left(\vectorr
        x\right)u_0^2v_0-u_0\right)+\beta\gamma u_0\right]=1+
 \mathcal O\left(\varepsilon^2\right)\,.
\end{equation}

Next, we calculate the effect of the localized spots on the outer
equation for $v$. Upon assuming $u\sim u_j$ and $v\sim v_j$
  near the $j$-th spot, we approximate the singular terms in the sense
  of distributions as Dirac masses. Indeed, labelling the reactive
  terms by $\mathcal R_j\equiv \tau\gamma\left(\alpha\left(\vectorr
  x\right)u_j^2v_j-u_j\right)+\beta\gamma u_j$, we use \eqref{eq:ident}
and the Mean Value Theorem to calculate, in the sense of distributions, that
\begin{gather}\label{eq:vout_jump}
	\varepsilon^{-2}\mathcal
        R_j\left(\varepsilon^{-1}\left(\vectorr x-\vectorr
        x_j\right)\right) \to 2\pi
        \left(\int_{0}^{\infty}\mathcal{R}_{jj}(\rho)\rho\;d\rho \right)
        \, \updelta(\vectorr x-\vectorr x_j) =2 \pi D_0 S_j
        \updelta(\vectorr x-\vectorr x_j)\,,
\end{gather} 
as $\varepsilon\to 0$, where
$\mathcal{R}_{jj}(\rho)\equiv\tau\gamma\left( \alpha\left(\vectorr
x_j\right)u_{0j}^2v_{0j}-u_{0j}\right)+ \beta\gamma u_{0j}$.  

Combining \cref{eq:vout}, \cref{eq:vout_jump} and
\cref{eq:uvspots2} we obtain that the leading-order outer solution
$v_0$, in the region away from ${\mathcal O}(\eps)$ neighborhoods of
the spots, satisfies
\begin{subequations}\label{eq:v0prob}
\begin{equation}\label{eq:v0field}
 \Delta v_0+\frac{1}{D_0}=2\pi \sum\limits_{i=1}^NS_i\updelta
 \left(\vectorr x-\vectorr x_i\right)\,, \quad \vectorr x \in \Omega\,;
 \qquad \partial_n v_0 =0 \,, \quad \vectorr x \in
 \partial\Omega\,.
\end{equation}
By using \cref{eq:bcond} to match the outer and inner solutions for
$v$, we must solve \cref{eq:v0field} subject to the following
singularity behavior as $\vectorr x\to\vectorr x_j$:
\begin{gather}\label{eq:v0field3}
  v_{0}\sim S_j\log\left|\vectorr x-\vectorr
 x_j\right|+\frac{S_j}{\upsigma}+\chi(S_j)\,, \quad 
 \textrm{for $j=1,\ldots,N$, where $\upsigma\equiv -\frac{1}{\log\varepsilon}$,}
\end{gather}
which results from writing~\eqref{eq:bcond} as $\rho=\varepsilon^{-1}
|\vectorr x - \vectorr x_j|$. 
\end{subequations}

We now proceed to solve \cref{eq:v0prob} for $v_{0}$ and derive a
nonlinear algebraic system for the source parameters
$S_{c1},\ldots,S_{cN}$. To do so, we introduce the unique Neumann
Green's function $G(\vectorr x;\vectorr y)$ satisfying (cf.~\cite{kolo01})
\[
    \Delta G=\frac{1}{d_y}-\updelta\left(\vectorr x-\vectorr y\right)\,, \quad
    \int_{\Omega}G\left(\vectorr x;\vectorr y\right)\:d\vectorr y=0\,, \quad
    \vectorr x \in \Omega\,, \qquad
    \partial_n G = 0\,, \quad
    \vectorr x \in \partial\Omega\,.
\]
This Green's function satisfies the reciprocity relation
$G\left(\vectorr x;\vectorr y\right)= G\left(\vectorr y;\vectorr
x\right)$, and can be globally decomposed as $G\left(\vectorr
x;\vectorr y\right)=-1/(2\pi) \log\left|\vectorr x-\vectorr
y\right|+R\left(\vectorr x;\vectorr y\right) $, where $R\left(\vectorr
x;\vectorr y\right)$ is the smooth regular part of $G\left(\vectorr
x;\vectorr y\right)$. For $\vectorr x\to \vectorr y$, this Green's
function has the local behavior
\[
 G\left(\vectorr x;\vectorr y\right)\sim -\frac{1}{2\pi}
\log\left|\vectorr x-\vectorr y\right|+
R\left(\vectorr y;\vectorr y\right)+\nabla_\vectorr x 
R\left(\vectorr y;\vectorr y\right)\cdot\left(\vectorr x-\vectorr y\right)
+\mathcal{O}\left(\left|\vectorr x-\vectorr y\right|^2\right)\,.
\]
An explicit infinite series solution for $G$ and $R$ in
the rectangle $\Omega$ is available (cf.~\cite{kolo01}).

In terms of this Green's function, the solution to \cref{eq:v0field} is 
\begin{equation}\label{eq:v0Si}
  v_0=-2\pi\sum\limits_{i=1}^NS_iG\left(\vectorr x;\vectorr x_i\right)+
\bar v_0\,, \qquad \textrm{where} \qquad 
\sum\limits_{i=1}^NS_i=\frac{d_y}{2\pi D_0}\,.
\end{equation}
Here $\bar v_0$ is a constant to be found. The condition on $S_i$ in
\cref{eq:v0Si} arises from the Divergence Theorem applied to
\cref{eq:v0prob}. We then expand $v_0$ as $\vectorr x\to\vectorr
x_j$ and enforce the required singularity behavior
\cref{eq:v0field3}. This yields that
\begin{gather}\label{eq:nas01}
 \frac{S_j}{\upsigma}+\chi\left(S_j\right)=-2\pi S_jR_{j}-
2\pi\sum\limits_{i=1}^NS_iG_{ji}+\bar v_0\,, \qquad \textrm{for} \quad
  j=1,\ldots\,N\,.
\end{gather}
Here $G_{ji}\equiv G\left(\vectorr x_j;\vectorr x_i\right)$ and
$R_{j}\equiv R\left(\vectorr x_j;\vectorr x_j\right)$. The system
\cref{eq:nas01}, together with the constraint in \cref{eq:v0Si},
defines a nonlinear algebraic system for the $N+1$ unknowns
$S_1,\ldots,S_N$ and $\bar v_0$. To write this system in matrix form,
we define $\vectorr e\equiv\left(1,\ldots,1\right)^T$, $\vectorr S
\equiv (S_1,\ldots,S_N)^T$, $\boldsymbol\chi \equiv
(\chi(S_1),\ldots,\chi(S_N)^T$, $E\equiv{\vectorr e\vectorr e^T/N}$,
and the symmetric Green's matrix $\vectorr G$ with entries
\begin{equation*}
    (\vectorr G)_{ii}=R_i \,, \qquad   (\vectorr G)_{ij}=G_{ij} \,,
 \quad \mbox{where} \quad G_{ij}=G_{ji}\,.
\end{equation*}
Then, \cref{eq:nas01}, together with the
constraint in \cref{eq:v0Si}, can be written as
\begin{equation}\label{eq:nas02}
 \vectorr S+2\pi\upsigma\vectorr G\vectorr S=\upsigma
\left(\bar v_0\vectorr e-\boldsymbol\chi\right)\,, \qquad
 \vectorr e^T\vectorr S=\frac{d_y}{2\pi D_0} \,.
\end{equation}
Left multiplying the first equation in \cref{eq:nas02} by 
$\vectorr e^T$, and using the second equation in \cref{eq:nas02}, we 
find an expression for $\bar v_0$ 
\begin{gather}\label{eq:vobar}
	\bar v_0=\frac{1}{N\upsigma}\left(\frac{d_y}{2\pi
          D_0}+2\pi\upsigma\vectorr e^T\vectorr G\vectorr
        S+\upsigma\vectorr e^T\boldsymbol\chi\right)\,.
\end{gather}
Substituting $\bar v_0$ back into the first equation of
\cref{eq:nas02}, we obtain a nonlinear system for $S_1,\ldots,S_N$ given
by
\begin{gather}\label{eq:voese}
 \vectorr S+\upsigma\left(\vectorr I -\vectorr E\right)
\left(2\pi\vectorr G\vectorr S+\boldsymbol\chi\right)=
\frac{d_y}{2\pi ND_0}\vectorr e\,.
\end{gather}
In terms of $\vectorr S$, $\bar v_0$ is given in \cref{eq:vobar}.
To relate $S_j$ and $\chi_j$ with the SCCP~\cref{full:core}, we define
\[
  \label{eq:corevar02}
  \boldsymbol\upalpha\equiv
  \diag\left({\frac{1}{ \sqrt{\alpha\left(\vectorr x_1\right)}},\ldots,
  \frac{1}{\sqrt{\alpha\left(\vectorr x_N\right)}}}\right),
  \quad
  \vectorr S_c\equiv
  \begin{pmatrix}
      S_{c1}\\
      \vdots\\
      S_{cN}
  \end{pmatrix},
  \quad
  \boldsymbol\chi_c\equiv
  \quad
  \begin{pmatrix}
    \chi_{c}\left(S_{c1}\right)\\
    \vdots\\
    \chi_{c}\left(S_{cN}\right)
  \end{pmatrix}.
\]
Then, by using \cref{eq:scjxcj}, we find
\begin{gather}\label{eq:corevar03}
 \vectorr S=\upomega\boldsymbol\upalpha\vectorr S_{c}\,, \qquad
 \boldsymbol\chi=\upomega\boldsymbol\upalpha\boldsymbol\chi_c\,,
 \qquad \upomega\equiv\sqrt{\frac{\beta\gamma}{D_0}}\,.
\end{gather}
Finally, upon substituting \cref{eq:corevar03} into \cref{eq:voese}
we obtain our main result regarding quasi steady-state spot patterns.

\begin{proposition}\label{prop:multispotprof}
Let $0<\varepsilon\ll1$ and assume $D_0={\mathcal O}(1)$ in
\cref{eq:uvspots} so that $D=\mathcal O\left(\varepsilon^{-2}\right)$
in \cref{eq:spotsystem}. Suppose that the nonlinear algebraic system
for the source parameters, given by
\begin{subequations}\label{eq:uvprofile}
	\begin{gather}
 \boldsymbol\upalpha\vectorr S_c+\upsigma\left(\vectorr I -\vectorr
 E\right)\left(2\pi\vectorr G\boldsymbol\upalpha\vectorr
 S_c+\boldsymbol\upalpha\boldsymbol\chi_c\right)= \frac{d_y}{2\pi
   \upomega ND_0}\vectorr e\,, \label{eq:uvprofilea}\\ \bar
 v_0=\frac{1}{N\upsigma}\left(\frac{d_y}{2\pi
   D_0}+2\pi\upomega\vectorr e^T\vectorr G\boldsymbol\upalpha\vectorr
 S_c+\upsigma\upomega\vectorr
 e^T\boldsymbol\chi_c\right)\,,\label{eq:uvprofileb}
	\end{gather}
\end{subequations}
has a solution for the given spatial configuration $\vectorr
x_1,\ldots,\vectorr x_N$ of spots.  Then, there is a quasi
steady-state solution for \cref{eq:spotsystem} with
$U=\varepsilon^{-2} u$ and $V=\varepsilon^{2} v$. In the outer region
away from the spots, we have $u= {\mathcal O}(\varepsilon^4)$ and
$v\sim v_0$, where $v_0$ is given asymptotically in terms of $\vectorr
S=\upomega\boldsymbol \upalpha\vectorr S_{c}$ by \cref{eq:v0Si}.  In
contrast, in the $j$-th inner region near $\vectorr x_j$, we have
\begin{gather}\label{prop:corechvar}
 u \sim \sqrt{\frac{D_0}{\beta\gamma\alpha\left(\vectorr x_j\right)}}
u_c\,, \qquad v \sim\sqrt{\frac{\beta\gamma}
{D_0\alpha\left(\vectorr x_j\right)}}v_c\,,
\end{gather}
where $u_c$ and $v_c$ is the solution of the SCCP~\cref{full:core} in terms of
$S_{cj}$ satisfying
\cref{eq:uvprofilea}.
\end{proposition} 

\begin{figure}[t!]
	\begin{center}
\vspace*{-0.2cm}
		\centering
		\subfigure[]{\includegraphics[height=0.2\textheight]{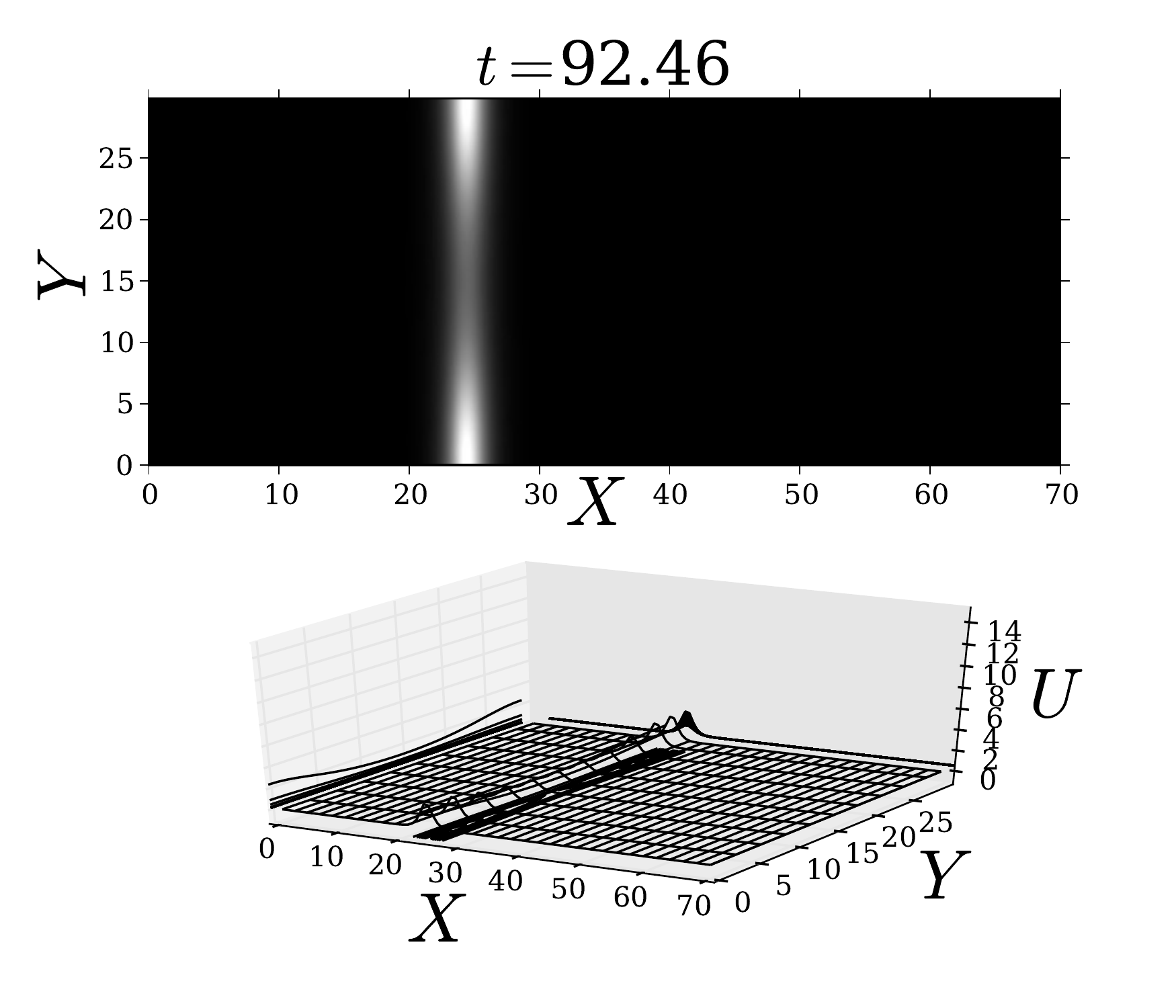}\label{sf:breakupT3a}}
\vspace*{-0.2cm}
		\centering
		\subfigure[]{\includegraphics[height=0.2\textheight]{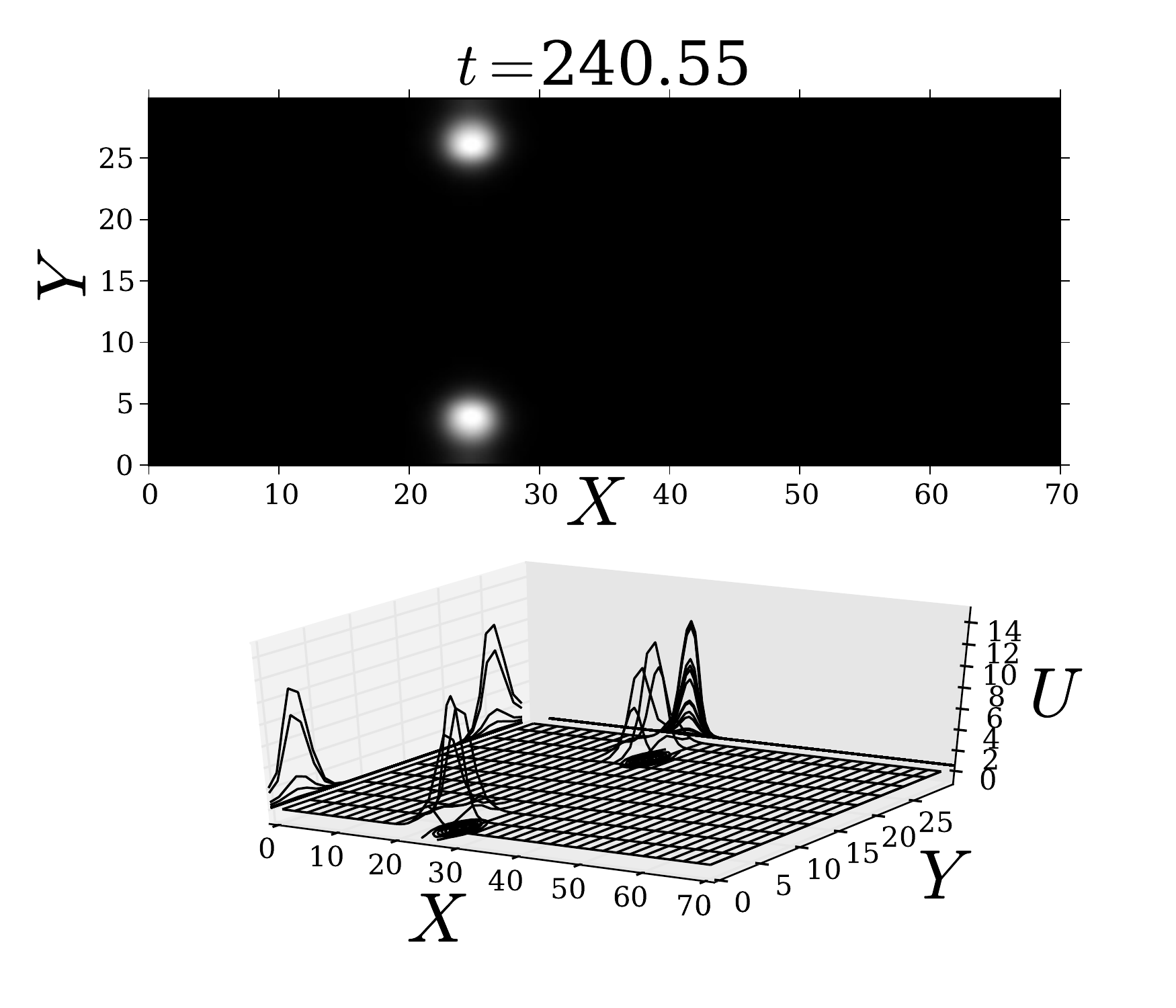}\label{sf:breakupT3b}}
\vspace*{-0.2cm}
		\centering
		\subfigure[]{\includegraphics[height=0.2\textheight]{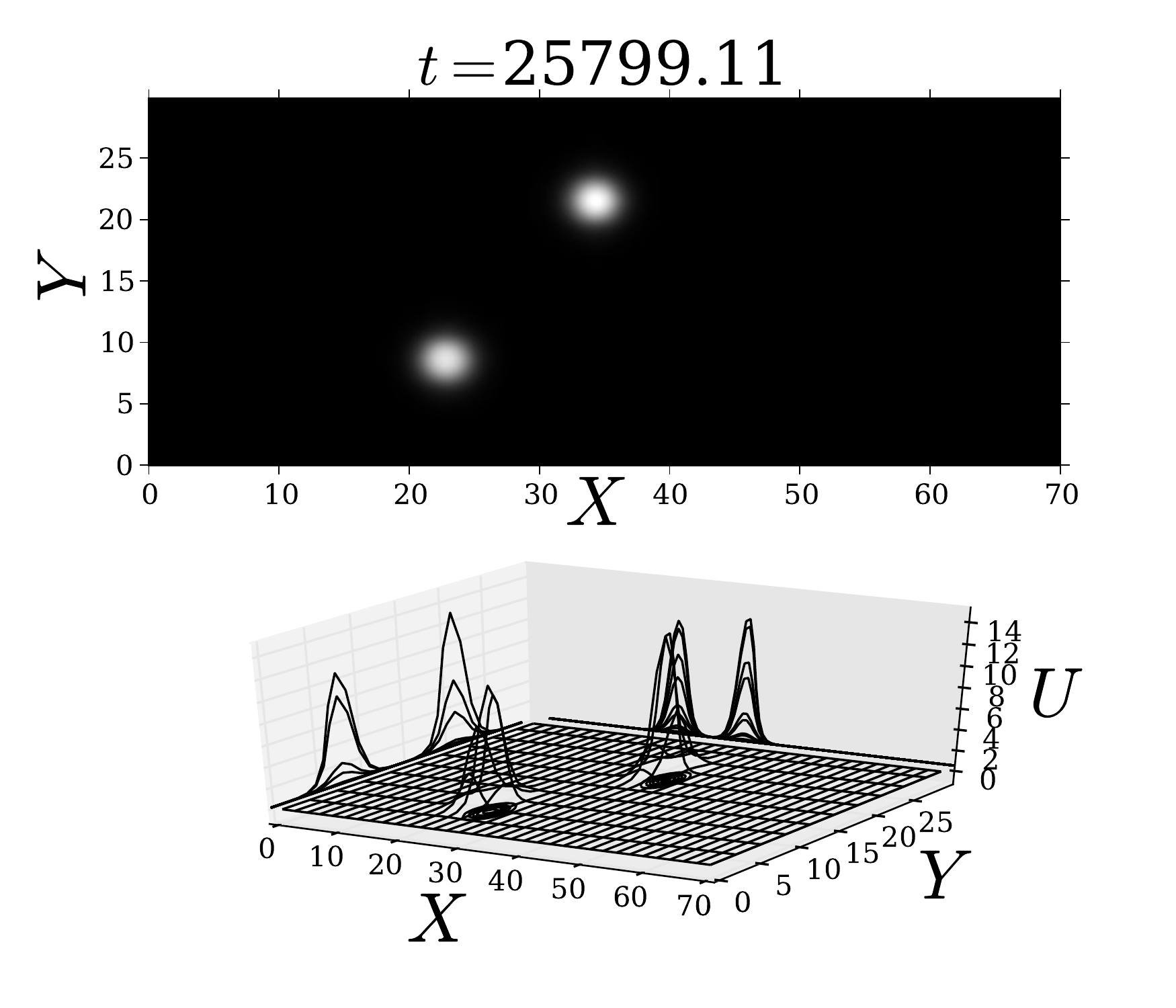}\label{sf:breakupT3c}}
\vspace*{-0.2cm}
		\centering
		\subfigure[]{\includegraphics[height=0.2\textheight]{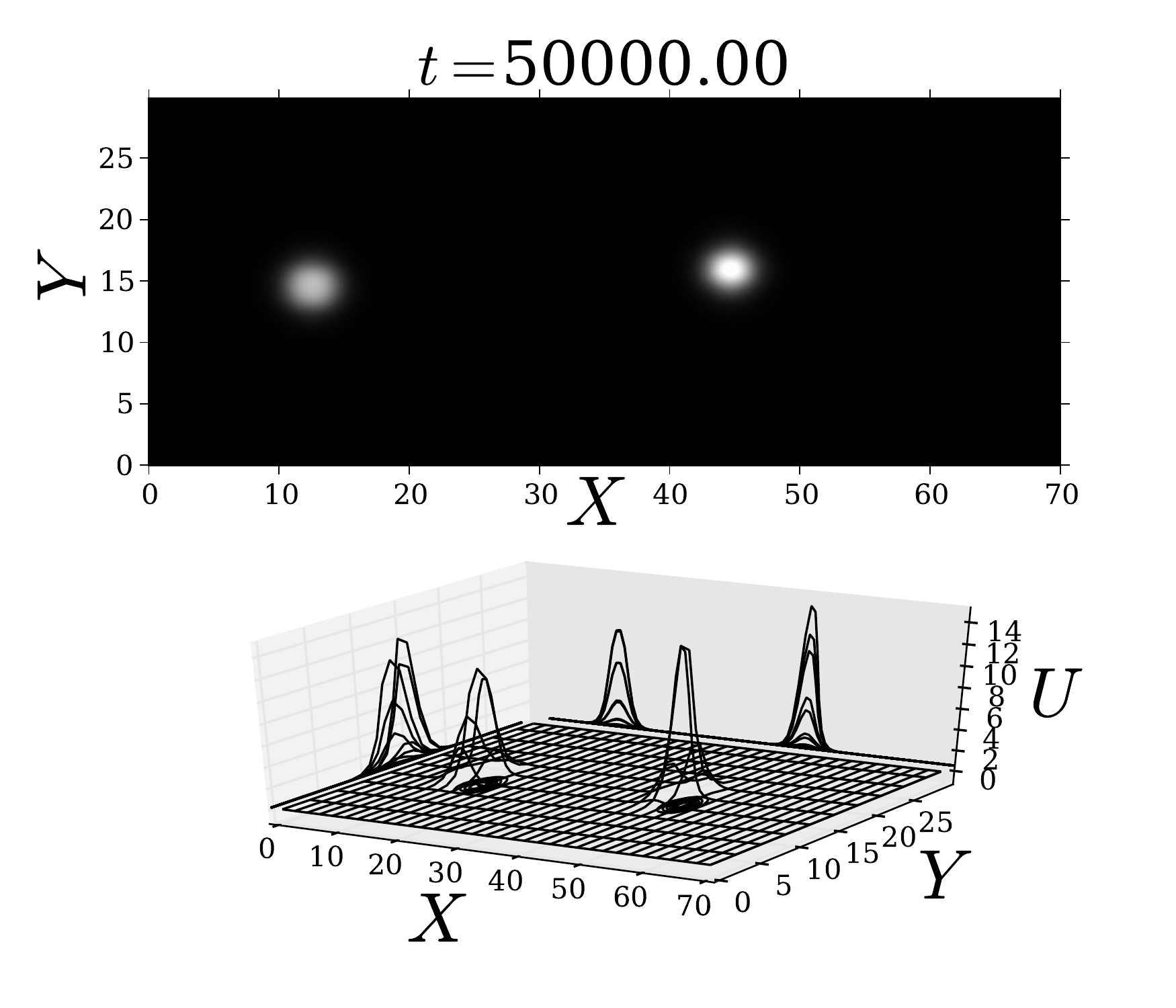}\label{sf:breakupT3d}}
	\end{center}
	\caption{Breakup instability of an interior localized
          homoclinic stripe for $U$, and the resulting slow spot
          dynamics. (a)~The localized stripe initially breaks into two
          spots; (b)~once formed, the spots migrate from the boundary
          towards each other in the vertical direction, and then
          (c)~rotate until they get aligned with the longitudinal
          direction. (d)~Finally, they get pinned far from each other
          along the $y$-midline by the auxin gradient. Parameter Set
          A from \cref{tab:tab} was used with $k_{20}=0.5$. For
          these values the initial steady-state stripe is centered at
          $X_0=24.5$. Figure from~\cite{bacw}.}
	\label{fig:breakupT3}
\end{figure}

\subsection{Spots and Baby Droplets}
\label{subsec:spobabydrop}

\begin{figure}[t!]
	\begin{center}
\vspace*{-0.2cm}
		\centering
		\subfigure[]{\includegraphics[height=0.2\textheight]{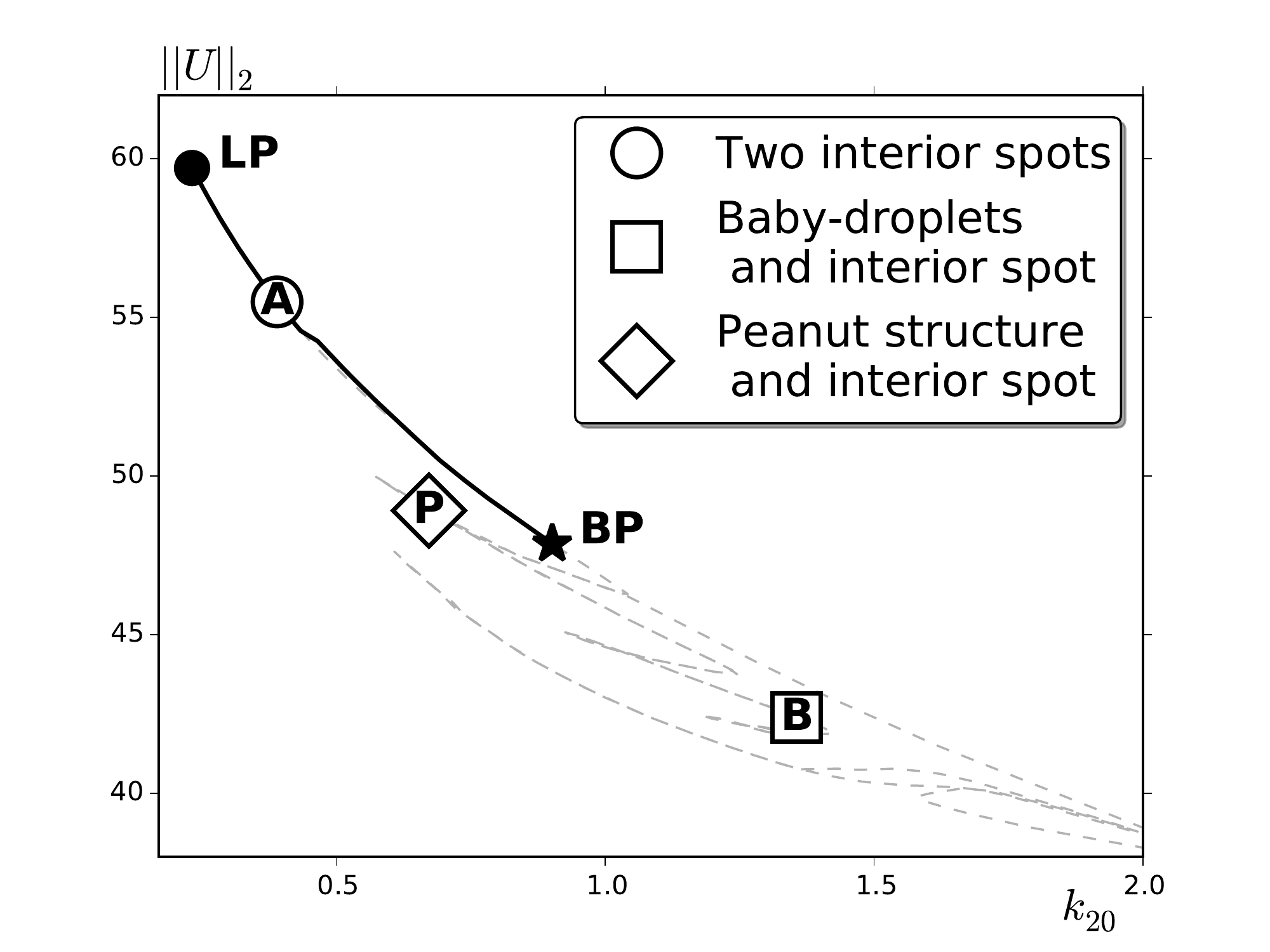}\label{sf:expgrada}}
\vspace*{-0.2cm}
		\centering
		\subfigure[]{\includegraphics[height=0.2\textheight]{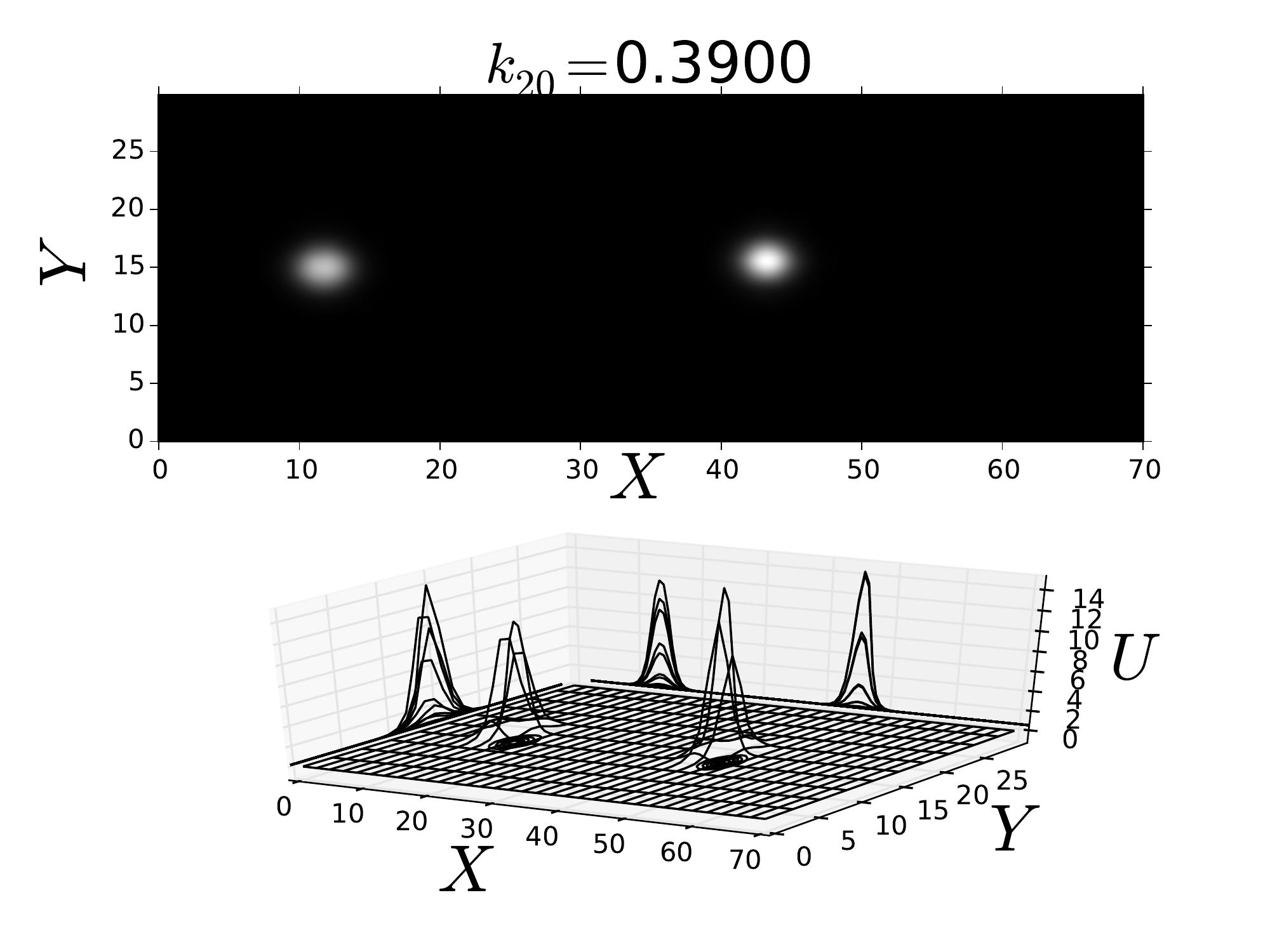}\label{sf:expgradb}}
\vspace*{-0.2cm}
		\centering
		\subfigure[]{\includegraphics[height=0.2\textheight]{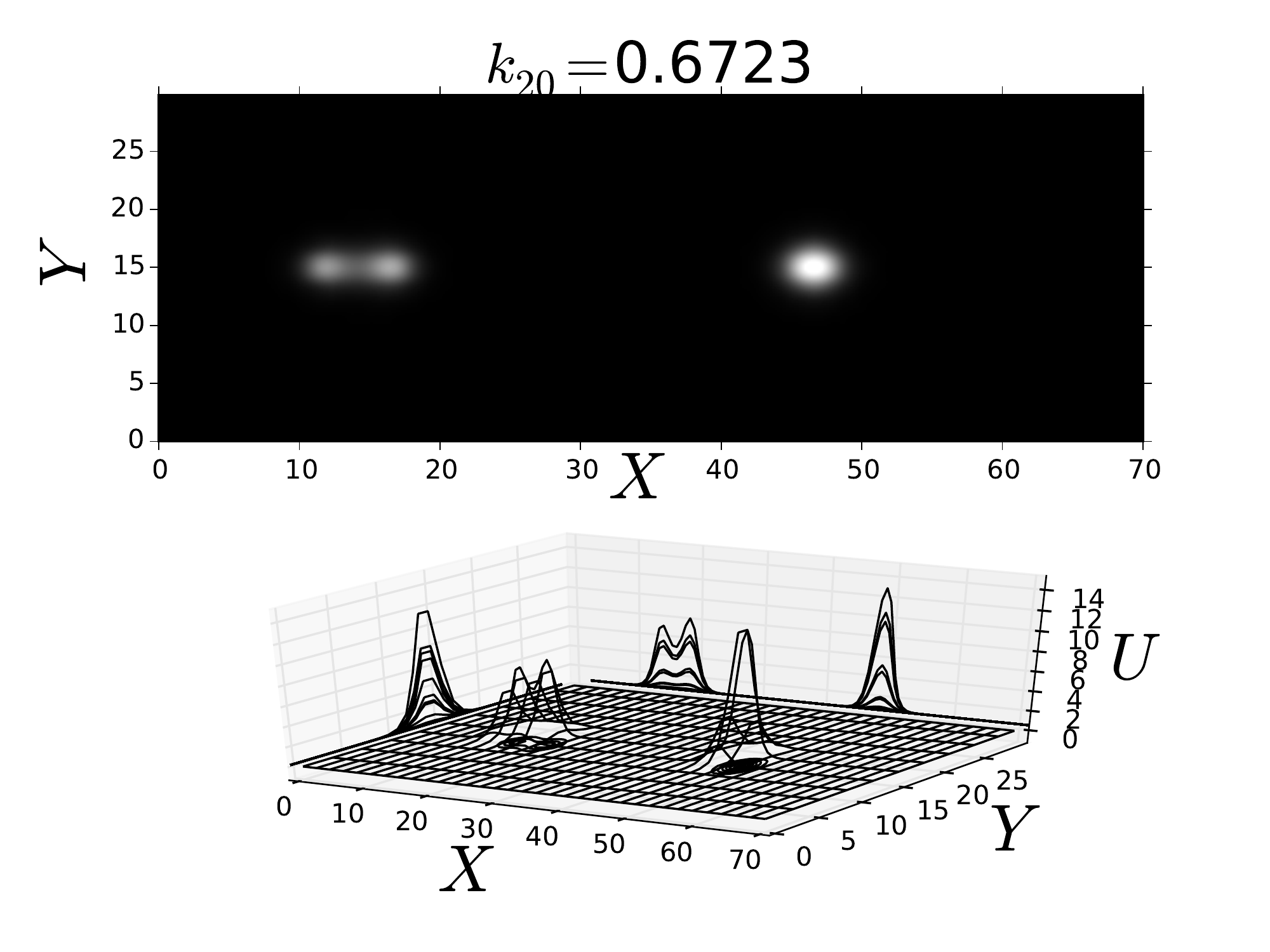}\label{sf:expgradd}}
\vspace*{-0.2cm}
		\centering
		\subfigure[]{\includegraphics[height=0.2\textheight]{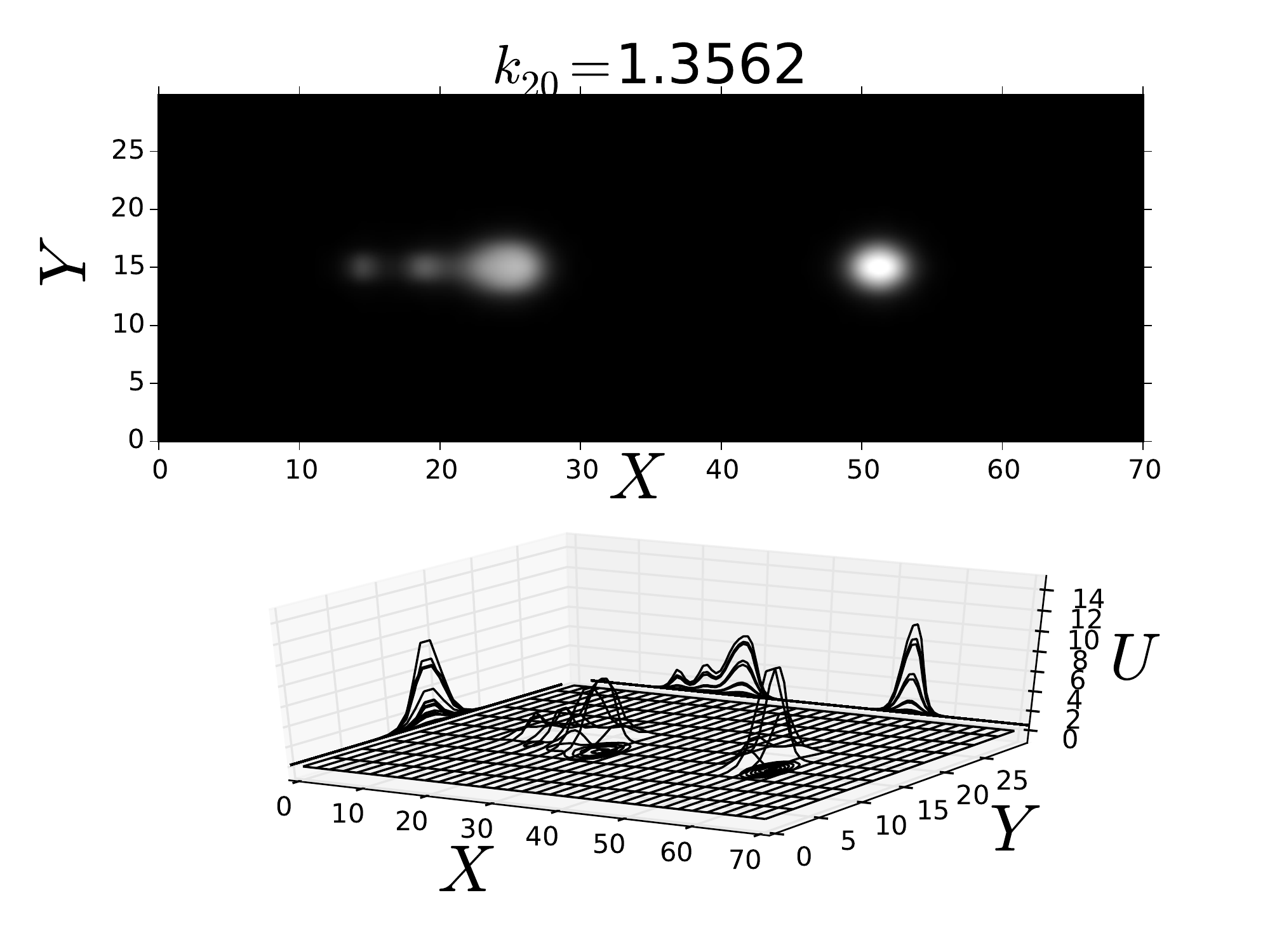}\label{sf:expgradc}}
	\end{center}
	\caption{(a) Bifurcation diagram as $k_{20}$ is varied showing
          a stable branch, labelled~A, of (b)~two spots, and unstable
          branches (light gray-dashed curves). Filled circle at
          $k_{20}^{o}=0.2319$ is a fold bifurcation (LP) and the
          filled star at $k_{20}^{*}=0.90161$ is a pitchfork
          bifurcation (BP). (c) A peanut structure and one spot
          steady-state, from label P in~(a). (d) A two-spot with baby
          droplets steady-state solution, from label B
          in~(a). Parameter Set A was used, as given in
          \cref{tab:tab}. The $k_{20}$ values are shown on top of
          upper panels in (b)-(d).}
	  \label{fig:xexpgrad}
\end{figure}

\cref{prop:multispotprof} provides an asymptotic characterization of a
$N$-spot quasi steady-state solution.  As has been previously studied
in the previous paper \cite{bacw}, localized spots can be generated by
the breakup of a straight homoclinic stripe. The numerical
  simulations in \cref{fig:breakupT3}, for the 1-D auxin gradient in
  \cref{mod:aux}, illustrate how small spots are formed from the
  breakup of an interior homoclinic stripe and how they evolve slowly
  in time to a steady-state two-spot pattern aligned with the
  direction of the auxin gradient. The initial condition for these
  simulations is obtained by trivially extending a 1-D spike in the
  transversal direction to obtain a homoclinic stripe, which is a
  steady-state of~\cref{eq:spotsystem}, and then perturbing it with a
  transverse perturbation of small amplitude.  As shown in
  \cref{sf:breakupT3c}--\cref{sf:breakupT3d}, the amplitude of the
  left-most spot decreases as the spot drifts towards the left domain
  boundary at $X=0$, while the amplitude of the right-most spot grows
  as it drifts towards the right domain boundary at $X=L_x$.  The spot
  profiles are described by \cref{prop:multispotprof} for each fixed
  re-scaled variables $\vectorr x_j$. Our realizations of the
  asymptotic theory shown below in ~\cref{fig:loctwospot} confirm the
  numerical results in \cref{sf:breakupT3d} that the spots become
  aligned with the direction of the auxin gradient.

 In order to explore solutions that bifurcate from a multi-spot steady
 state, we adapted the \textsc{MATLAB} code of
 \cite{avitabile16,rankin01} and performed a numerical continuation
 for the RD system~\cref{eq:spotsystem} in the bifurcation parameter
 $k_{20}$, starting from the solution in \cref{sf:breakupT3d} (
   all other parameters as in Parameter Set A in \cref{tab:tab}). The
 results are shown in \cref{fig:xexpgrad}. In the bifurcation diagram
 depicted in \cref{sf:expgrada} the linearly stable branch is plotted
 as a solid curve and unstable ones by light gray-dashed curves. We
 found fold bifurcations and a pitchfork bifurcation, represented by a
 filled circle and a filled star, 
 respectively. To the right of the pitchfork bifurcation, the symmetric solution
 branch destabilises; the emerging branch of asymmetric states is fully unstable, and
 features a sequence of fold bifurcations.
 Overall, these results suggest that a snaking bifurcation structure occurs in
   2-D domains, similar to what was found in~\cite{bacw} for the corresponding 1-D
   system. This intricate branch features a set
   of fold points (not explicitly labelled in \cref{sf:expgrada}), and around each of
   them a new spot emerges.  %
 A pattern consisting of a spot and a peanut structure is found
 in the solution branch labelled as P. This solution is shown in \cref{sf:expgradd}.

 We found that solutions with peanut structures or baby droplets are unstable
steady-states in this region of parameter space, and so they do not
persist in time; see a sample of these in~\cref{sf:expgradc}. In
addition, as can be seen in \cref{sf:expgradb}, there are stable
multi-spot solutions when $k_{20}$ is small enough. Overall,
  this suggests that certain $\mathcal O(1)$ time-scale instabilities
  play an important role in transitioning between steady-state
  patterns (cf.~\cite{chen01,iron03}). Further,
  in~\cref{sec:instab2D}, peanut-splitting instabilities are
  addressed.

\section{The Slow Dynamics of Spots}
\label{sec:dynspot}

In \cref{subsec:multispots} we constructed an $N$-spot quasi
steady-state solution for an arbitrary, but ``frozen'', spatial
configuration $\vectorr x_1,\ldots,\vectorr x_N$ of localized patches
of active-ROP. In this section, we derive a differential-algebraic
system of ODEs (DAE) for the slow time-evolution of $\vectorr
x_1,\ldots,\vectorr x_N$ towards their equilibrium locations in
$\Omega$. This system will consist of a constraint that involves the
slow time evolution of the source parameters $
S_{c1},\ldots,S_{cN}$. The derivation of this DAE system is
an extension of the 1-D analysis of \cite{bcwg}.

To derive the DAE system, we must extend the analysis in
\cref{subsec:multispots} by retaining the ${\mathcal O}(\varepsilon)$
terms in \cref{eq:asympexpan} and \cref{eq:gradexp}.  By allowing the
spot locations to depend on a distinguished limit for the slow time
scale $\eta=\varepsilon^2 t$ as $\vectorr x_j=\vectorr
x_j\left(\eta\right)$, we obtain in the $j$-th inner region that the
corrections terms $u_{1j}$ and $v_{1j}$ to the SCCP \cref{full:core}
satisfy
\begin{subequations}\label{eq:spotsdyn}
 \begin{gather}\label{eq:dyn01}
 \Delta\vectorr Q_1+\vectorr M\vectorr Q_1= \vectorr F \equiv 
\vecol{-\frac{d\vectorr x_j}{d\eta}\cdot\nabla u_{0j}}{0}+
\left( \boldsymbol\xi\cdot\nabla\alpha_j \right) u_{0j}^2v_{0j}
\vecol{-1}{\frac{\tau\gamma}{D_0}}\,,
\end{gather}
where the vector $\vectorr Q_1$ and the matrix $\vectorr M$ are defind by
\begin{gather}\label{eq:w1m}
	\vectorr Q_1\equiv
\left(\begin{array}{c}
		u_{1j}\\
		v_{1j} \end{array}\right)\,,
\qquad \vectorr M\equiv
\mados{2\alpha_ju_{0j}v_{0j}-1}{\alpha_ju_{0j}^2}{-\frac{\tau\gamma}{D_0}
\left(2\alpha_ju_{0j}v_{0j}-1\right)-\frac{\beta\gamma}{D_0}}
 {-\frac{\tau\gamma}{D_0}\alpha_ju_{0j}^2}\,.
\end{gather}
Here $\alpha_j\equiv\alpha\left(\vectorr x_j\right)$ and
$\nabla\alpha_j\equiv\nabla\alpha(\vectorr x_j)$. To determine
the far-field behavior
for $\vectorr Q_1$, we expand the outer solution $v_0$ in
\cref{eq:v0Si} as $\vectorr x\to\vectorr x_j$ and retain the
${\mathcal O}(\varepsilon)$ terms. This yields
\begin{equation}\label{eq:dyn02}
 \vectorr Q_1\to\vecol{0}{\boldsymbol\upzeta_j\cdot\boldsymbol\xi}, \quad
 \textrm{as $\quad \rho\equiv |\boldsymbol \xi| \to\infty\,,$}
\end{equation}
where, in terms of the source strengths $S_j$ defined from
\cref{eq:voese}, we have
\begin{equation}
\quad \boldsymbol\upzeta_j\equiv-2\pi\left(S_j\nabla_\vectorr x R_j+
\sum\limits_{i\neq j}^NS_i\nabla_\vectorr xG_{ji}\right)\,,
\qquad
j = 1,\ldots,N\,.
\end{equation}
\end{subequations}
Here $ \nabla_\vectorr x R_j\equiv\nabla_\vectorr xR\left(\vectorr
x_j;\vectorr x_j \right)$ and $\nabla_\vectorr x
G_j\equiv\nabla_\vectorr xG\left(\vectorr x_j; \vectorr x_j\right)$
for $j = 1,\ldots,N$.

To impose a solvability condition on \cref{eq:spotsdyn}, which will
lead to ODEs for the spot locations, we first rewrite \cref{eq:spotsdyn}
in terms of the canonical variables \cref{eq:corechvar} of
the SCCP \cref{full:core}.  Using this transformation, the
right-hand side $\vectorr F$ of \cref{eq:dyn01} becomes
\begin{equation}\label{eq:dyn03}
 \vectorr F = \sqrt{\frac{D_0}{\beta\gamma\alpha_j}}
\left[\vecol{-\frac{d\vectorr x_j}{d\eta}\cdot\nabla u_{c}}{0}+
\boldsymbol\xi\cdot\frac{\nabla\alpha_j}{\alpha_j} u_{c}^2v_{c}
\vecol{-1}{\frac{\tau\gamma}{D_0}}\right] \,.
\end{equation}
In addition, the matrix $\vectorr M$ on the left-hand side of
\cref{eq:dyn01} becomes
\begin{gather}\label{eq:dyn04}
 \vectorr M=\mados{2u_cv_c-1}{\frac{D_0}{\beta\gamma}u_c^2}
{-\frac{1}{D_0}\left(\tau\gamma\left(2u_cv_c-1\right)+\beta\gamma\right)}
{-\frac{\tau}{\beta}u_c^2}\,.
\end{gather}
Together with \cref{eq:dyn03} this suggests that we define
new variables $\hat u_1$ and $\hat v_1$ by
\begin{gather*}
 u_{1j}\equiv\sqrt{\frac{D_0}{\beta\gamma\alpha_j}}\hat u_1\,, 
\qquad v_{1j}\equiv \frac{\beta \gamma}{D_0} 
\sqrt{\frac{D_0}{\beta\gamma\alpha_j}}\hat v_1\,.
\end{gather*}
In terms of these new variables, and upon substituting
\cref{eq:dyn03} and \cref{eq:dyn04} into \cref{eq:spotsdyn}, we
obtain that $\hat{\vectorr Q}_1\equiv(\hat{u}_1,\hat{v}_1)^T$
in $\boldsymbol\xi\in\mathbb{R}^2$ satisfies
\begin{subequations}\label{eq:dynam}
 \begin{gather}
{\mathcal L} \hat{\vectorr Q}_1 \equiv 
  \Delta\hat{\vectorr Q}_1+\vectorr M_c\hat{\vectorr Q}_1=
\vecol{-\frac{d\vectorr x_j}{d\eta}\cdot\nabla u_{c}}{0}+
\boldsymbol\xi\cdot\frac{\nabla\alpha_j}{\alpha_j}u_{c}^2v_{c}\vecol{-1}
{\frac{\tau}{\beta}}\,,  \label{eq:dynama}\\
 \hat{\vectorr Q}_1\to \vecol{0}{\hat{\boldsymbol\upzeta}_j\cdot
 \boldsymbol\xi} \, \quad \textrm{as} \quad \rho\equiv |\boldsymbol\xi|\to
\infty  \,; \quad \hat{\boldsymbol\upzeta}_j \equiv
-2\pi\left(S_{cj}\nabla_\vectorr x R_j+
\sum\limits_{i\neq j}^NS_{ci}\nabla_\vectorr xG_{ji}\right)\,,
 \label{eq:dynamb}
\end{gather}
and $S_{cj}$ for $j=1,\ldots,N$ satisfies the nonlinear algebraic
system \cref{eq:uvprofilea}. In addition, in \cref{eq:dynama}
$\vectorr M_c$ is defined by
\begin{gather}
 \vectorr M_c\equiv\mados{2u_cv_c-1}{u_c^2}{-\frac{\tau}{\beta}
\left(2u_cv_c-1\right)-1}{-\frac{\tau}{\beta} u_c^2}\,,
\end{gather}
\end{subequations} 
where $u_c$ and~$v_c$ satisfy the SCCP~\cref{full:core}, which is
parametrized by $S_{cj}$. 

In \cref{eq:dynamb}, the source parameters $S_{cj}$ depend indirectly
on the auxin distribution through the nonlinear algebraic system
\cref{eq:uvprofile}, as summarized in \cref{prop:multispotprof}. We
further observe from the second term on the right-hand side of
\cref{eq:dynama} that its coefficient $\nabla\alpha_j/\alpha_j$ is
independent of the magnitude of the spatially inhomogeneous
distribution of auxin, but instead depends on the direction of the
gradient.

\subsection{Solvability Condition}
\label{subsec:solvacond}

We now derive a DAE system for the spot dynamics by applying a
solvability condition to \cref{eq:dynam}. We label $\boldsymbol
\xi=(\xi_1,\xi_2)^T$ and ${\mathbf u}_c\equiv (u_c, v_c)^T$.
Differentiating the core problem~\cref{eq:core} with respect to the 
$\xi_i$, we get
\begin{equation*}
     {\mathcal L} \left( \partial_{\xi_i} {\mathbf u}_{c} \right)= \vectorr 0\,,
     \qquad \mbox{where} \qquad \partial_{\xi_i} {\mathbf u}_{c}
     \equiv \vecol{u_c^{\prime}(\rho)}{v_c^{\prime}(\rho)} \, \frac{\xi_i}{\rho}
     \,, \quad \mbox{for} \quad i=1,2 \,.
\end{equation*}
This demonstrates that the dimension of the nullspace of ${\mathcal
  L}$ in \cref{eq:dynam}, and hence its adjoint ${\mathcal
  L}^{\star}$, is at least two-dimensional.  Numerically, for
  our two parameter sets in \cref{tab:tab}, we have checked that the
  nullspace of ${\mathcal L}$ is exactly two-dimensional, provided
  that $S_j$ does not coincide with the spot self-replication
  threshold given in \cref{sec:split}.

There are two independent nontrivial solutions to the homogeneous
adjoint problem ${\mathcal L}^{\star} \vectorr \Psi \equiv \Delta
\vectorr \Psi +\vectorr M_c^T \vectorr \Psi=\vectorr 0$ given by
$\vectorr \Psi_i \equiv {\vectorr P}(\rho){\xi_i/\rho}$ for $i=1,2$,
where $\vectorr P$ satisfies
\begin{equation}\label{eq:adjspots}
 \Delta_\rho \vectorr P -\frac{1}{\rho^2} \vectorr P +
 \vectorr M_c^T \vectorr P = \vectorr{0}\,, \qquad 0<\rho<\infty\,;
 \qquad \vectorr P \sim \vecol{0}{{1/\rho}} \quad \textrm{as} 
\quad \rho \to \infty\,.
\end{equation}
Here the condition at infinity in \cref{eq:adjspots} is used as a
normalization condition for $\vectorr P$.

Next, to derive our solvability condition we use Green's identity over
a large disk $\Omega_{\rho_0}$ of radius $|\boldsymbol \xi|=\rho_0\gg
1$ to obtain for $i=1,2$ that
\begin{equation}\label{d:fred_1}
   \lim_{\rho_0\to\infty} \int_{\Omega_{\rho_0}} \Big( 
   \vectorr \Psi_i^T {\mathcal L} \hat{\vectorr Q}_1  
   - \hat{\vectorr Q}_1
   {\mathcal L}^{\star} \vectorr \Psi_i \, \Big) d\boldsymbol\xi = 
    \lim_{\rho_0\to\infty} \int_{\partial\Omega_{\rho_0}}
   \left(  \vectorr \Psi_i^T \partial_{\rho} \hat{\vectorr Q}_1 - 
   \hat{\vectorr Q}_1 \partial_{\rho} \vectorr \Psi_i \right) 
\Big{\vert}_{\rho=\rho_0} \, dS \,.
\end{equation}
With $\mathbf{\Psi}_i\equiv \vectorr P(\rho) {y_i/\rho}$, for $i\in
\lbrace{1,2\rbrace}$, we first calculate the left-hand side (LHS) of
(\ref{d:fred_1}) using~\cref{eq:dynama} to obtain with $\vectorr
P\equiv (P_1(\rho),P_2(\rho))^T$ and $\vectorr x_j^{\prime}\equiv
(x_{j1}^{\prime},x_{j2}^{\prime})^T$ that
\begin{flalign*}\label{adj:LHS}
 \textrm{LHS} = & \lim_{\rho_0\to\infty} \int_{\Omega_{\rho_0}} \left(
 -\vectorr \Psi_i^T \vecol{\vectorr x_j^{\prime} \cdot \nabla u_c}{0}
 + \vectorr \Psi_i^T \left(\boldsymbol \xi \cdot \frac{\nabla
   \alpha_j}{\alpha_j} \right) \vecol{-1}{{\tau/\beta}} u_c^2v_c
 \right) \, d\boldsymbol\xi \nonumber \\ 
   =& \lim_{\rho_0\to\infty}
 \left[- \int_{\Omega_{\rho_0}} P_1
   \frac{\xi_i}{\rho} \sum_{k=1}^{2} x_{jk}^{\prime} u_c^{\prime}(\rho)
   \frac{\xi_k}{\rho} \, d\boldsymbol \xi + \frac{\nabla
     \alpha_j}{\alpha_j} \cdot \int_{\Omega_{\rho_0}} \boldsymbol \xi
   \left(-P_1+\frac{\tau}{\beta} P_2 \right)
   \frac{\xi_i}{\rho} u_c^2 v_c \, d\boldsymbol\xi\right] \,,\nonumber\\ 
  = & \lim_{\rho_0\to\infty} \left[-x_{ji}^{\prime} 
   \int_{\Omega_{\rho_0}} P_1 \frac{\xi_i^2}{\rho^2}
   u_c^{\prime}(\rho) \, d\boldsymbol \xi + \frac{(\nabla
     \alpha_j)_i}{\alpha_j} \int_{\Omega_{\rho_0}} \frac{\xi_i^2}{\rho}
     \left(-P_1+\frac{\tau}{\beta} P_2 \right)
     u_c^2 v_c \, d\boldsymbol\xi\right] \,,\nonumber \\ 
= & -x_{ji}^{\prime} \pi \int_0^\infty P_1 u_{c}^{\prime}(\rho) \rho \,
  d\rho + \frac{\pi (\nabla \alpha_j)_i }{\alpha_j} \int_{0}^{\infty}
    \left( -P_1 + \frac{\tau}{\beta} P_2 \right) u_c^2 v_c \rho^2
   \, d\rho \,, 
\end{flalign*}
where $\vectorr x^\prime_j=d\vectorr x_j/d\eta$. In deriving the result above, we used the identity
$\int_{\Omega_{\rho_0}} \xi_i \xi_k f(\rho)\,d\boldsymbol \xi =
\delta_{ik}\pi \int_{0}^{\rho_0} \rho^3 f(\rho) \,d\rho $,
for any radially symmetric function $f(\rho)$, where
$\delta_{ik}$ is the Kronecker delta.

Next, we calculate the right-hand side (RHS) of \cref{d:fred_1} using the
far-field behaviors of $\hat{\vectorr Q}_1$ and $\vectorr P$ as $\rho\to
\infty$. We derive that
\begin{flalign*}\label{adj:RHS}
  \textrm{RHS} &= \lim_{\rho_0\to\infty} \int_{\partial\Omega_{\rho_0}} \left(
  P_2 \frac{y_i}{\rho} \partial_\rho \left(
\hat{\boldsymbol\upzeta}_j \cdot \boldsymbol\xi \right) -
\hat{\boldsymbol\upzeta}_j \cdot \boldsymbol\xi \, \partial_\rho \left(
  P_2 \frac{\xi_i}{\rho} \right) \right) \, dS\,,\nonumber \\
   &= \lim_{\rho_0\to\infty} \int_{\partial\Omega_{\rho_0}} \left( P_2 \frac{\xi_i^2}
 {\rho^2} \hat{\boldsymbol\upzeta}_{ji} - 
\left(\hat{\boldsymbol\upzeta}_j \cdot \boldsymbol\xi \right)
  \frac{\xi_i}{\rho} \partial_\rho P_2 \, \right) dS =
  \lim_{\rho_0\to\infty} \int_{\partial\Omega_{\rho_0}}  \frac{2\xi_i^2}
  {\rho^3} \hat{\boldsymbol\upzeta}_{ji} \, dS = 2\pi 
  \hat{\boldsymbol\upzeta}_{ji} \,. 
\end{flalign*}
In the last passage, we used $dS=\rho_0 d\theta$ where $\theta$ is the
polar angle.  Finally, we equate LHS and RHS for $i=1,2$, and write
the resulting expression in vector form:
\begin{equation}
  -\vectorr x_j^{\prime}\pi \int_0^{\infty} P_1 u_c^{\prime}\rho \, d\rho + 
 \pi \frac{\nabla \alpha_j}{\alpha_j} \int_{0}^{\infty} \left(
   \frac{\tau}{\beta} P_2 - P_1 \right) u_c^2 v_c \rho^2 \, d\rho = 2\pi
   \hat{\boldsymbol\upzeta}_{j} \,. \label{ode:old}
\end{equation}
We summarize our main, formally-derived, asymptotic result for slow spot
dynamics as follows:

\begin{proposition}\label{prop:spotsdynamics}
Under the same assumptions as \cref{prop:multispotprof},
and assuming that the $N$-spot quasi steady-state solution is stable
on an ${\mathcal O}(1)$ time-scale, the slow dynamics on the long
time-scale $\eta=\varepsilon^2t$ of this quasi steady-state spot
pattern consists of the constraints \cref{eq:uvprofile} coupled to
the following ODEs for $j=1,\ldots,N$:
\begin{subequations} \label{eq:final}
 \begin{equation}\label{eq:thesystem}
  \frac{d\vectorr x_j}{d\eta}= n_1\hat{\boldsymbol\upzeta}_j+
 n_2\frac{\nabla\alpha_j}{\alpha_j}, 
 \qquad
 \hat{\boldsymbol\upzeta}_j\equiv-2\pi\left(S_{cj}\nabla_\vectorr x
 R_j+\sum\limits_{i\neq j}^NS_{ci}\nabla_\vectorr xG_{ji}\right)\,.
\end{equation}
The constants $n_1$ and $n_2$, which depend on $S_{cj}$ and the ratio
${\tau/\beta}$ are defined in terms of the solution to the
SCCP~\cref{full:core} and the homogeneous adjoint
solution~\cref{eq:adjspots} by
\begin{gather}\label{eq:integrals}
	n_1\equiv-\frac{2}{\displaystyle \int_0^\infty
  P_1 u_{c}^\prime \rho \, d\rho}\,,\qquad  n_2\equiv 
\frac{\displaystyle\int_{0}^{\infty}
\left(\frac{\tau}{\beta} P_2 - P_1 \right) u_c^2 v_c \rho^2 \, d\rho}
{\displaystyle  \int_{0}^{\infty} P_{1} u_c^{\prime} \rho \, d\rho} \,.
\end{gather}
\esub 
In \cref{eq:thesystem}, the source parameters $S_{cj}$ satisfy
the nonlinear algebraic system \cref{eq:uvprofile}, which depends on
the instantaneous spatial configuration of the $N$ spots. Overall,
this leads to a DAE system characterizing slow spot dynamics.
\end{proposition}

\cref{prop:spotsdynamics} describes the slow dynamics of a collection
of $N$ localized spots under an arbitrary, but smooth,
spatially-dependent auxin gradient. It is an extension of the 1-D
analysis of spike evolution, considered in \cite{bcwg}. The dynamics
in \cref{eq:thesystem}, shows that the spot locations depend on the
gradient of the Green's function, which depends on the domain
$\Omega$, as well as the spatial gradient of the auxin
distribution. In particular, the spot dynamics depends only indirectly
on the magnitude of the auxin distribution $\alpha(\vectorr x_j)$
through the source parameter $S_{cj}$. The auxin gradient $\nabla
\alpha_j$, however, is essential to determining the true steady-state
spatial configuration of spots. In addition, the spatial interaction
between the spots arises from the terms in the finite sum of
\cref{eq:thesystem}, mediated by the Green's function.  Since the
Green's function and its regular part can be found analytically for a
rectangular domain (cf.~\cite{kolo01}), we can readily use
\cref{eq:final} to numerically track the slow time-evolution of a
collection of spots for a specified auxin gradient.

\begin{figure}[t!]
 \begin{center}
 \centering
 \subfigure[]{\includegraphics[height=0.2\textheight]{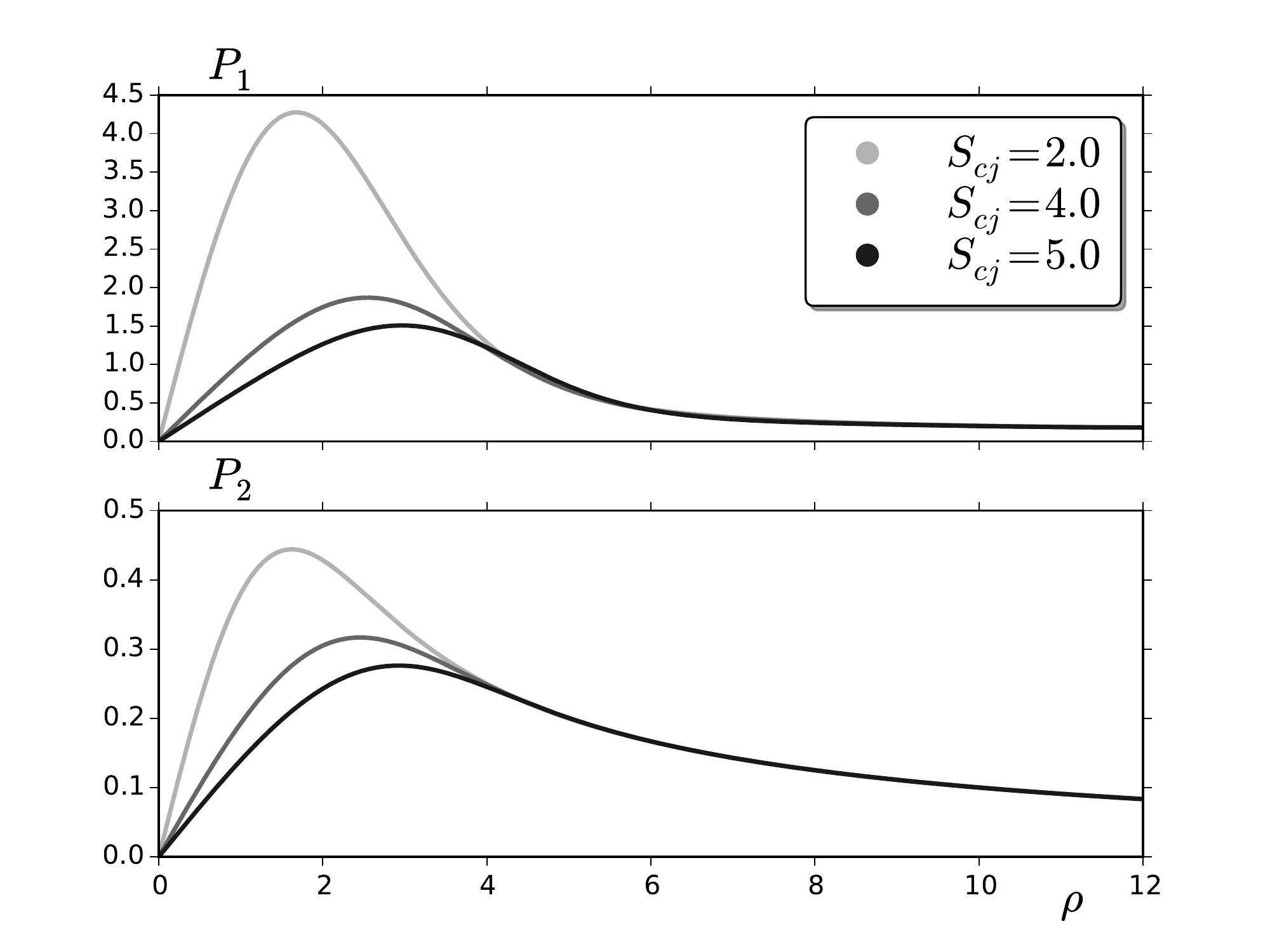}\label{sf:p1p2}}
 \centering
 \subfigure[]{\includegraphics[height=0.2\textheight]{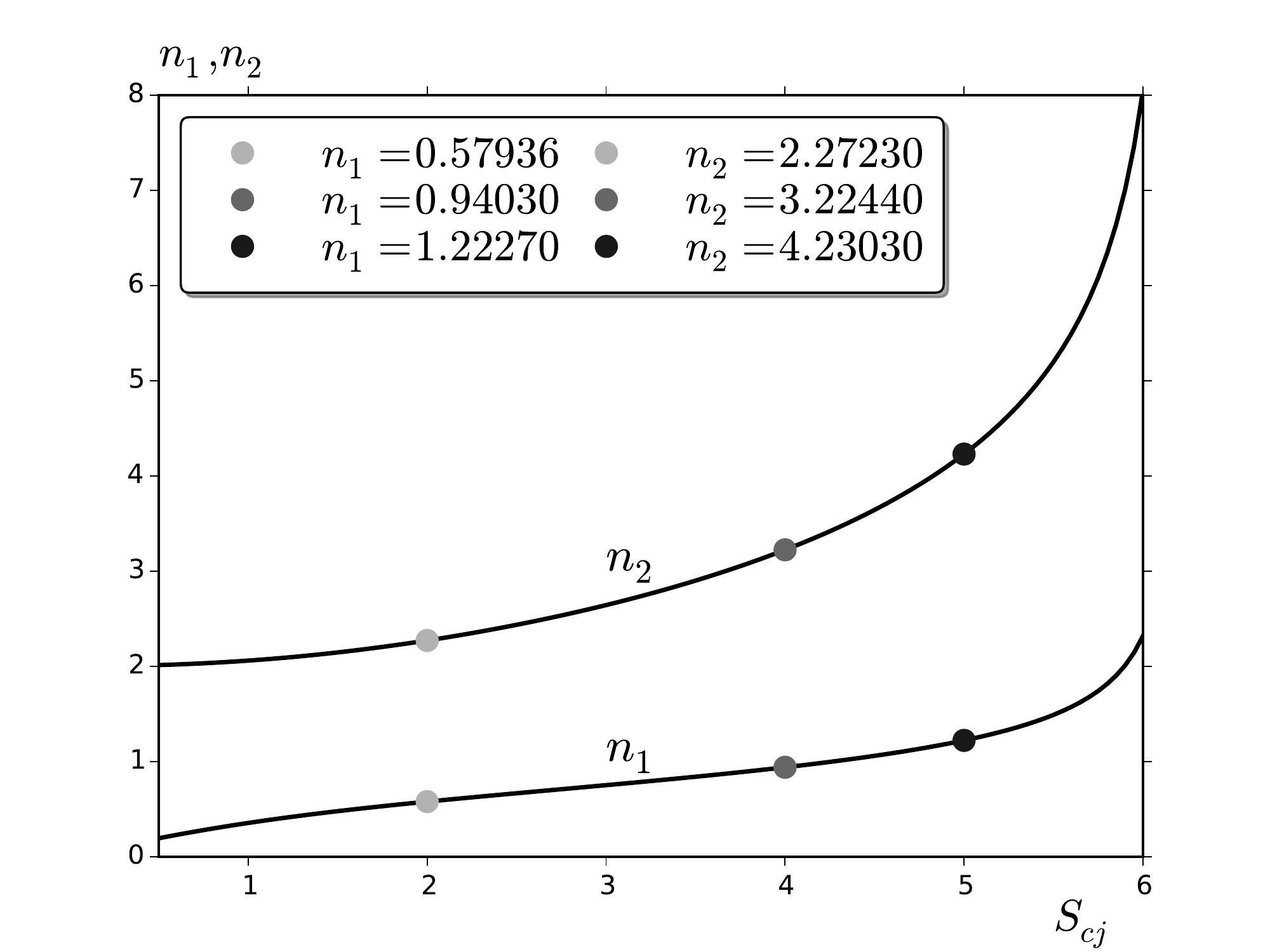}\label{sf:n1n2}}
 \end{center}
\caption{(a)~Numerical solution of the adjoint solution satisfying
  \cref{eq:adjspots} for $\vectorr P = (P_1,P_2)^T$ for values of
  $S_{cj}$ as shown in the legend; $P_1$~(top panel) and $P_2$ (bottom
  panel). (b)~Numerical results for the solvability condition
  integrals $n_1$ and $n_2$, defined in~\cref{eq:integrals}, as
  $S_{cj}$ increases. The filled circles correspond to $S_{cj}$ values
  in~(a). We use Parameter Set A from \cref{tab:tab}, for which
  ${\tau/\beta}=3$. }\label{fig:p1p2n1n2}
\end{figure}

Before illustrating results from the DAE dynamics, we must determine
$n_1$ and $n_2$ as a function of $S_{cj}$ for a prescribed ratio
${\tau/\beta}$. This ratio is associated with the linear terms in the
kinetics of the original dimensional system~\cref{eq:spotsystem},
which are related to the deactivation of ROPs and production of other
biochemical complexes which promote cell wall softening
(cf.~\cite{bcwg,payne01}). To determine $n_1$ and $n_2$, we first
solve the adjoint problem \cref{eq:adjspots} numerically using the
\textsc{MATLAB} routine \textsc{BVP4C}. This is done by enforcing the
local behavior that $\vectorr P = {\mathcal O}(\rho)$ as $\rho\to 0$
and by imposing the far-field behavior for $\vectorr P$, given
in~\cref{eq:adjspots}, at $\rho=\rho_0=12$. In \cref{sf:p1p2} we plot
$P_1$ and $P_2$ for three values of $S_{cj}$, where ${\tau/\beta}=3$,
and Parameter Set A in \cref{tab:tab} was used.  For each of the three
values of $S_{cj}$, we observe that the far-field behavior
$P_2(\rho)\sim {1/\rho}$ and $P_{1}(\rho)\sim
\left(\tau/\beta-1\right)/\rho$ as $\rho\to \infty$, which is readily
derived from \cref{eq:adjspots}, is indeed satisfied. Upon performing
the required quadratures in \cref{eq:integrals}, in \cref{sf:n1n2} we
plot $n_1$ and $n_2$ versus $S_{cj}$.  These numerical results
  show that $n_1>0$ and $n_2>0$ for $S_{cj}>0$, which will ensure
  existence of stable fixed points of the DAE dynamics. These stable
  fixed points correspond to realizable steady-state spot
  configurations for our two specific forms for the auxin gradient.

\subsection{Comparison Between Theory and Asymptotics for Slow Spot Dynamics}
\label{subsec:exper_slow}

In this subsection we compare predictions from our asymptotic theory
for slow spot dynamics with corresponding full numerical results
computed from~\cref{eq:uvspots} using a spatial mesh with 500 and 140
gridpoints in the $x$ and $y$ directions, respectively. For the
time-stepping a modified Rosenbrock formula of order 2 was used.

\begin{figure}[t!]
 \begin{center}
 \includegraphics[height=0.29\textheight]{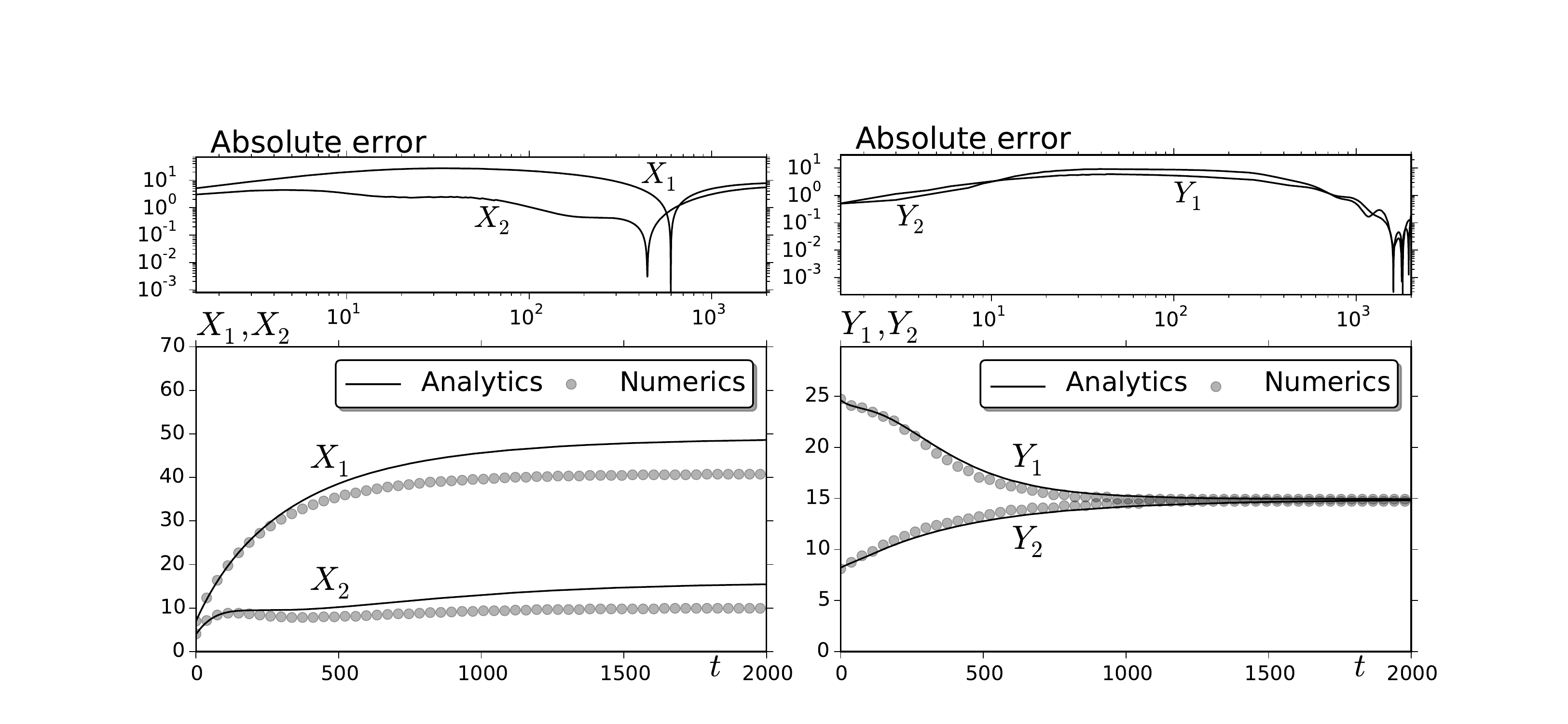}\label{sf:loctwospota}
\end{center}
	\caption{The time-dependent location of two spots for an auxin
          gradient of the type~(i) in~\cref{mod:aux} as obtained from
          the DAE system (solid curves) in \cref{prop:spotsdynamics}
          with $k_{20}=0.3$ and with parameters of
          Parameter Set A of \cref{tab:tab}.  The initial locations
          for the two spots are $X_1(0)=7.0$, $Y_{1}(0)=8.21$ and
          $X_2(0)=3.97$, $Y_2(0)=24.63$. The full numerical results
          computed from~\cref{eq:spotsystem}~ are the filled
          circles; absolute error in the top panels. Observe that the DAE system predicts that the two
          spots become aligned along the longitudinal midline of the
          cell.  The $x$-coordinate (left-hand panels) and the $y$-coordinate
          (right-hand panels) of the two spots.}
	\label{fig:loctwospot}
\end{figure}

The procedure to obtain numerical results from our asymptotic DAE
system in \cref{prop:spotsdynamics} is as follows. This DAE
system is solved numerically by using Newton's method to solve the
nonlinear algebraic system \cref{eq:uvprofile} together with a
Runge--Kutta ODE solver to evolve the dynamics in
\cref{eq:thesystem}. The solvability integrals $n_1(S_{cj})$ and $n_2(S_{cj})$
in the dynamics~\cref{eq:thesystem} and the function~$\boldsymbol \chi_c$ in the
nonlinear algebraic system~\cref{eq:uvprofile} are pre-computed at
$200$ grid points in~$S_{cj}$ and a cubic spline interpolation is fitted to
the discretely sampled functions to compute them at arbitrary values
of~$S_{cj}$. For the rectangular domain, explicit expressions for the
Green's functions $G$ and $R$, together with their gradients, as
required in the DAE system, are calculated from the expressions in \S
4 of \cite{kolo01}.

To compare our results for slow spot dynamics we use Parameter Set A
of \cref{tab:tab} and take $k_{20}=0.3$. For the auxin gradient we
took the monotone 1-D gradient in type~(i) in \cref{mod:aux}. By a
small perturbation of the unstable 1-D stripe solution, our full
numerical computations of \cref{eq:uvspots} lead to the creation of
two localized spots. Numerical values for the centers of the two spots
are calculated (see the caption of \cref{fig:loctwospot}) and these
values are used as the initial conditions for our numerical solution
of the DAE system in \cref{prop:spotsdynamics}.  In
\cref{fig:loctwospot} we compare our full numerical results for the
$x$ and~$y$ coordinates for the spot trajectories, as computed from
the RD system \cref{eq:uvspots}, and those computed from the
corresponding DAE system. The key distinguishing feature in these
results is that the spots become aligned to the longitudinal midline
of the domain.  We observe that the $y$-components of the spot
  trajectories are predicted rather accurately over long time
  intervals by the DAE system. However, although the $x$-components of
  the trajectories are initially close, they deviate somewhat as $t$
  increases. Since we only have a formal asymptotic theory, with no
  error bounds, and because it is computationally expensive to perform
  more refined numerical simulations over very-long time scales from
  the full RD system, we are not able to precisely identify why the
  agreement in the $y$-coordinate is better than for the
  $x$-coordinate. However, overall, the asymptotic theory accurately
  identifies the time-scale over which the two localized spots become
  aligned with the auxin gradient along the mid-line of the cell.

\section{Fast $\mathcal O(1)$ Time-Scale Instabilities}\label{sec:instab2D}

We now briefly examine the stability properties of the
$N$-spot quasi-equilibrium solution of \cref{prop:multispotprof} to ${\mathcal O}(1)$
time-scale instabilities, which are fast relative to the slow dynamics. We first
consider spot self-replication instabilities associated with
non-radially symmetric perturbations near each spot.

\subsection{The Self-Replication Threshold}\label{sec:split}

Since the speed of the slow spot drift is ${\mathcal O}(\eps^2)\ll 1$,
in our stability analysis below we will assume that the spots are
``frozen'' at some configuration $\vectorr x_1,\ldots,\vectorr x_N$.
We will consider the possibility of instabilities that are locally
non-radially symmetric near each spot. In the inner region near the
$j$-th spot at $\vectorr x_j$, where~$\boldsymbol
\xi=\eps^{-1}(\vectorr x-\vectorr x_j)$, we linearize
\cref{eq:uvspots} around the leading-order core solution $u_{0j}$,
$v_{0j}$, satisfying \cref{eq:uvcore}, by writing $u = u_{0j} +
e^{\lambda t} \phi_{0j}$ and $v = v_{0j} + e^{\lambda t} \psi_{0j}$.
From \cref{eq:uvspots}, we obtain to leading order that
\begin{align*}
  & \Delta_{\boldsymbol \xi}\phi_{0j} + \alpha(\vectorr x_j) \left(
 u_{0j}^2 \psi_{0j} + 2 u_{0j} v_{0j} \phi_{0j}\right) - \phi_{0j}=
\lambda \phi_{0j}, \\
  & D_0 \Delta_{\boldsymbol \xi}\psi_{0j} - \tau \gamma \left[ \alpha(\vectorr x_j) 
  \left(u_{0j}^2 \psi_{0j} + 2u_{0j} v_{0j} \phi_{0j}\right) - \phi_{0j}
  \right] - \beta \gamma \phi_{0j} = 0,
\end{align*}
where $\Delta_{\boldsymbol \xi}$ is the Laplacian in the local
variable $\boldsymbol \xi$. Then, upon relating $u_{0j}$, $v_{0j}$ to
the SCCP by using (\ref{eq:corechvar}), and defining $\psi_{0j}\equiv
{\beta\gamma\tilde{\psi}_{0j}/D_0}$, the system above reduces to
\begin{equation}\label{split:2}
  \begin{aligned}
  & 
  \Delta_{\boldsymbol \xi}\phi_{0j} + \tilde{\psi}_{0j} u_c^2 + 2 u_c v_c 
 \phi_{0j} - \phi_{0j} = \lambda \phi_{0j}\,, \qquad \\
 &
 \Delta_{\boldsymbol \xi}\tilde{\psi}_{0j} + \left(
  \frac{\tau}{\beta} - 1 - \frac{2\tau}{\beta} u_c v_c \right) 
  \phi_{0j} - \frac{\tau}{\beta} u_c^2 \tilde{\psi}_{0j} =0\,,
  \end{aligned}
\end{equation}
where $u_c$, $v_c$ is the solution to the SCCP \cref{full:core}.

We then look for an ${\mathcal O}(1)$ time-scale instability
associated with the local angular integer mode $m\geq 1$ by
introducing the new variables $\Phi_0(\rho)$ and $\Psi_0(\rho)$ defined by
\begin{equation}  \label{split:inn}
   \phi_{0j} = e^{i m \theta} \Phi_0(\rho)\,, 
   \quad
   \tilde{\psi}_{0j} = e^{im\theta} \Psi_{0}(\rho)\,, 
   \quad \text{where $\rho=|\boldsymbol \xi|$, $\boldsymbol \xi=\eps^{-1}(\vectorr x-\vectorr x_j)\,,$}
\end{equation}
and $\boldsymbol \xi^T=\rho(\cos\theta,\sin\theta)$. Substituting
\cref{split:inn} into \cref{split:2}, we obtain the eigenvalue
problem:
\begin{equation}\label{eq:eig_split}
  \begin{aligned}
  & {\cal L}_m \Phi_0 + \left(2 u_c v_c -1\right) \Phi_0 + u_c^2 \Psi_0 = 
\lambda \Phi_0\,, \\
  & {\cal L}_m \Psi_0  + \left(\frac{\tau}{\beta}-1 - 
\frac{2\tau}{\beta} u_c v_c \right) \Phi_0 - 
\frac{\tau}{\beta}u_c^2 \Psi_0 = 0 \,,
  \end{aligned}
  \qquad 0 \leq \rho < \infty\,.
\end{equation}
Here we have defined ${\cal
  L}_m\Upsilon\equiv\partial_{\rho\rho}\Upsilon+\rho^{-1}\partial_{\rho}\Upsilon
-m^2\rho^{-2}\Upsilon$. We impose the usual regularity condition for
$\Phi_0$ and $\Psi_0$ at $\rho=0$.  The appropriate far-field boundary
conditions for~\cref{eq:eig_split} is discussed below.

\begin{figure}[t!]
\begin{center}
 \centering
 \subfigure[]{\includegraphics[height=0.2\textheight]{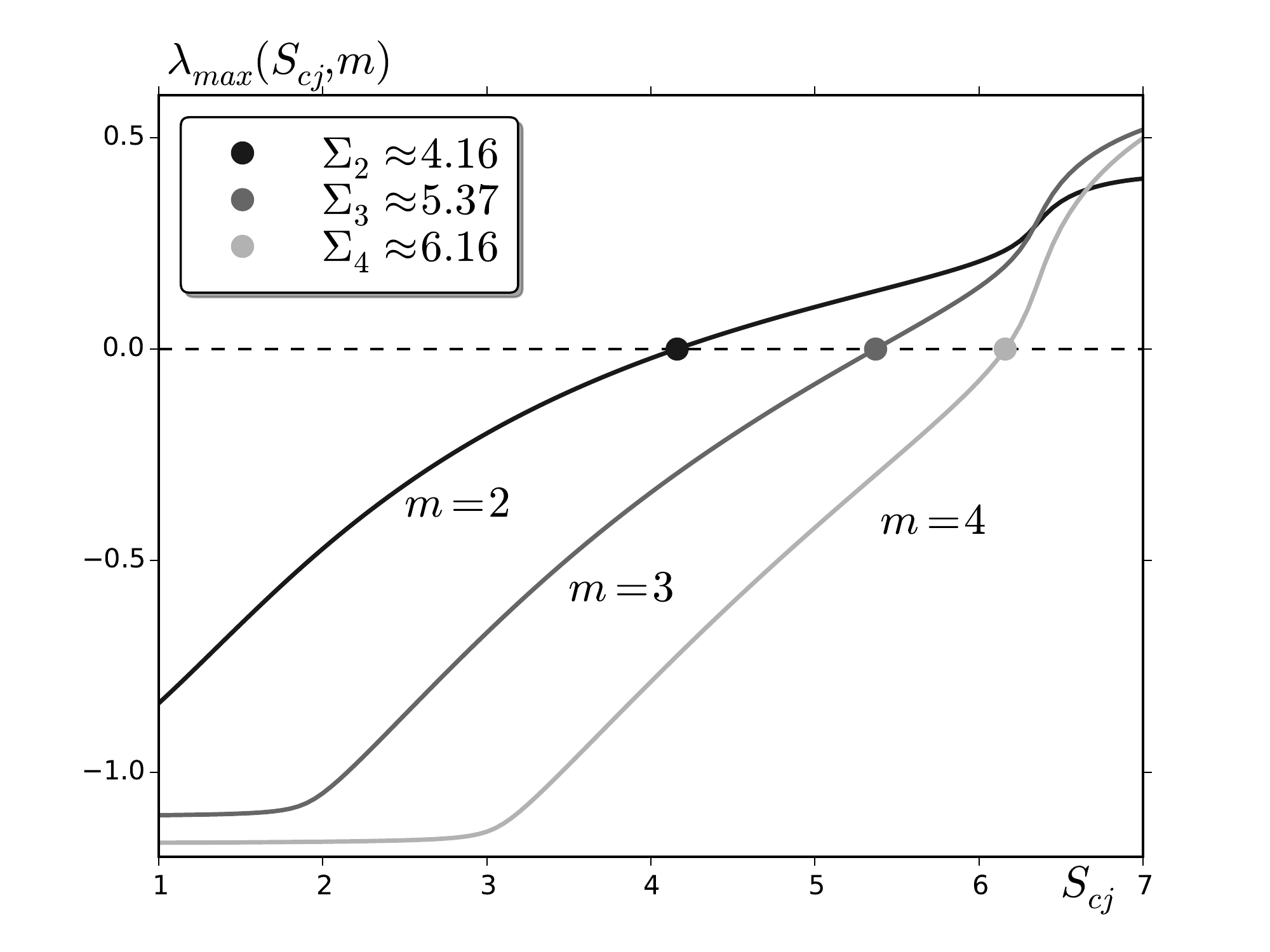}
\label{sf:eig}}
 \centering
 \subfigure[]{\includegraphics[height=0.2\textheight]{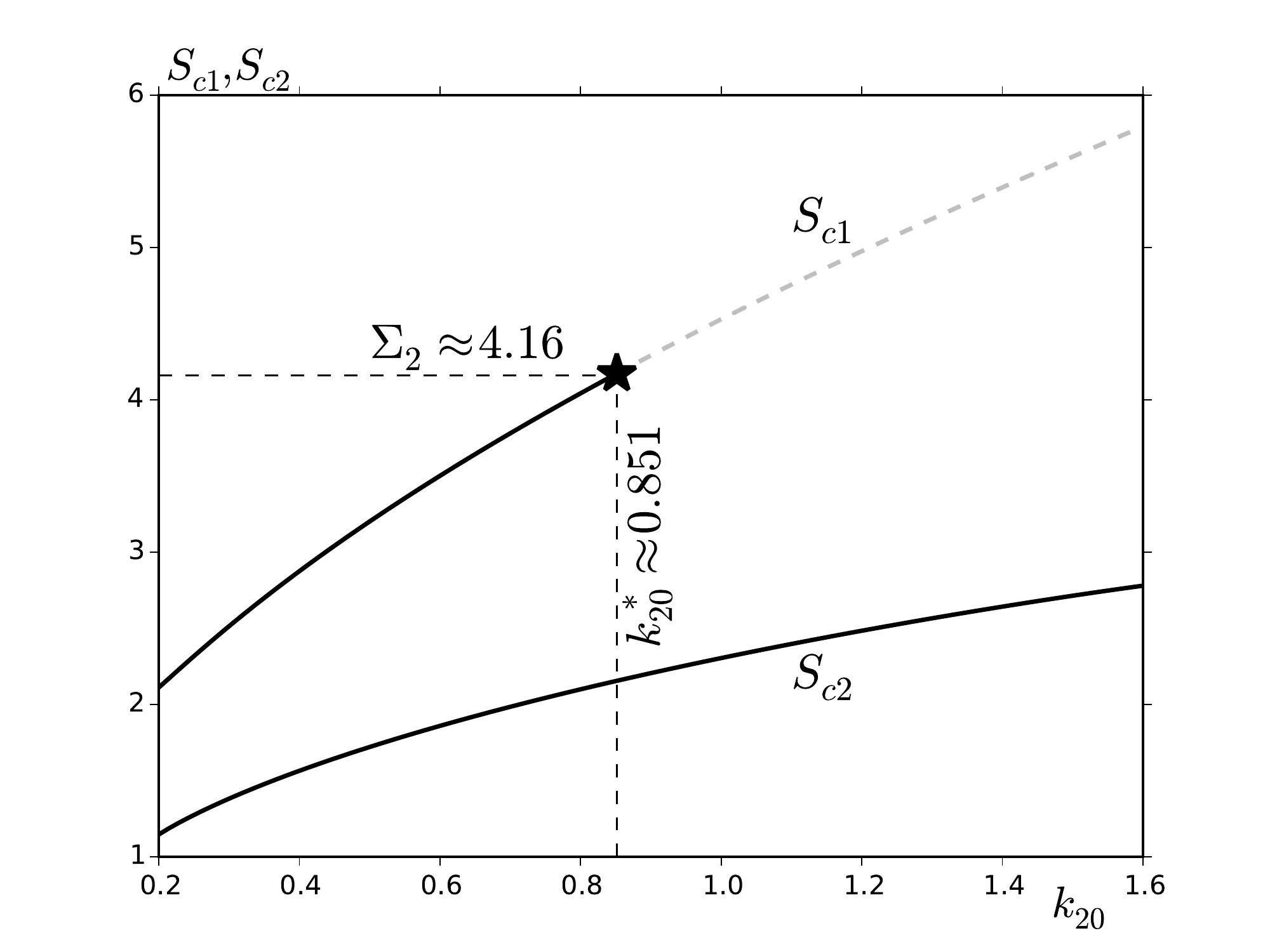}
\label{sf:s1s2k20}}
 \end{center}
\caption{(a)~The largest real-valued eigenvalue
  $\lambda_{\textrm{max}}$ of \cref{eq:eig_split} versus $S_{cj}$ for
  different angular modes $m$. The peanut-splitting
  thresholds~$\Sigma_m$ for $m=2,3,4$ are indicated by filled
  circles. (b)~Source parameters for a two-spot steady-state solution
  of the DAE system \cref{eq:final} as $k_{20}$ is varied. Stable
  profiles are the solid curves, and the spot closest to
  the left boundary at $x=0$ is unstable to shape deformation on the
  dashed curve, which begins at $k_{20}^* \approx 0.851$. This value
  is the asymptotic prediction of the pitchfork bifurcation point
  in~\cref{sf:expgrada} drawn by a filled star. Parameter Set A
  of~\cref{tab:tab} was used.}
\label{fig:split}
\end{figure}

 Since the eigenvalue problem \cref{eq:eig_split} is difficult to
study analytically, we solve it numerically for various integer
values of $m$. We denote $\lambda_{\textrm{max}}$ to be the eigenvalue
of \cref{eq:eig_split} with the largest real part. Since $u_c$ and
$v_c$ depend on $S_{cj}$ from the SCCP~\cref{full:core}, we have
implicitly that
$\lambda_{\textrm{max}}=\lambda_{\textrm{max}}(S_{cj},m)$.  To
determine the onset of any instabilities, for each $m$ we compute the
smallest threshold value $S_{cj}=\Sigma_m$ where
$\mbox{Re}(\lambda_{\textrm{max}}(\Sigma_m,m))=0$. In our
computations, we only consider $m=2,3,4,\ldots$, since
$\lam_{\textrm{max}}=0$ for any value of $S_{cj}$ for the
translational mode $m=1$. Any such instability for $m=1$ is reflected
in instability in the DAE system~\cref{eq:thesystem}.

For $m\geq 2$ we impose the far-field behavior that $\Phi_0$ decays
exponentially as $\rho\to \infty$ while $\Psi_0\sim {\mathcal
  O}(\rho^{-m})\to 0$ as~$\rho\to \infty$. With this limiting
behavior, \cref{eq:eig_split} is discretized with centered
differences on a large but finite domain. We then determine
$\lambda_{\textrm{max}}(S_{cj},m)$ by computing the eigenvalues of the discretized
eigenvalue problem, in matrix form. For $m\geq 2$ our computations show that
$\lambda_{\textrm{max}}(S_{cj},m)$ is real and that
$\lambda_{\textrm{max}}(S_{cj},m)>0$ if and only if $S_{cj}>\Sigma_m$.  In
our computations we took $400$ meshpoints on the interval $0\leq \rho<
15$. For the ratio ${\tau/\beta}=3$, corresponding to Parameter Set A of \cref{tab:tab}, the results for the threshold values
$\Sigma_m$ for $m=2,3,4$ given in \cref{sf:eig}
are insensitive to increasing either the domain length or the number
of grid points. Our main conclusion is that as $S_{cj}$ is increased,
the solution profile of the $j$-th spot first becomes unstable to a
non-radially symmetric peanut-splitting mode, corresponding to $m=2$,
as $S_{cj}$ increases above the threshold $\Sigma_2\approx 4.16$ when
${\tau/\beta}=3$. As a remark, if we take ${\tau/\beta}=11$, corresponding
to Parameter Set B of \cref{tab:tab}, we compute instead that
$\Sigma_2\approx 3.96$.

To illustrate this peanut-splitting threshold, we consider a pattern
with a single localized spot. Using \cref{eq:v0Si} we find
$S_1={d_y/(2\pi D_0)}$. Then, \cref{eq:scjxcj} yields the following
expression for the source parameter for the SCCP \cref{full:core}
\begin{equation}\label{split:test}
 S_{c1} = \frac{d_y}{2\pi} \, \sqrt{ \frac{ \alpha(x_1)}{D_0\beta
 \gamma}} \,.
\end{equation}
When ${\tau/\beta}=3$, corresponding to Parameter Set A of
\cref{tab:tab}, we predict that the one-spot pattern will first
undergo a shape-deformation, due to the $m=2$ peanut-splitting mode,
when $S_{c1}$ exceeds the threshold $\Sigma_2\approx 4.16$. Since
$\gamma$ is inversely proportional to the parameter $k_{20}$, as shown
in \cref{eq:gabe}, which determines the auxin level, we predict that
as the auxin level is increased above a threshold a peanut structure will
develop from a solitary spot. As shown below in \cref{sec:comp} for an
auxin distribution of the type~(i) in \cref{mod:aux}, this linear
peanut-splitting instability triggers a nonlinear spot replication
event.

In addition, in~\cref{sf:s1s2k20} we plot the source parameters
corresponding to the two-spot steady-state aligned along the midline
$Y = L_y/2$ as the parameter $k_{20}$ is varied. These asymptotic
results, which asymptotically characterize states in branch A
of~\cref{sf:expgrada}, are computed from the steady-state of the DAE
system~\cref{eq:final}. Notice that the spot closest to the left-hand
boundary loses stability at $k_{20}^*\approx 0.851$, when $S_{c1}$
meets the peanut-splitting threshold $\Sigma_2$.  On the other
  hand, $S_{c2}$ values corresponding to the spot closest to the
  right-hand boundary are always below $\Sigma_2$ for the parameter
  range considered. Hence, no shape change for this spot is
  observed. This result rather accurately predicts the pitchfork
  bifurcation in~\cref{sf:expgrada} portrayed by a filled star at
  $k_{20}^*\approx 0.902$.

\subsection{Numerical Illustrations of ${\mathcal O}(1)$ Time-Scale 
Instabilities}\label{sec:comp}

\begin{figure}[t!]
	\begin{center}
\vspace*{-0.2cm}
		\centering
                \subfigure[]{\includegraphics[height=0.2\textheight]{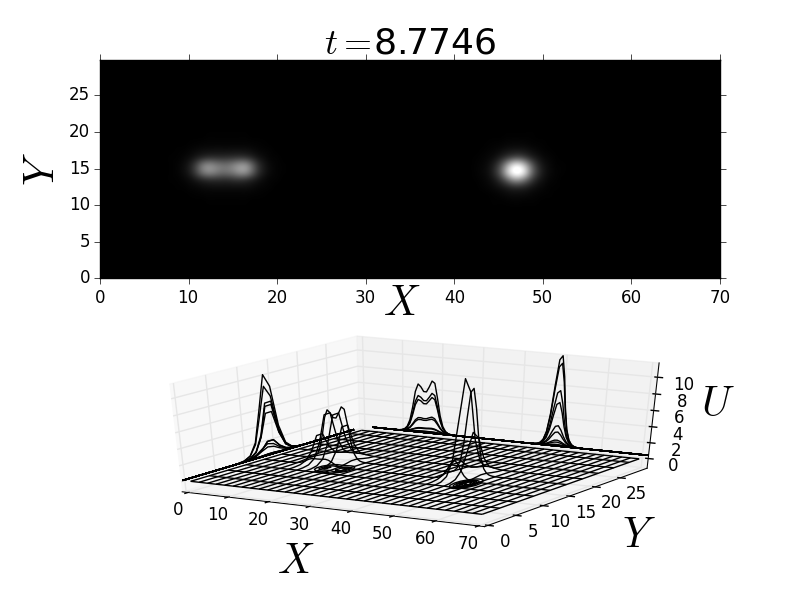}\label{sf:spotmergeda}}
\vspace*{-0.2cm}
                \centering
                \subfigure[]{\includegraphics[height=0.2\textheight]{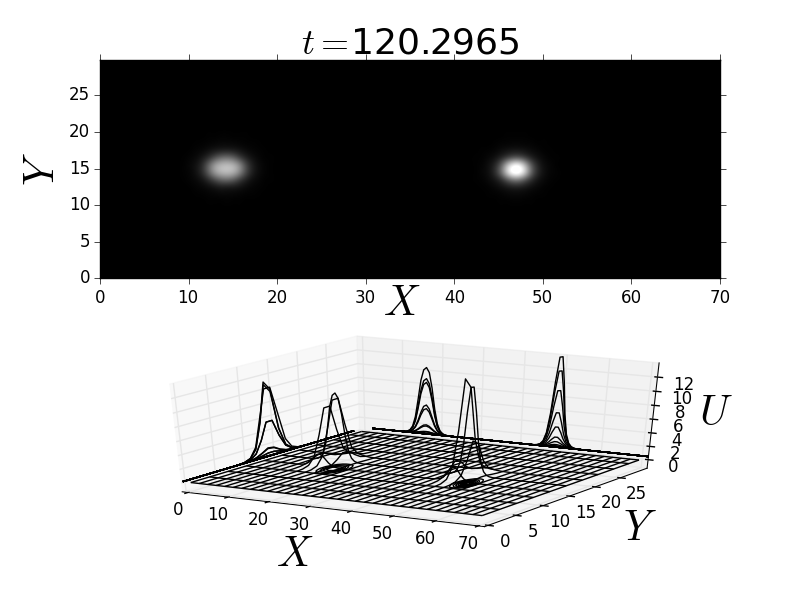}\label{sf:spotmergedb}}
	\end{center}
	\caption{Two snapshots of a time-dependent numerical
          simulation where a peanut structure merges into a spot. (a)
          Peanut structure and a spot. (b) Two spots. Parameter Set A as
          given in \cref{tab:tab} and $k_{20}=0.6723$ for an auxin gradient of the type~(i) in~\cref{mod:aux} }
	\label{fig:spotmerged}
\end{figure}

In \cref{sf:expgradd} we showed a steady-state solution consisting of
a spot and a peanut structure, which is obtained from numerical
continuation in the  parameter~$k_{20}$. This solution
belongs to the unstable branch labelled by P in \cref{sf:expgrada},
which is confined between two fold bifurcations (not explicitly
shown). We take a solution from this branch as an initial condition in
\cref{sf:spotmergeda} for a time-dependent numerical simulation of the
full RD model \cref{eq:spotsystem}-\cref{eq:uvspots} using a
centered finite-difference scheme. Our numerical results in
\cref{sf:spotmergedb} show that this initial state evolves dynamically
into a stable two-spot solution. This is a result of the overlapping
of the stable branch~A with an unstable one in the bifurcation diagram.

\begin{figure}[t!]
	\begin{center}
\vspace*{-0.2cm}
		\centering
		\subfigure[]{\includegraphics[height=0.2\textheight]{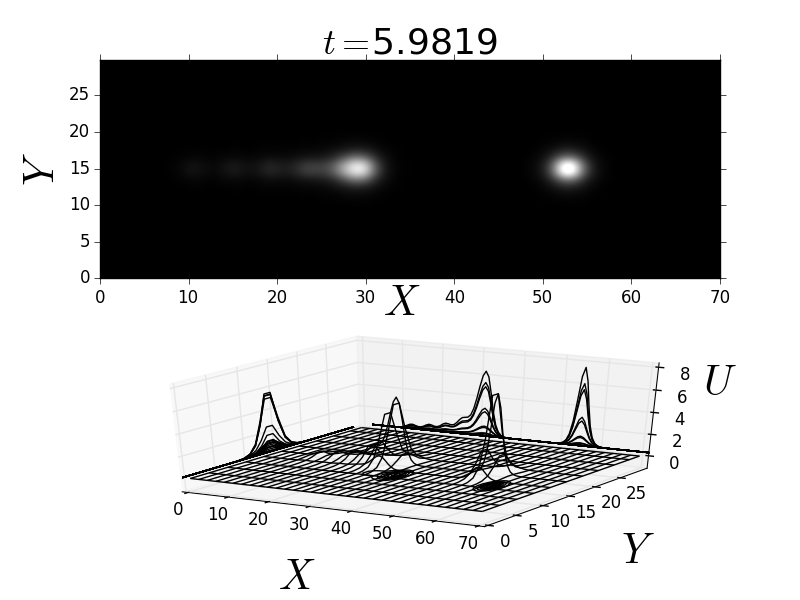}\label{sf:peanut01aa}}
\vspace*{-0.2cm}
		\centering
		\subfigure[]{\includegraphics[height=0.2\textheight]{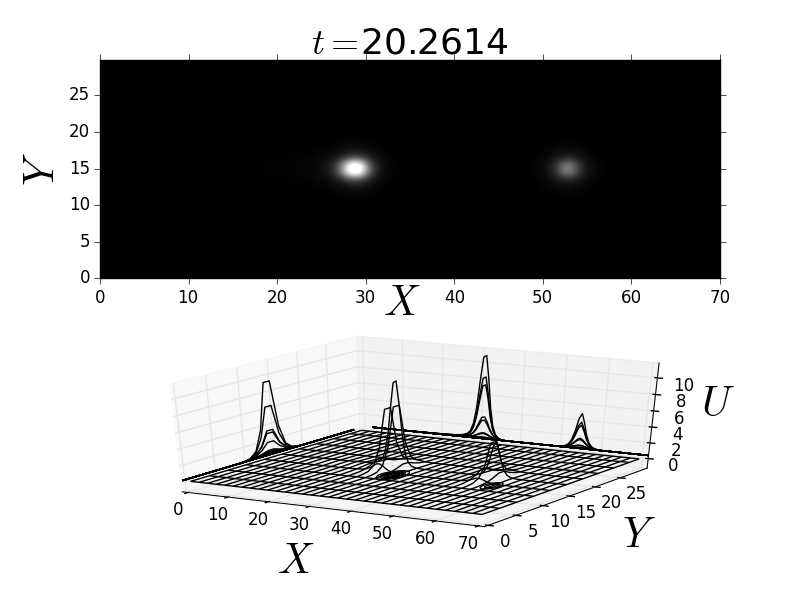}\label{sf:peanut01ab}}
\vspace*{-0.2cm}
		\centering
		\subfigure[]{\includegraphics[height=0.2\textheight]{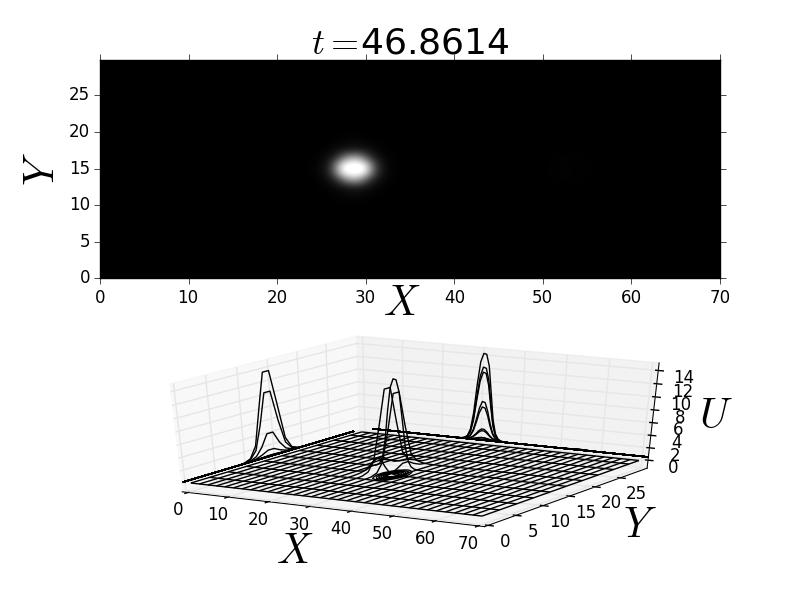}\label{sf:peanut01a}}
\vspace*{-0.2cm}
		\centering
		\subfigure[]{\includegraphics[height=0.2\textheight]{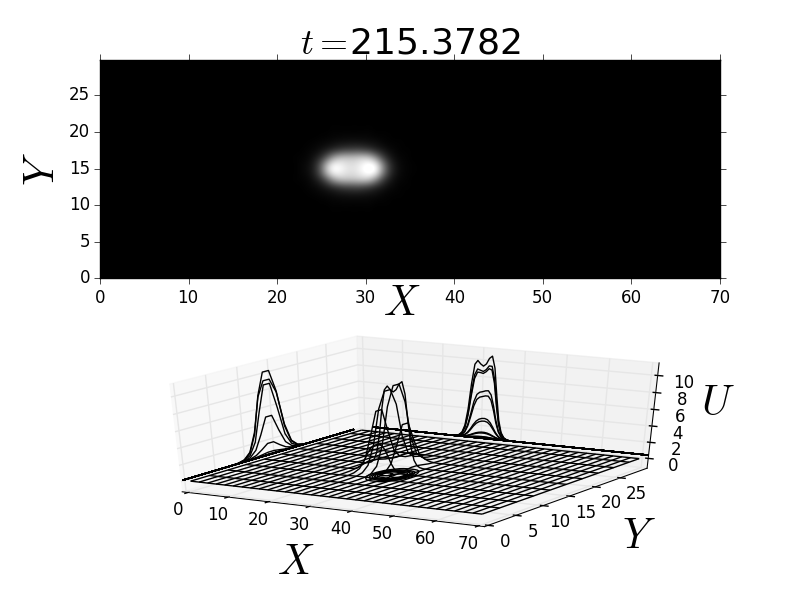}\label{sf:peanut01b}}
		\centering
		\subfigure[]{\includegraphics[height=0.2\textheight]{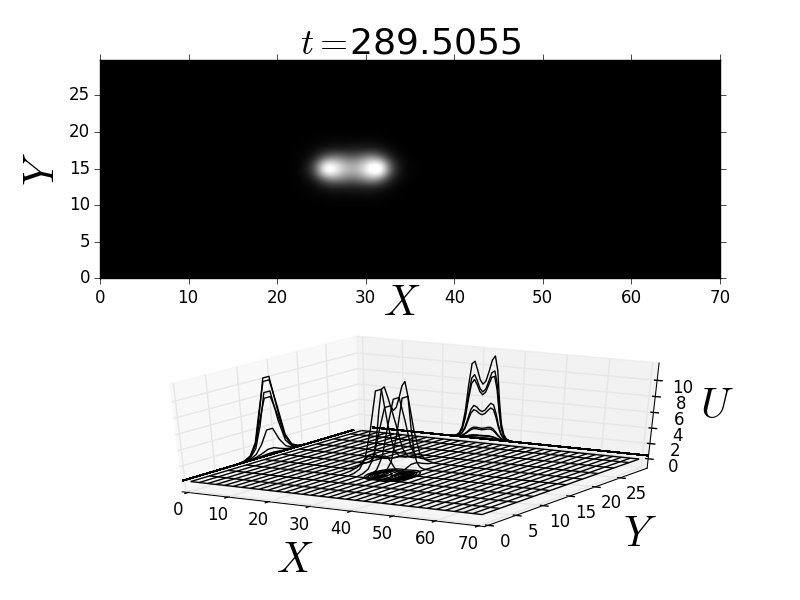}\label{sf:peanut01c}}
		\centering
		\subfigure[]{\includegraphics[height=0.2\textheight]{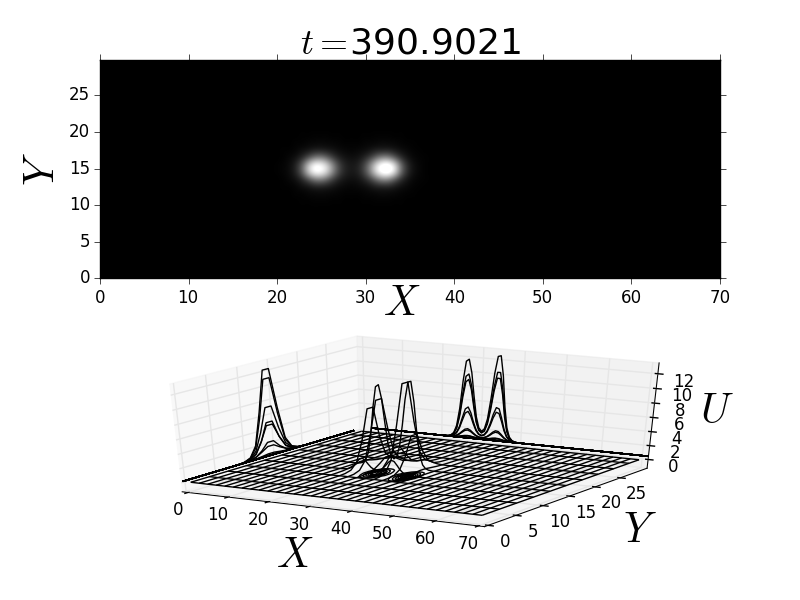}\label{sf:peanut01d}}
	\end{center}
	\caption{Numerical simulations of the full RD system
          illustrating ${\mathcal O}(1)$ time-scale
          instabilities. Competition instability: (a)~baby droplets
          and (b) a spot gets annihilated. Spot self-replication:
          (c)~one spot, (d)~early stage of a self-replication process,
          (e)~a clearly visible peanut structure, and (f)~two distinct
          spots moving away from each other. Parameter Set A as
          given in \cref{tab:tab} and $k_{20}=1.6133$.}
	\label{fig:peanut01}
\end{figure}

Other initial conditions and parameter ranges also exhibit ${\mathcal
  O}(1)$ time-scale instabilities. To illustrate these instabilities,
and see whether a self-replication spot process is triggered by a
linear peanut-splitting instability, we perform a direct numerical
simulation of the full RD model~\cref{eq:spotsystem} taking as initial
condition the unstable steady-state solution labelled by B
in~\cref{sf:expgrada}. This steady-state, shown
in~\cref{sf:peanut01aa}, consists of two spots, with one having four
small droplets associated with it. This particular steady-state
  is chosen since both competition and self-replication instabilities
  can be seen in the time evolution of this initial condition in the
  panels of~\cref{fig:peanut01}. Firstly, we observe in
\cref{sf:peanut01ab} that the small droplets for the left-most spot in
\cref{sf:peanut01aa} are annihilated in $\mathcal O(1)$ times. 
  Then, the right-most spot, which has a relatively low source
  parameter owing to its distance away from the left-hand boundary
  (see~\cref{eq:scjxcjA}), is annihilated on an ${\mathcal O}(1)$
time-scale triggered by a linear competition instability. The solitary
remaining spot at $(X_1,Y_1)=(28,14.9240)$, shown in
\cref{sf:peanut01a}, now has a source parameter $S_{c1}=9.1403$, given
in \cref{split:test} of \cref{sec:split}, that exceeds the
peanut-splitting threshold $\Sigma_2\approx 4.16$. This linear
shape-deforming instability gives rise to the peanut structure shown
in \cref{sf:peanut01b}, and the direction of spot-splitting is aligned
with the direction of the monotone decreasing auxin gradient, which
depends only on the longitudinal direction. In \cref{sf:peanut01c} and
\cref{sf:peanut01d} we show that this linear peanut-splitting
instability has triggered a nonlinear spot-splitting event that
creates a second localized spot. Finally, these two spots drift away
from each other and their resulting slow dynamics is characterized by
the DAE dynamics in \cref{prop:spotsdynamics}.

Competition and self-replication instabilities, such as illustrated
above, are two types of fast ${\mathcal O}(1)$ time-scale
instabilities that commonly occur for localized spot patterns in
singularly perturbed RD systems. Although it is beyond the scope of
this paper to give a detailed analysis of a competition instability
for our plant RD model~\cref{eq:spotsystem}, based on analogies with
other RD models (cf.~\cite{xchen01,kolok02,kolo01,rozada02}), this
instability typically occurs when the source parameter of a particular
spot is below some threshold or, equivalently, when there are too many
spots that can be supported in the domain for the given substrate
diffusivity. In essence, a competition instability is an overcrowding
instability. Alternatively,  as we have unravelled
  in~\cref{sec:split}, the self-replication instability is an
  undercrowding instability and is triggered when the source parameter
  of a particular spot exceeds a threshold, or equivalently when there
  are too few spots in the domain. For the standard Schnakenberg model
  with spatially homogeneous coefficients, in \cite{kolo01} it was
  shown through a center-manifold type calculation that the direction
  of splitting of a spot is always perpendicular to the direction of
  its motion. However, with an auxin gradient the direction of
  spot-splitting is no longer always perpendicular to the direction of
  motion of the spot. In our plant RD model, the auxin gradient not
only enhances robustness of solutions, which results in the
overlapping of solution branches and is suggestive that a strong
slanted homoclinic snaking mechanism can occur (see the next section),
but it also controls the location of steady-state spots (see
\cref{subsec:spobabydrop} and \cref{sec:dynspot}).

\section{More Robust 2D Patches and Auxin Transport}\label{chap:c07}

\begin{figure}[t!]
	\begin{center}
\vspace*{-0.2cm} 
		\centering
		\subfigure[]{\includegraphics[height=0.2\textheight]{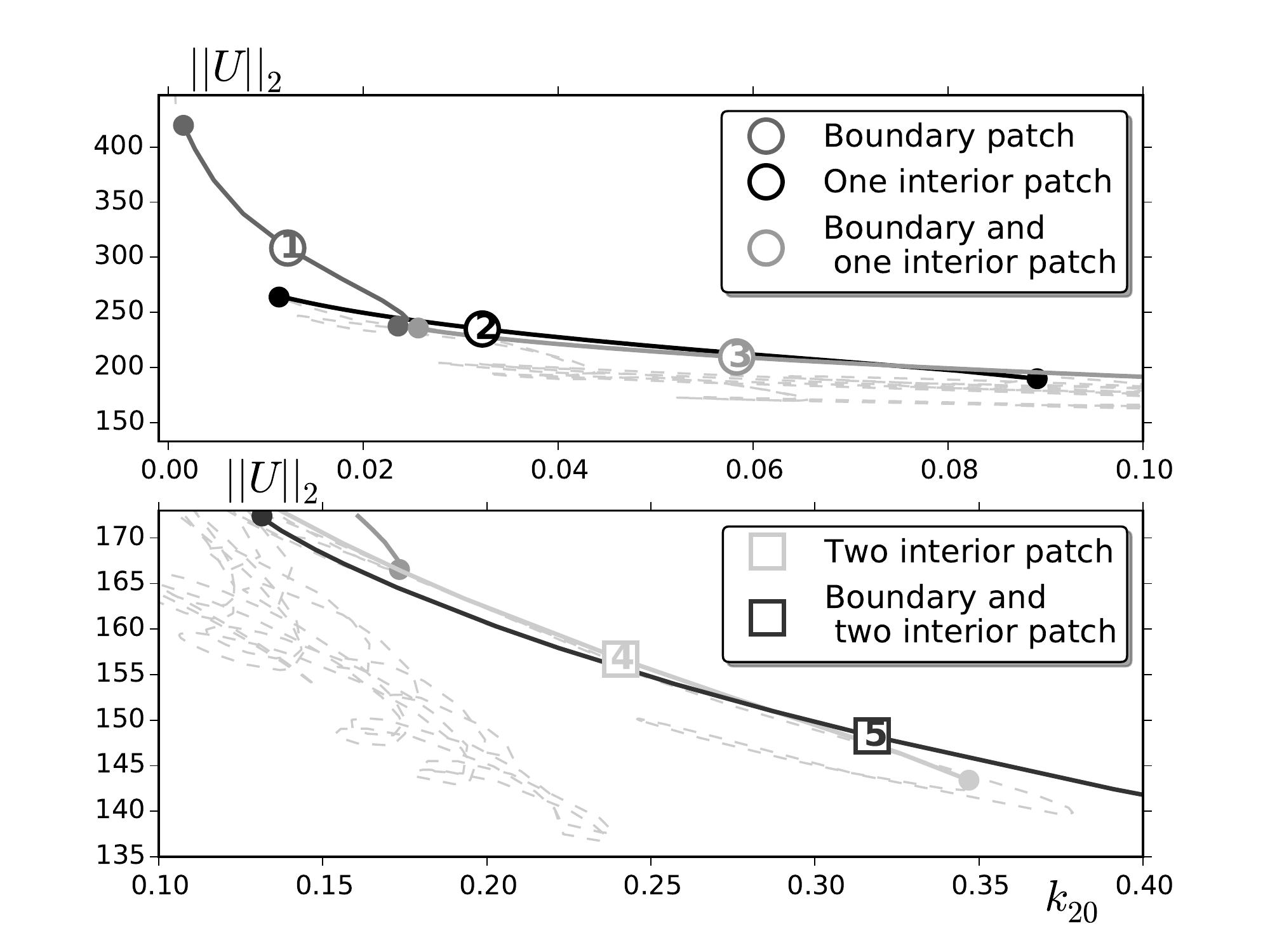}\label{sf:xygrada}}
\vspace*{-0.2cm} 
		\centering
		\subfigure[]{\includegraphics[height=0.2\textheight]{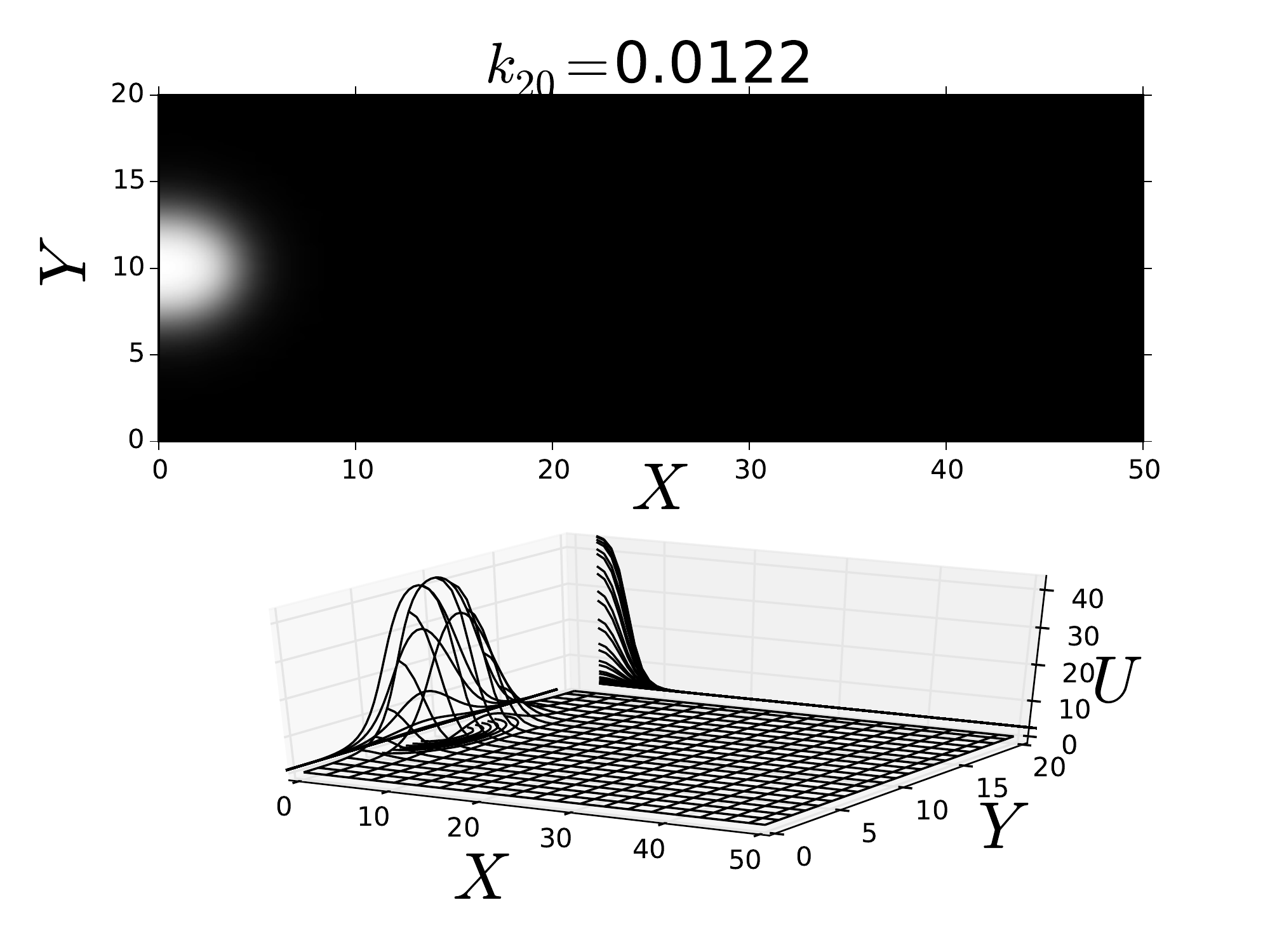}\label{sf:xygradb}}
\vspace*{-0.2cm} 
		\centering
		\subfigure[]{\includegraphics[height=0.2\textheight]{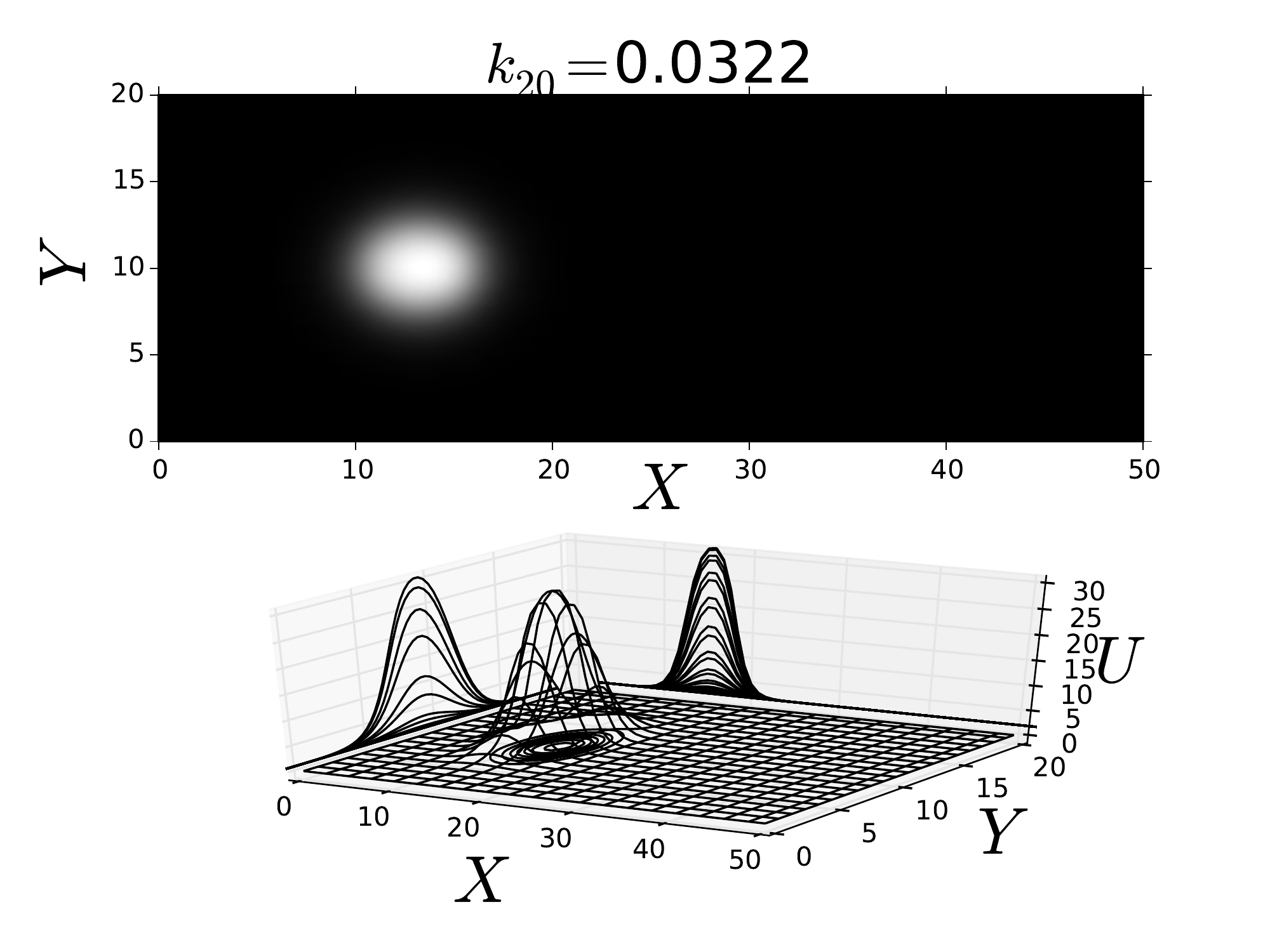}\label{sf:xygradc}}
\vspace*{-0.2cm} 
		\centering
		\subfigure[]{\includegraphics[height=0.2\textheight]{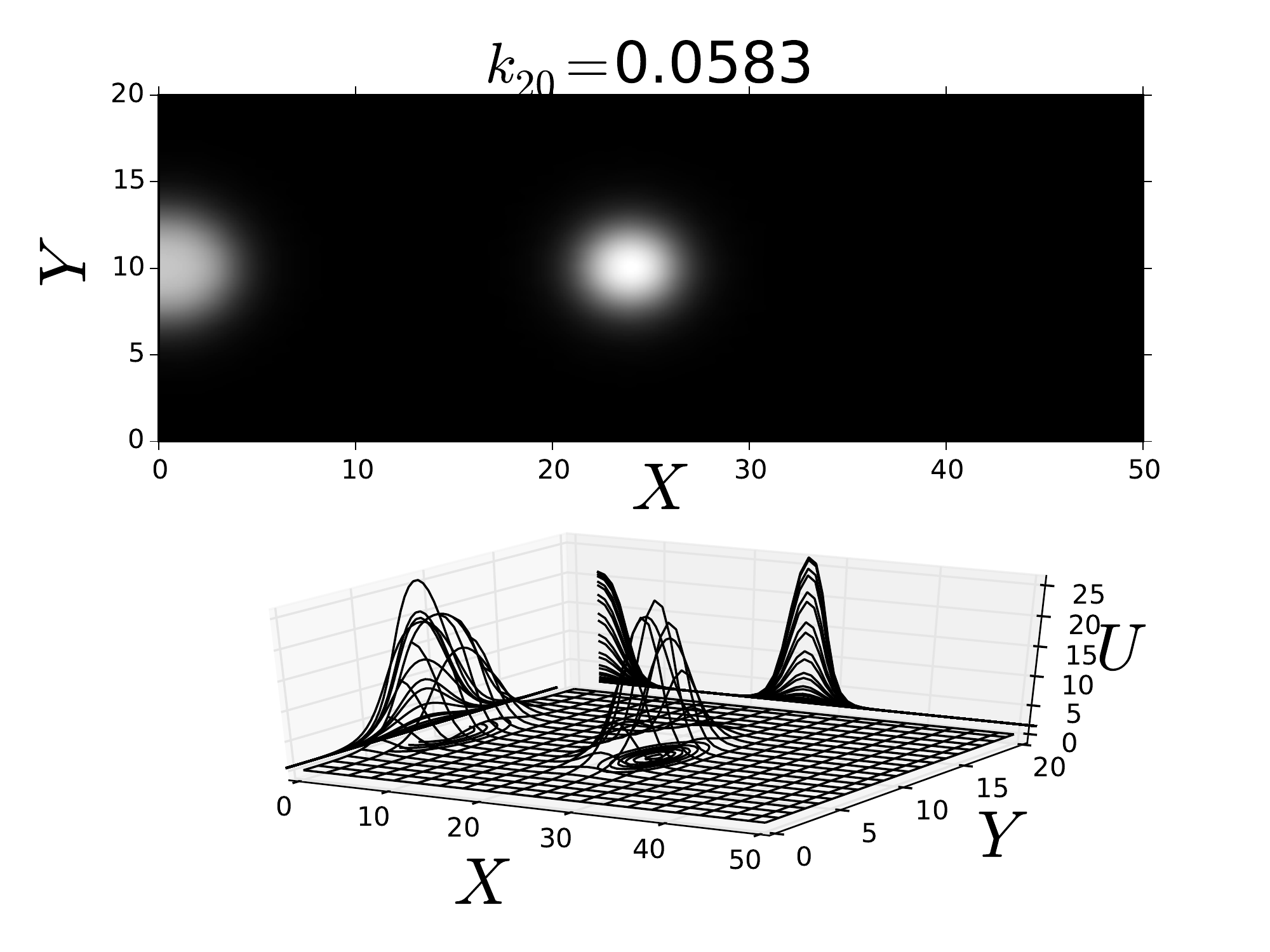}\label{sf:xygradd}}
\vspace*{-0.2cm} 
		\centering
		\subfigure[]{\includegraphics[height=0.2\textheight]{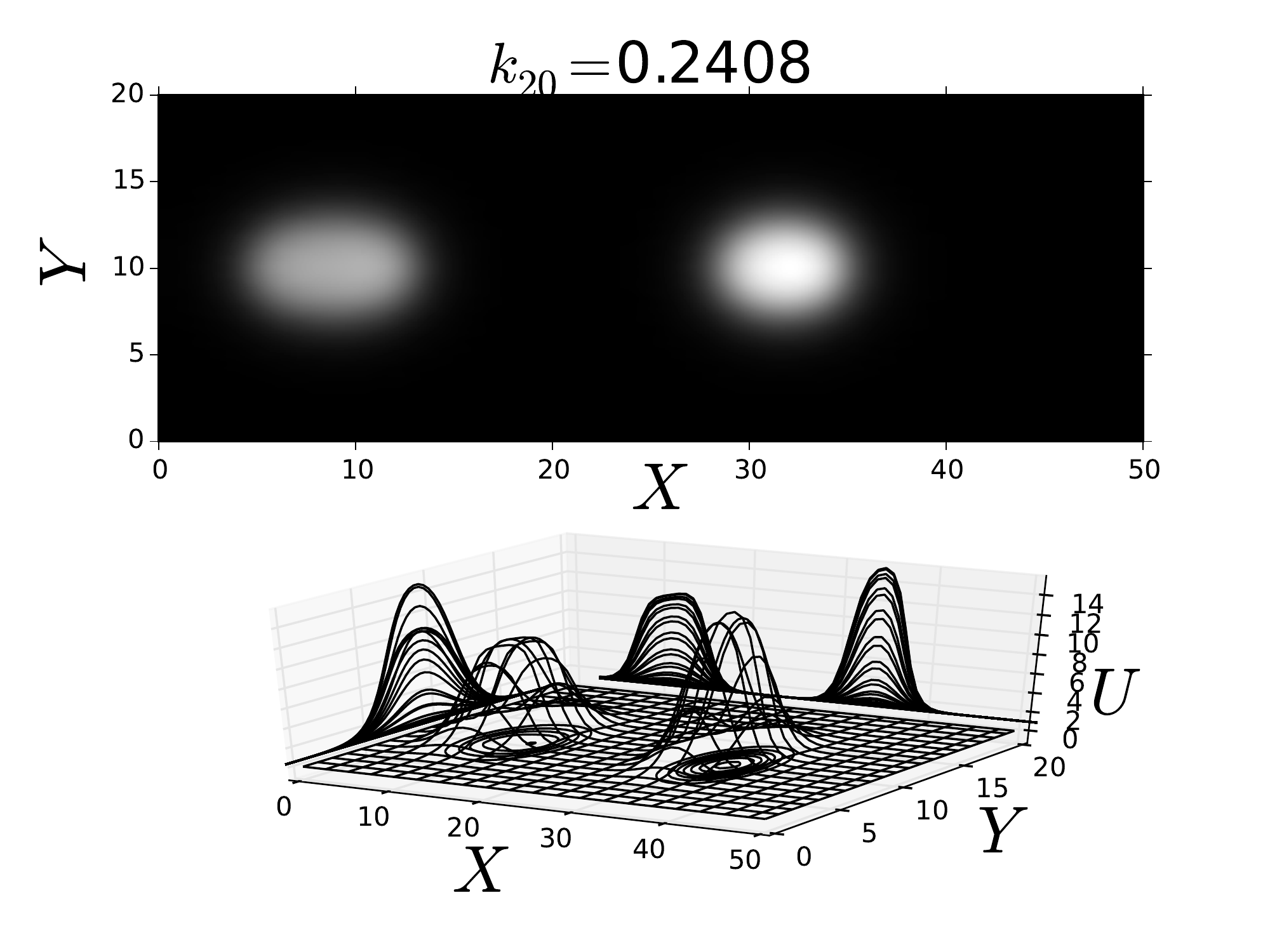}\label{sf:xygrade}}
\vspace*{-0.2cm} 
		\centering
		\subfigure[]{\includegraphics[height=0.2\textheight]{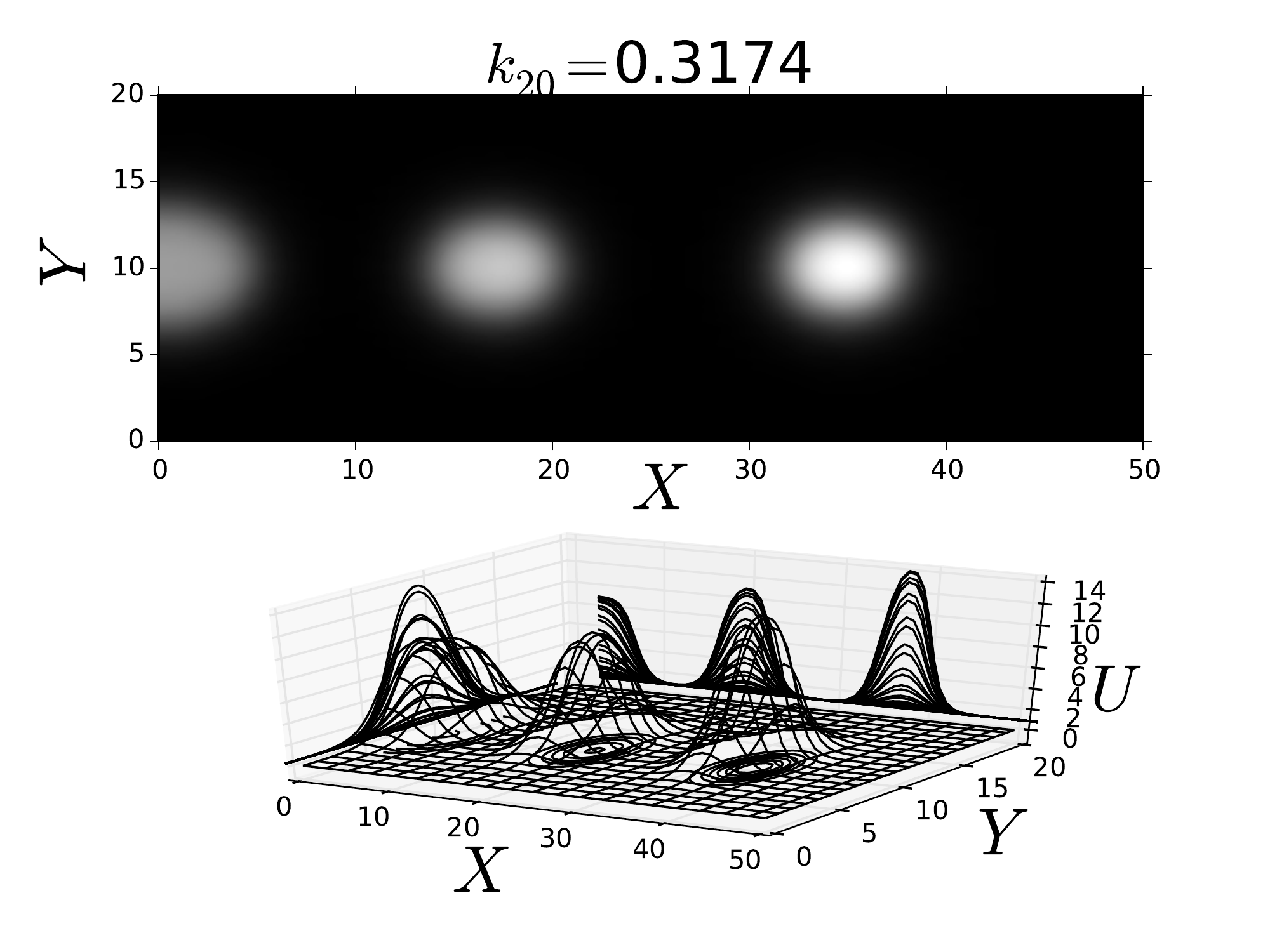}\label{sf:xygradf}}
	\end{center}
	\caption{Bifurcation diagram of the RD system
          \cref{eq:spotsystem} in terms of the original parameters
          while varying $k_{20}$. Here $\alpha=\alpha\left(\vectorr
          X\right)$ as given in (ii) of \cref{mod:aux} in a
          2-D-rectangular domain. (a) Stable branches are the solid
          lines, and filled circles represent fold points; top and
          bottom panels show overlapping of branches of each steady
          solution: (b) boundary spot, (c) one interior spot,
          (d)~boundary and one interior spot, (e) two interior
          spots, and (f) boundary and two interior
          spots. Parameter Set B, as given in \cref{tab:tab}, was
          used. The $k_{20}$ values are given on top of the upper
          panels for each steady-state in (b)-(f).  }
	\label{fig:xygrad}
\end{figure}

We now consider a more biologically plausible model for the auxin
transport, which initiates the localization of ROP. This model allows
for both a longitudinal and transverse spatial dependence of auxin.
The original ROP model, derived in~\cite{bcwg,payne01} and analyzed in
\cite{bacw,bcwg} and the sections above, depends crucially on the
spatial gradient of auxin. The key assumption we have made above was
to assume a decreasing auxin distribution along the
$X$-direction. Indeed, such an auxin gradient controls the
$x$-coordinate spot location in such a way that the larger the
overall auxin parameter $k_{20}$, the more spots are likely to
occur. However, as was discussed in \cref{sec:instab2D}, there are
instabilities that occur when an extra spatial dimension is
present. In other words, when an RH cell is modelled as a
two-dimensional flat and oblong cell, certain pattern formation
attributes become relevant that are not present in a 1-D setting. In
particular, 1-D stripes generically break up into localized spots,
which are then subject to possible secondary instabilities.  We now
explore the effect that a 2-D spatial distribution of auxin has on
such localized ROP spots.

\subsection{A 2-D Auxin Gradient}\label{sec:xydependentgrad}

Next, we perform a numerical bifurcation analysis of the RD system
\cref{eq:spotsystem} for an auxin distribution of the form given in
(ii) of~\cref{mod:aux}.  Such a distribution represents a decreasing
concentration of auxin in the $X$-direction, as is biologically
expected, but with a greater longitudinal concentration of auxin along
the midline $Y={L_y/2}$ of the flat rectangular cell than at the edges
$Y=0,L_y$.

To perform a numerical bifurcation study we discretize
\cref{eq:spotsystem} using centered finite differences and we adapt
the 2-D continuation code written in \textsc{MATLAB} given
in~\cite{rankin01}. We compute branches of steady-state solutions of
this system using pseudo arclength continuation, and the stability of
these solutions is computed a posteriori using \textsc{MATLAB}
eigenvalue routines. The resulting bifurcation diagram is shown in
\cref{sf:xygrada}. We observe that there are similarities between this
bifurcation diagram and the one for the 1-D problem shown in Figure~6
of~\cite{bcwg} in that only fold-point bifurcations were found for the
branches depicted. As similar to bifurcation diagrams associated with
homoclinic snaking theory on a finite domain
(cf.~\cite{Avitabile:2015wg,acvf,burke02,Draelants:2015kv,mccalla01}),
we observe a key distinguishing feature: 
solutions belong to a single branch of steady states, undergoing a sequence of
  fold bifurcations and, in some cases, a change in stability. In large intervals of
  $k_{20}$ we observe multistability which, in turn, indicates that hysteretic
  transitions between solutions with varying number of spots can occur.

To explore fine details of this bifurcation behavior, the bifurcation
diagram in \cref{sf:xygrada} has been split into two parts, with the
top and bottom panels for lower and higher values of $k_{20}$,
respectively.  From \cref{sf:xygradb} we observe that a boundary spot
emerges for very low values of the overall auxin rate. This
corresponds to branch 1 of \cref{sf:xygrada}.  As $k_{20}$ is
increased, stability is lost at a fold-point bifurcation which gives
rise to branch 2 of \cref{sf:xygrada}, which gathers a family of
single interior spot steady-states, as shown in
\cref{sf:xygradc}. As~$k_{20}$ increases further, this interior spot
solution persists until stability is lost at a fold-point
bifurcation. Branch~3 in \cref{sf:xygrada} consists of stable
steady-states of one interior spot and one boundary spot, as shown in
\cref{sf:xygradd}.  Furthermore, \cref{sf:xygrade} shows that
  steady-states consisting of two interior spots occur at even
  larger values of $k_{20}$. This corresponds to branch~4 of
\cref{sf:xygrada}. Finally, at much larger values of $k_{20}$,
\cref{sf:xygradf} shows that an additional spot is formed at the
domain boundary. This qualitative behavior associated with increasing
$k_{20}$ continues, and leads to a creation-annihilation cascade
similar to that observed for 1-D pulses and stripes in \cite{bcwg} and
\cite{bacw}, respectively. In other words, overlapping stable branches
yield steady-states consisting of interior and/or boundary spots that
appear or disappear as $k_{20}$ is either slowly increased or
decreased, respectively.

\subsection{Instabilities with a 2-D Spatially-Dependent Auxin Gradient}
\label{subsec:instabilities}

\begin{figure}[t!]
	\begin{center}
\vspace*{-0.2cm} 
		\centering
		\subfigure[]{\includegraphics[height=0.2\textheight]{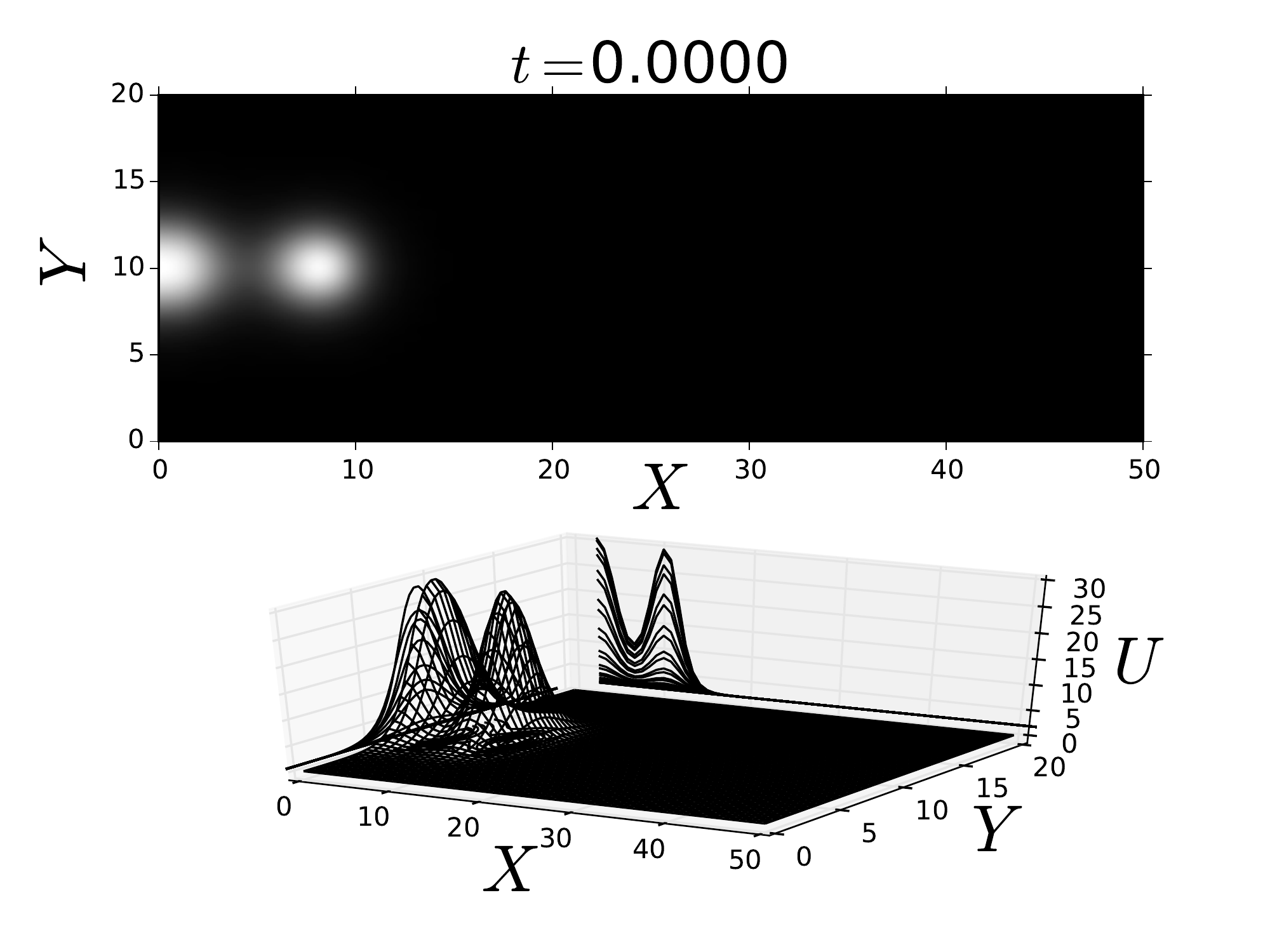}\label{sf:xygradinsta}}
\vspace*{-0.2cm} 
		\centering
		\subfigure[]{\includegraphics[height=0.2\textheight]{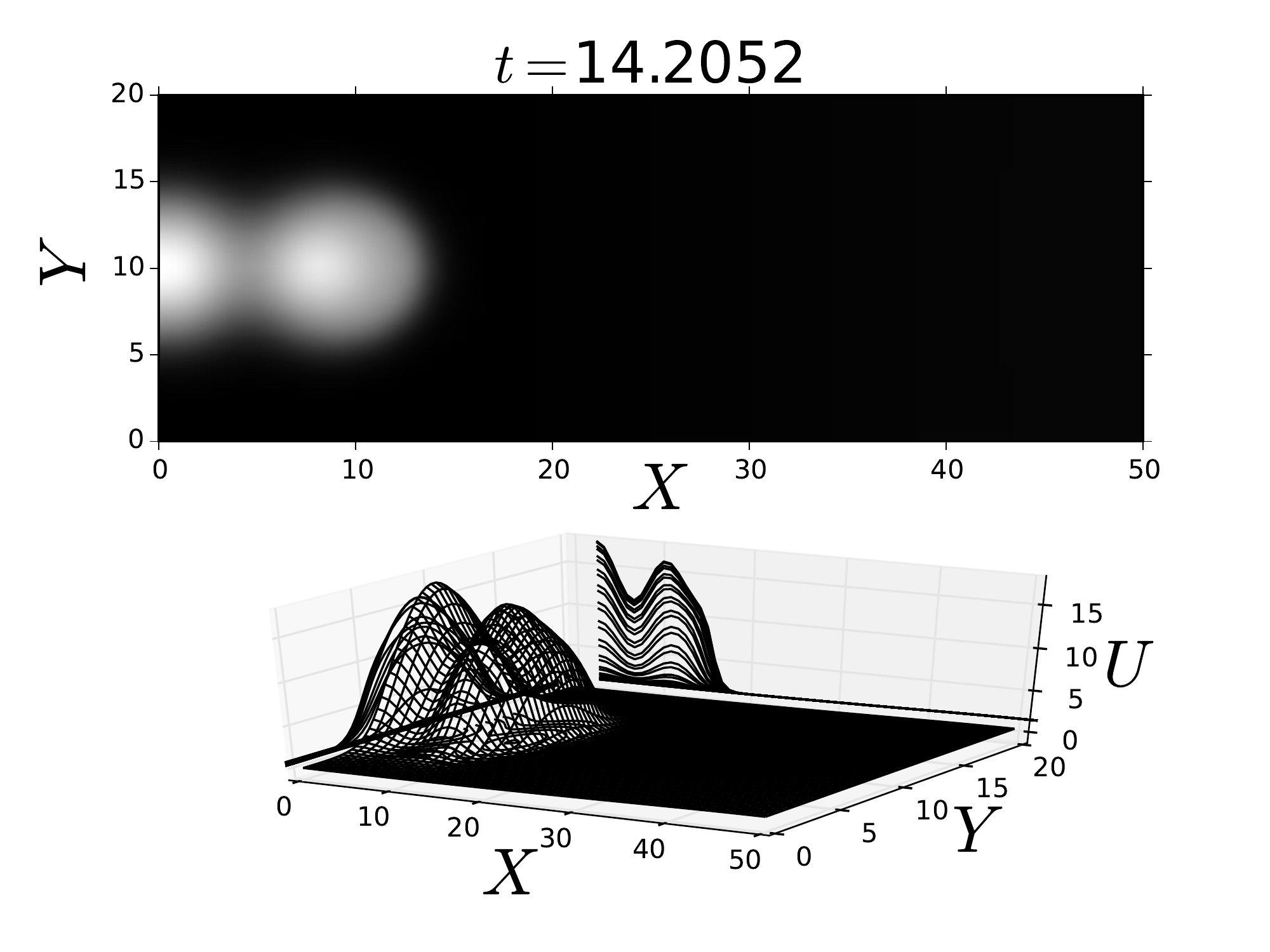}\label{sf:xygradisntb}}
\vspace*{-0.2cm} 
		\centering
		\subfigure[]{\includegraphics[height=0.2\textheight]{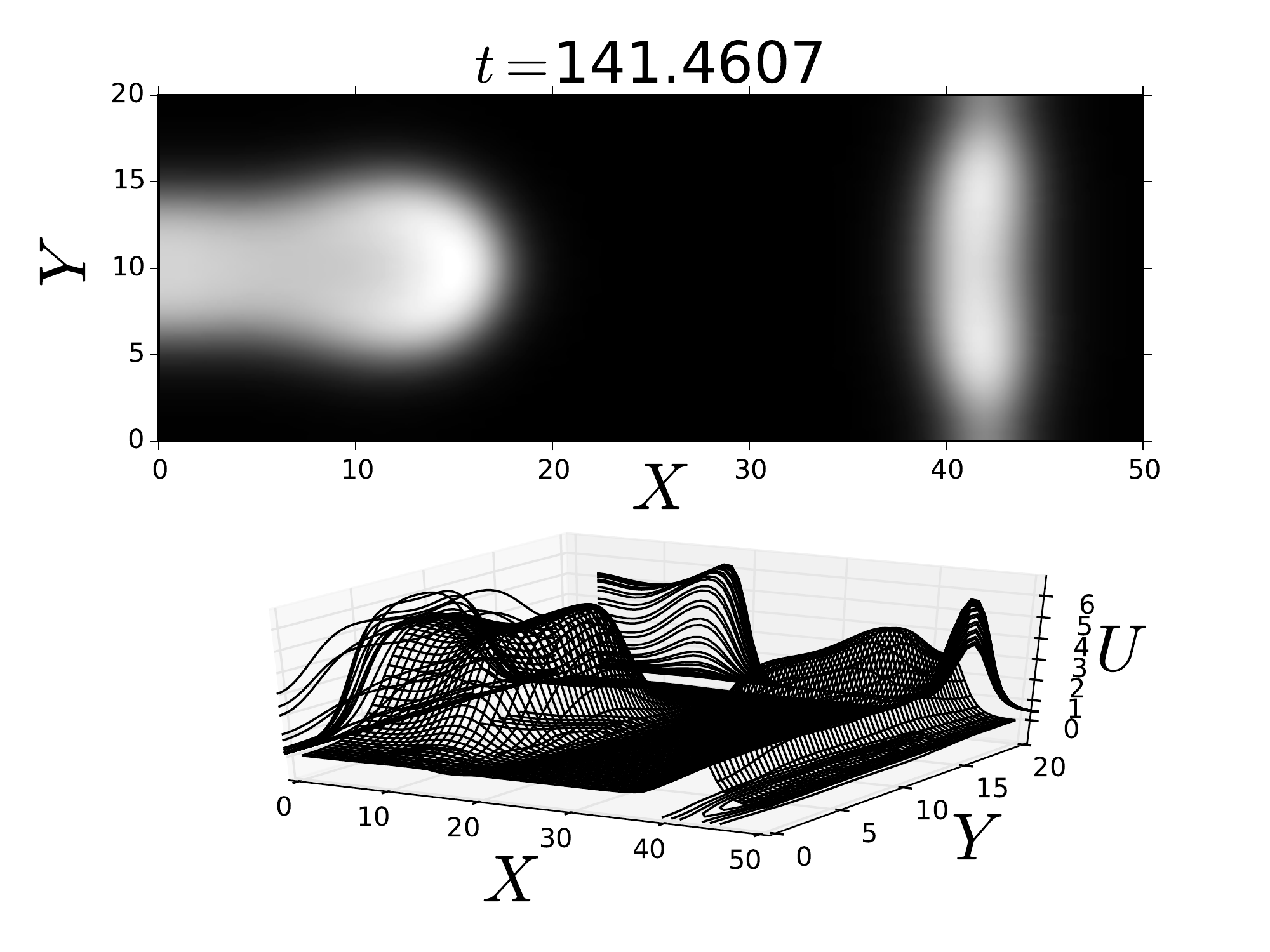}\label{sf:xygradisntc}}
\vspace*{-0.2cm} 
		\centering
		\subfigure[]{\includegraphics[height=0.2\textheight]{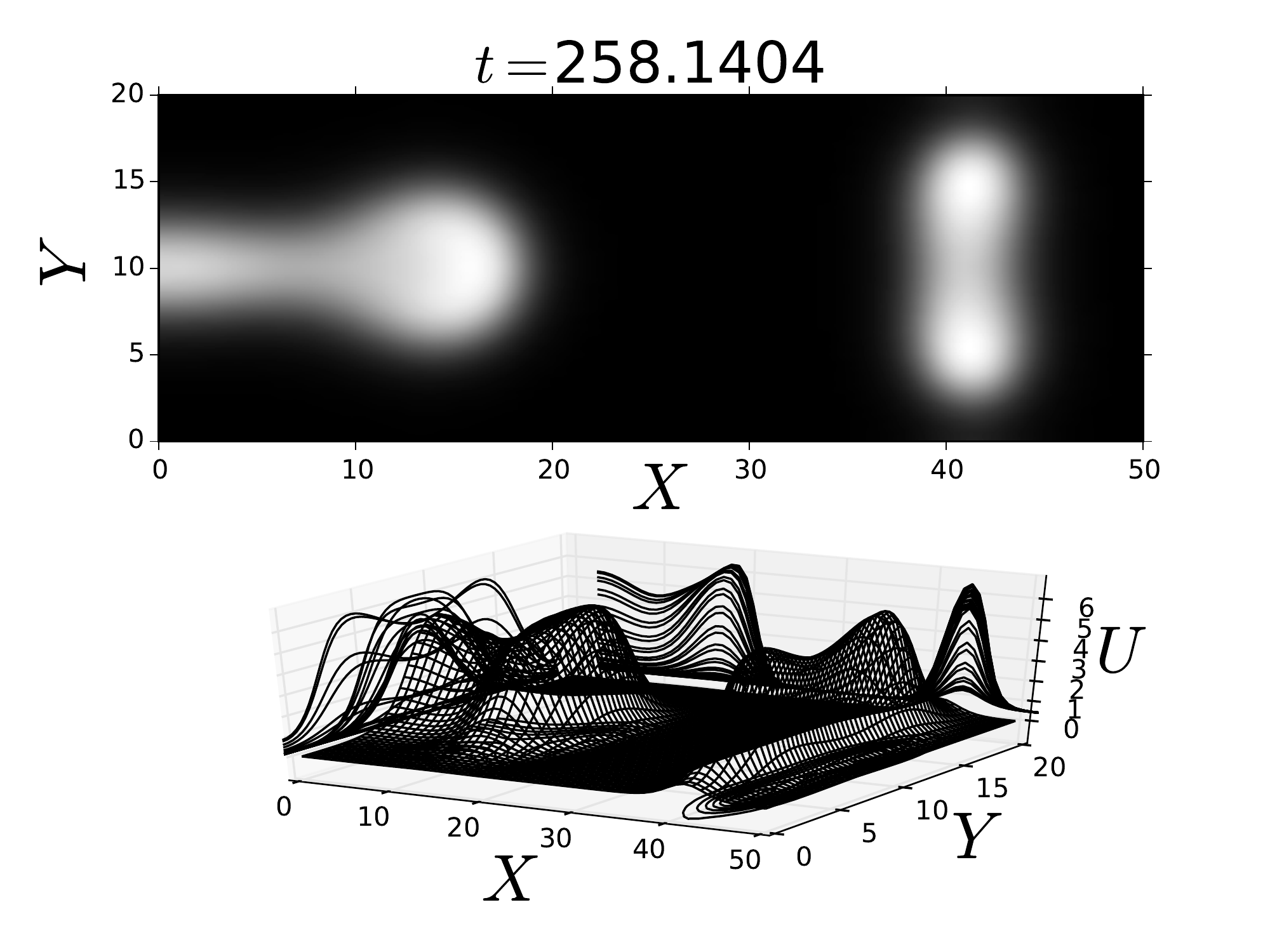}\label{sf:xygradinstd}}
\vspace*{-0.2cm} 
		\centering
		\subfigure[]{\includegraphics[height=0.2\textheight]{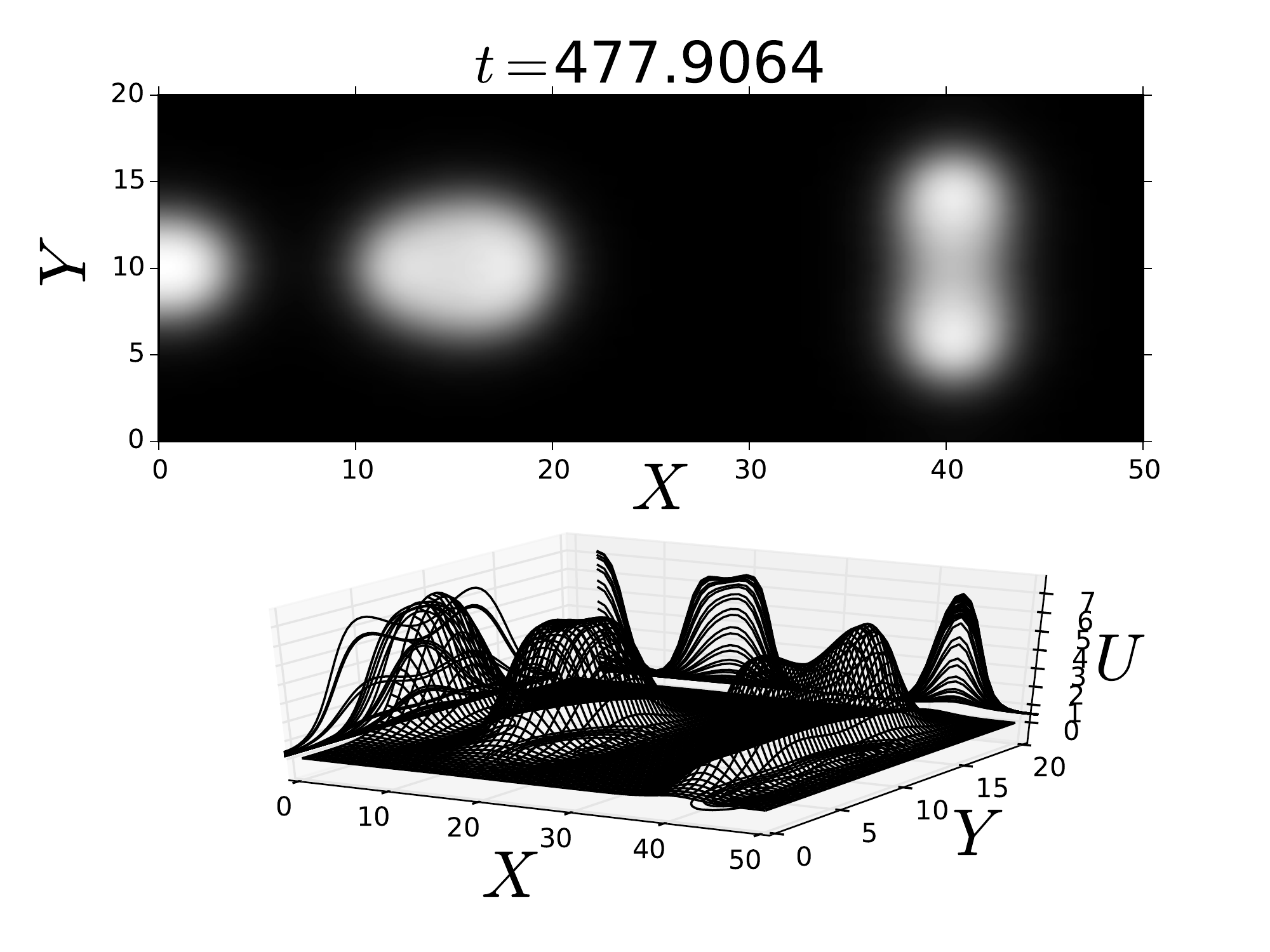}\label{sf:xygradinste}}
\vspace*{-0.2cm} 
		\centering
		\subfigure[]{\includegraphics[height=0.2\textheight]{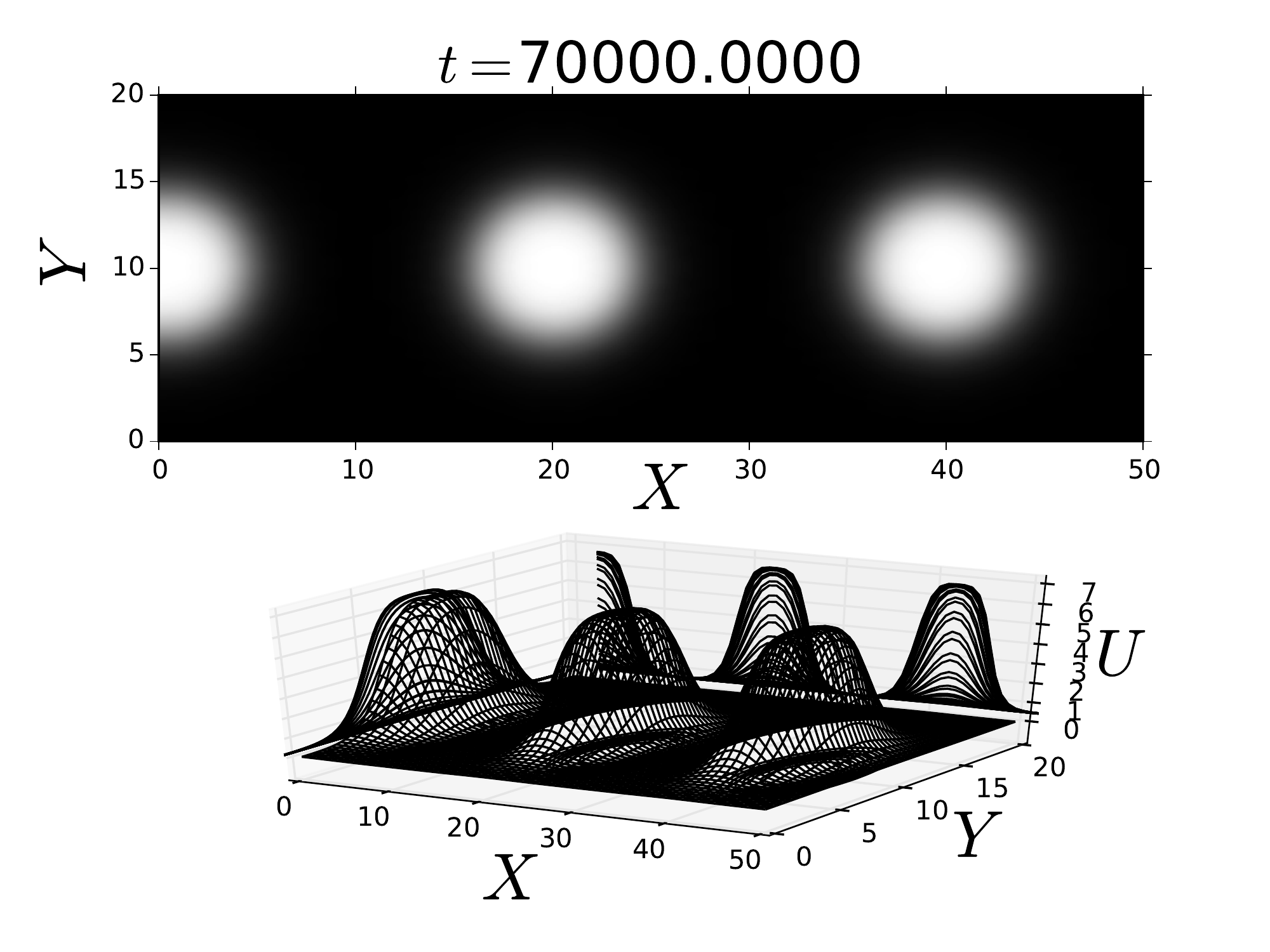}\label{sf:xygradinstf}}
	\end{center}
	\caption{Numerical simulations of the full RD system with
          the 2-D spatial auxin distribution given by
         type~(ii) of~\cref{mod:aux}. The initial condition, shown in (a),
          is a boundary and interior spot unstable steady-state. The
          subsequent time-evolution of this steady-state is: (b) spots
          merging; (c) a new homoclinic stripe is born; (d)~a
          peanut structure emerges; (e) an interior spot arises from a
          collapsed peanut structure, and (f) finally, a two interior and one
          boundary spot stable steady-state. Parameter Set B, as
          given in \cref{tab:tab} with $k_{20}=0.0209$, was
          used.}	\label{fig:xygradinst}
\end{figure}

Similar to the 1-D studies in \cite{bacw,bcwg}, the auxin gradient
controls the location and number of 2-D localized regions of active
ROP.  As the level of auxin increases in the cell, an increasing
number of active ROPs are formed, and their spatial locations are
controlled by the spatial gradient of the auxin distribution.
  Moreover, in analogy with the theory of homoclinic snaking,
  overlapping of stable solution branches occurs, and this leads to a
  wide range of different observable steady-states in the RD
  system. These signature features of the bifurcation structure
  suggest that homoclinic snaking can occur for \cref{eq:spotsystem},
  similar to that observed in \cite{acvf} for a RD system with
  spatially homogeneous coefficients. 
  In passing, we note that the snaking observed in our system is slanted. Slanted
  snaking has been reported in systems with conserved quantities (see for
instance~\cite{dawes01,Thiele:2013er}). The system under consideration does not have
a conserved quantity, and the slanting is caused by the auxin gradient.
  In other words, the
  gradient gives rise to a strongly slanted homoclinic snaking behavior.

 Transitions from unstable to stable branches are determined through
 fold bifurcations, and are controlled by $\mathcal O(1)$ time-scale
 instabilities. To illustrate this behavior, we perform a direct
 numerical simulation of the full RD model \cref{eq:spotsystem} taking
 as initial condition an unstable steady-state solution consisting of
 two rather closely spaced spots, as shown in
 \cref{sf:xygradinsta}. The time-evolution of this initial condition
 is shown in the different panels of \cref{fig:xygradinst}.  We first
 observe from \cref{sf:xygradisntb} that these two spots begin to
 merge at the left domain boundary. As this merging process continues,
 a new-born stripe emerges near the right-hand boundary, as shown in
 \cref{sf:xygradisntc}. This interior stripe is weakest at the top and
 bottom boundaries, due to the relatively low levels of auxin in these
 regions. In \cref{sf:xygradinstd}, the interior stripe is observed to
 give rise to a transversally aligned peanut structure, which remains
 centered at the $Y$-midline for the same reasons as above at the top
 and bottom boundaries, whilst another peanut structure in the
 longitudinal direction occurs near the left-hand boundary. The
 right-most peanut structure slowly collapses to a solitary spot,
 while the left-most form undergoes a breakup instability, yielding
 two distinct spots, as shown in \cref{sf:xygradinste}. 
   Finally, \cref{sf:xygradinstf} shows a steady-state solution with
   three localized spots that are spatially aligned along the
   $Y$-midline.

In the instabilities and dynamics described above, the 2-D auxin
gradient apparently play a very important role, and leads to ROP
activation in the cell primarily where the auxin distribution is
higher, rather than at the transversal boundaries. In particular, although
a homoclinic-type transverse interior stripe is formed, it quickly 
collapses near the top and bottom domain boundaries where the auxin 
gradient is weakest. This leads to a peanut structure centered near the 
$Y$-midline, which then does not undergo a self-replication process
but instead leads to the merging, or aggregation, of the peanut structure
into a solitary spot. The transverse component of the 2-D auxin 
distribution is essential to this behavior. 

We remark that the asymptotic analysis in \cref{sec:spots} and
\cref{sec:dynspot} for the existence and slow dynamics of 2-D quasi
steady-state localized spot solutions can also readily be implemented
for the 2-D auxin gradient type~(ii) of \cref{mod:aux}. The
self-replication threshold in \cref{sec:split} also applies to a 2-D
spatial gradient for auxin. However, for an auxin gradient that
depends on both the longitudinal and transverse directions, it is
difficult to analytically construct a 1-D quasi steady-state stripe
solution such as that done in~\cite{bcwg}. As a result, for a 2-D
auxin gradient, it is challenging to obtain a dispersion relation
predicting the linear stability properties of a homoclinic stripe (see
for instance \cite{doelman01,kolok01}).

\section{Discussion}\label{sec:sum06}

We have used a combination of asymptotic analysis, numerical
bifurcation theory, and full time-dependent simulations of the
governing RD system to study localized 2-D patterns of active ROP
patches, referred to as spots, in a two-component RD model for the
initiation of RH in the plant cell {\em Arabidopsis thaliana}. This
2-D study complements our previous analyses in~\cite{bacw,bcwg} of
corresponding 1-D localized patterns. In our RD model, where the 2-D
domain is a projection of a 3-D cell membrane and cell wall, as shown
in \cref{fig:auxincell3D}, the amount of available plant hormone auxin
has been shown to play a key role in both the formation and number of
2-D localized ROP patches observed, while the spatial gradient of the
auxin distribution is shown to lead to the alignment of these patches
along the transversal midline of the cell. The interaction between
localized active ROP patches and our specific 2-D domain geometry is
mediated by the inactive ROP concentration, which has a long range
diffusion. This long-range spatial interaction can trigger the
formation of localized patches of ROP, which are then bound to the
cell membrane. In addition, as is illustrated
in~\cref{fig:loctwospot}, our results in~\cref{prop:spotsdynamics}
suggest that once ROP patches consisting of localized multi-spots are
quickly formed, their location is controlled by the auxin gradient and
RH cell shape, where the latter seems to play a more relevant part
than in a less realistic 1-D setting (cf.~\cite{bcwg}). In this way, a
sensitive interplay between overall auxin levels and geometrical
properties is crucial to regulate separation of active-ROP patches,
which promotes the onset of RHs in mutants. This may qualitatively
describe findings on the {\em rhd6} mutation of {\em Arabidopsis
  thaliana} reported in~\cite{massuci01}.

From a mathematical viewpoint, our hybrid asymptotic-numerical
analysis of the existence and slow dynamics of localized active ROP
patches, in the absence of any ${\mathcal O}(1)$ time-scale
instabilities, extends the previous analyses of
\cite{xchen01,kolok01,kolok02,rozada02} of spot patterns for
prototypical RD systems by allowing for a spatially-dependent
coefficient in the nonlinear kinetics, representing the spatially
heterogeneous auxin distribution. In particular, we have derived an
explicit DAE system for the slow dynamics of active ROP patches that
includes typical Green's interaction terms that appear in other models
(cf.~\cite{xchen01,kolok01,kolok02,rozada02}), but that now includes
a new term in the dynamics associated with the spatial gradient of the
auxin distribution. This new term contributing to spot dynamics is
shown to lead to a spot-pinning effect (cf.~\cite{mori02}) whereby
localized active ROP patches become aligned along the transverse
midline of the RH cell. We have also determined a specific criterion
that predicts when an active ROP patch will become linearly unstable
to a shape deformation. This criterion, formulated in terms of the
source parameter $S_{j}$ for an individual spot encodes the domain
geometry and parameters associated with the inactive ROP density. In
addition, the effect of auxin on instability and shape of active-ROP
patches is captured by this parameter, and the nonlinear algebraic
system for the collection of source parameters encodes the spatial
interaction between patches via a Green's matrix. When $S_j$ exceeds a
threshold, the predicted linear instability of peanut-splitting type
is shown to lead to a nonlinear spot self-replication event.

Although our asymptotic analysis of localized active ROP patches has
been developed for an arbitrary spatial distribution of auxin, we have
focused our study to two specific, biologically reasonable, forms for
the auxin distribution given in \cref{mod:aux}. The first form assumes
that the auxin concentration is monotone decreasing in the
longitudinal direction, as considered in \cite{bcwg} and \cite{bacw}
and motivated experimentally, with no spatially transverse
component. The second form is again monotone decreasing in the
longitudinal direction, but allows for a spatially transverse
component. This second form is motivated by the fact that our 2-D
model is a projection of a 3-D cell membrane and cell wall. As a
result, it is biologically plausible that some auxin can diffuse out
of the transverse boundaries of our 2-D projected domain, leading to
lower auxin levels near the transverse boundaries than along the
midline of the cell.

We performed a numerical bifurcation analysis on the full RD model for
both specific forms for auxin to study how the level of auxin
determines both the number of ROP patches that will appear and the
increasing complexity of the spatial pattern.  This study showed
similar features as in the 1-D studies in \cite{bacw} in that there
are stable solution branches that overlap (see \cref{sf:expgrada} and
\cref{sf:xygrada}), which allows for hysteretic transitions between
various 2-D pattern types through fold-point bifurcations. In
addition, owing to a homoclinic snaking type structure, a
creation-annihilation cascade yields 2-D spatial patterns where both
spots, baby droplets, and peanut structures can occur. Starting from
unstable steady-states, our full time-dependent simulations
have shown that $\mathcal O(1)$ time-scale competition and
self-replication instabilities can occur leading to transitions
between various spatial patterns. For the auxin distribution model
with a transverse spatial dependence, we showed numerically that the
active ROP patches are more confined to the cell midline and
that there are no active ROP homoclinic stripes.

As an extension to our analysis, it would be worthwhile to consider a
more realistic domain geometry rather than the projected 2-D geometry
considered herein. In particular, it would be interesting to study the
effect of transversal curvature on the dynamics and stability of
localized active ROP patches. In addition, in~\cite{bacw,bcwg} and
here, transport dynamics of ROPs have been modeled under the assumption
of a standard Brownian diffusion process. Even though the experimental
observations of pinning in \cite{dolan01,fischer01} seem to
reproduced, at least qualitatively, by our model, it would be
interesting to extend our model to account for, possibly, more
realistic diffusive processes. In particular, it would be interesting
to consider a hyperbolic-type diffusive process having a finite
propagation speed for signals, as governed by the Maxwell--Cattaneo
transfer law (cf.~\cite{lattanzio,zemskov02}), or to allow for
anomalous diffusion (cf.~\cite{hernandez}).

\vspace*{-0.23cm}
\section*{Acknowledgments}
D.~A., V.~F.~B.--M. and M.~J.~W. were supported by EPSRC grant
EP/P510993/1 (United Kingdom), UNAM--PAPIIT grant IA104316--RA104316
(Mexico) and NSERC Discovery Grant 81541 (Canada), respectively.


\vspace*{-0.23cm}
\bibliographystyle{siamplain}
\bibliography{2Dspotdynbiblio_aug17}
\end{document}